\documentclass[twocolumn,superscriptaddress,showpacs,nofootinbib,notitlepage,preprintnumbers,secnumarabic,amssymb, nobibnotes, aps, prd]{revtex4-2}

\usepackage[utf8]{inputenc}
\usepackage{graphicx}
\usepackage{latexsym,amsmath,amssymb,amsthm,lmodern,float,url,bm}
\usepackage{natbib}
\usepackage{color}
\usepackage{microtype}
\usepackage{import}
\usepackage{bbold}
\usepackage[plain]{fancyref}
\usepackage{varioref}
\usepackage{slashed}
\usepackage{tikz}
\usepackage{braket}
\usetikzlibrary{shapes}
\usetikzlibrary{positioning}
\usepackage[normalem]{ulem}
\usepackage{qcircuit}

\usepackage{graphics}
\usepackage{amsfonts}
\usepackage{multirow}
\usepackage{ulem}

\usepackage{bbm}
\usepackage{mathrsfs}

\newcommand{\one}{\mathbbm{1}}
\newcommand{\add}{\mathfrak{add}}
\newcommand{\mul}{\mathfrak{mul}}

\newcommand{\obs}{\mathcal{O}}
\newcommand{\avg}[1]{\langle #1 \rangle}
\newcommand{\prep}{\mathcal{A}}
\newcommand{\grov}{\mathcal{Q}}
\newcommand{\po}{\mathbbm{U}_{\varphi}}
\newcommand{\iam}{S_{\prep}}
\newcommand{\snaught}{S_{0}}
\newcommand{\subo}{U_{\obs}}
\newcommand{\uobs}{U_{\varphi}}
\newcommand{\Tr}{\text{Tr}}

\usepackage[colorlinks=true,backref=false, linktocpage=true,
citecolor=blue,urlcolor=blue,linkcolor=blue,pdfpagemode=UseOutlines]{hyperref}
\hypersetup{%
  bookmarksnumbered=true,
  pdftitle = {},
  pdfsubject = {},
  pdfauthor = {},
  pdfkeywords = {}
}

\begin{document}

\preprint{FERMILAB-PUB-23-095-QIS-T}

\title{Quantum mean estimation for lattice field theory}
\author{Erik J. Gustafson}
\email{egustafs@fnal.gov}
\affiliation{Fermi National Accelerator Laboratory, Batavia,  Illinois, 60510, USA}
\author{Henry Lamm}
\email{hlamm@fnal.gov}
\affiliation{Fermi National Accelerator Laboratory, Batavia,  Illinois, 60510, USA}
\author{Judah Unmuth-Yockey}
\email[Corresponding author: ]{jfunmuthyockey@gmail.com}
\affiliation{Fermi National Accelerator Laboratory, Batavia,  Illinois, 60510, USA}
\begin{abstract}
    We demonstrate the quantum mean estimation algorithm on Euclidean lattice field theories. This shows a quadratic advantage over Monte Carlo methods which persists even in presence of a sign problem, and is insensitive to critical slowing down.  The algorithm is used to compute $\pi$ with and without a sign problem, a toy U(1) gauge theory model, and the Ising model. The effect of $R_{Z}$-gate synthesis errors on a future fault-tolerant quantum computer is investigated. 
\end{abstract}

\maketitle

\section{Introduction}

Computations in lattice field theory (LFT) are typically framed within the formalism of statistical mechanics~\cite{montvay_munster_1994,gattringer,kogut:1979}.  This is possible by analytically continuing the field theory from Minkowski to Euclidean spacetime, which transforms complex phases into probability weights.  This changes the quantum path integral into a classical partition function, $Z = \sum_{i} e^{-S(g_i)}$, which is a sum over all configurations of the degrees of freedom, $g_i$, weighted by the lattice action, $S \geq 0$.
Quantum expectation values of observables, $\obs$, then transform into statistical averages
\begin{align}
\label{eq:obsZ}
    \avg{\obs} = \frac{1}{Z} \sum_{i} \obs(g_i) \; e^{-S(g_i)}.
\end{align}
Summing over all configurations is, in general, impractical.  Instead, Monte Carlo methods sample a finite number, $N$, of them with an error, $\sigma$, on $\langle \mathcal{O}\rangle$ scaling asymptotically like $\sigma \sim 1/ \sqrt{N}$.  Often, only a few thousand samples yield precision predictions for complex observables like hadronic form factors~\cite{FermilabLattice:2021cdg} or QCD contributions to the muon $g-2$~\cite{Borsanyi:2020mff}.  

However, there are limitations.  These include situations where there is a sign problem~\cite{deForcrand:2009zkb,Tripolt:2018xeo}, and critical slowing down~\cite{Schaefer:2010hu}.  The sign problem crops up at finite fermion density~\cite{Philipsen:2019rjq} and when simulating real-time dynamics~\cite{Alexandru:2020wrj}.  Critical slowing down is a consequence of running Monte Carlo algorithms with updates that neglect long-distance correlations.  Such a situation appears as one tries to approach the continuum limit or when studying topological observables~\cite{DelDebbio:2004xh}. In the case of the sign problem, the required $N$ scales exponentially in model parameters, and with critical slowing down it scales as a power.
These obstacles have prompted new classical algorithms that attempt to address these issues including: cluster algorithms~\cite{swendsen-wang:87,wolff:89}, dual variables approaches~\cite{RevModPhys.52.453,Marchis:2018}, tensor networks~\cite{Orus2019,RevModPhys.94.025005}, complexification~\cite{Alexandru:2020wrj}, and density of states methods~\cite{klangfeld:2016}.  Despite these successes, it is unlikely that general solutions exist cf.~\cite{Troyer:2004ge}.

Through quantum computers it is possible to avoid some of these limitations.  This result comes fundamentally from the abilities of quantum computers to enumerate an exponential number of states via quantum superposition, and their capacity to generate entanglement.  Notable quantum algorithms which harness these properties include the quantum Fourier transform~\cite{Cleve_1998,PhysRevLett.76.3228,coppersmith2002approximate}, quantum phase estimation (QPE)~\cite{kitaev:1995,Chapeau-Blondeau2020,Smith:2022,nielsen_chuang_2010}, and Hamiltonian simulation.  This last class of algorithms could further allow for tremendous advances in simulating real-time dynamics for LFT~\cite{Feynman:1981tf,Lloyd1073,Jordan:2011ne,Jordan:2017lea,klco2021standard,Bauer:2022hpo,Zohar:2012ay,Zohar:2012xf,Zohar:2013zla,Zohar:2014qma,Zohar:2015hwa,Zohar:2016iic,Klco:2019evd,Ciavarella:2021nmj,Bender:2018rdp,Liu:2020eoa,Hackett:2018cel,Alexandru:2019nsa,Yamamoto:2020eqi,Haase:2020kaj,Armon:2021uqr,PhysRevD.99.114507,Bazavov:2015kka,Zhang:2018ufj,Unmuth-Yockey:2018ugm,Unmuth-Yockey:2018xak,Kreshchuk:2020dla,Kreshchuk:2020aiq,Raychowdhury:2018osk,Raychowdhury:2019iki,Davoudi:2020yln,Wiese:2014rla,Luo:2019vmi,Brower:2020huh,Mathis:2020fuo,Singh:2019jog,Singh:2019uwd,Buser:2020uzs,Bhattacharya:2020gpm,Barata:2020jtq,Kreshchuk:2020kcz,Ji:2020kjk,Bauer:2021gek,Gustafson:2021qbt,Hartung:2022hoz,Grabowska:2022uos,Murairi:2022zdg,Alexandru:2019nsa,Ji:2020kjk,Ji:2022qvr,Alexandru:2021jpm,Gustafson:2022xlj,Gustafson:2019mpk,Bender:2018rdp,Lamm:2019bik,Alam:2021uuq,Fromm:2022vaj,Yeter-Aydeniz:2018mix,Gustafson:2022xdt,Carena:2022kpg}.  Investigation of quantum algorithms have also lead to faster classical algorithms~\cite{10.1145/3313276.3316310,Arrazola2020quantuminspired}.

Another quantum algorithm relevant specifically to this work is quantum mean estimation (QME)~\cite{kothari:2022,Prasanth:2021,Ham21}.  Quantum mean estimation is capable of quadratically reducing the asymptotic scaling of $\sigma$ to $\sim 1/N$.  It achieves this using QPE, and by using superposition to incorporate a full probability distribution into calculations of $\braket{\obs}$.
It has been developed for boolean variables, positive bounded real variables~\cite{PhysRevA.76.030306,Grinko_2021}, bounded real variables~\cite{Prasanth:2021}, and unbounded real variables~\cite{montanaro:2015,kothari:2022,Ham21}.

In this article, we will demonstrate how and when QME can be used to improve classical LFT calculations that use Monte Carlo sampling.  In Sec.~\ref{sec:meqc}, we detail how to use QME on classical statistical models, and provide circuits to construct the QME algorithm.  In Sec.~\ref{sec:examples}, we provide numerical results including circumstances with sign problems, and investigate the effects of noise.  In Sec.~\ref{sec:cvq}, we compare traditional sampling methods to the QME algorithm.  We conclude with a discussion and directions for future work in Sec.~\ref{sec:conclusion}.

\section{Quantum Mean Estimation}
\label{sec:meqc}
In this section, we will demonstrate how a quantum computer using QME can provide an estimate of $\langle \mathcal{O} \rangle\in[-1,1]$ with fixed precision and quadratically fewer resources than traditional sampling using the method of Ref.~\cite{Prasanth:2021}.

\subsection{Overview of QME}
\label{sec:qme-overview}

Estimating $\braket{\obs}$ on a quantum computer uses QPE as a subroutine.  By encoding $\obs(g_i)$ into the phases of states and running QPE on a judiciously chosen unitary and starting state, a phase is returned approximating the mean.  Alternatively, the mean can be stored as the amplitude of some target state.  This amplitude can be approximated and returned---again using QPE---using quantum amplitude estimation~\cite{Brassard_2002,Grinko_2021,PhysRevA.76.030306,montanaro:2015,Ham21}.

\begin{figure}
    \centering
    \includegraphics[width=0.7\linewidth]{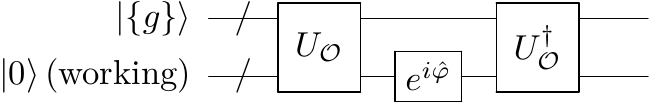}
    \caption{The circuit for $\uobs$ which applies the phases for QME by calculating the phases in a ``working'' register, and then extracting them with a diagonal operator.  $\subo$ is model-dependent however its general action is described in Eq.~\ref{eq:sudo-action}.}
    \label{fig:uobs}
\end{figure}

Following Ref.~\cite{Prasanth:2021}, for a given unitary---or phase oracle---in diagonal form,
\begin{align}
\label{eq:spectral-po}
    U_{\varphi} \equiv \sum_{i} e^{i \varphi(g_i)} \ket{g_i}\bra{g_i},
\end{align}
QME returns
\begin{align}
\label{eq:cosavg}
\nonumber
    \Re[\bra{\psi_{0}} U_{\varphi} \ket{\psi_{0}}] &= 
    \Re[\bra{\psi_{0}} \sum_{i} e^{i \varphi(g_i)} \ket{g_i}\braket{g_i | \psi_{0}}] \\ \nonumber
    &= \sum_{i} |\braket{\psi_{0} | g_i}|^{2}   \Re[e^{i \varphi(g_i)}] \\
    &= \sum_{i} |\braket{\psi_{0} | g_i}|^{2} \cos(\varphi(g_i)),
\end{align}
where $\ket{\psi_{0}}$ is a generic initial state.\footnote{One can also estimate $\bra{\psi_{0}} U_{\varphi} \ket{\psi_{0}}$, but we focus on the real part.}  For appropriate $\varphi$s, one can calculate averages of real numbers $\in [-1,1]$.
We provide here a method to construct $\uobs$ through qubit arithmetic, by computing the appropriate $\varphi_i$ into a register, applying a phase based on that register's value, and then uncomputing the phase register back to $|0\rangle$.  The computation of the $\varphi_i$ is done by a new oracle, $\subo$ which for a given configuration, $g_i$, places the state $\ket{\varphi}$ into a separate register.  In terms of states and operators,
\begin{align}
\label{eq:sudo-action}
    \subo \ket{g_i} \ket{0} = \ket{g_i} \ket{\varphi(g_i)}.
\end{align}
The new ``working'' register stores the value of the appropriate phases to apply to each configuration.
From that state, the proper phase can be applied via
\begin{align}
    \one \otimes e^{i \hat{\varphi}} \ket{g_i} \ket{\varphi(g_i)} = 
    e^{i \varphi(g_i)} \ket{g_i} \ket{\varphi(g_i)},
\end{align}
and the working register can be uncomputed,
\begin{align}
    e^{i \varphi(g_i)} \subo^{\dagger} \ket{g_i} \ket{\varphi(g_i)} = e^{i \varphi(g_i)} \ket{g_i} \ket{0}.
\end{align}
This entire circuit can be captured by a single unitary $\uobs = \subo^{\dagger} (\one \otimes e^{i \hat{\varphi}}) \subo$, as seen in Fig.~\ref{fig:uobs}.

Using QME corresponds to recasting Eq.~\eqref{eq:obsZ} in the form of Eq.~\eqref{eq:cosavg}.  The relevant identification is to see that the product $|\braket{\psi_0 | g_i}|^{2} \cos(\varphi(g_i)) \sim (e^{-S(g_i)}/Z) \mathcal{O}(g_i)$.  This identification in turn allows for flexibility in defining $\ket{\psi_{0}}$ and the phases $\varphi(g_i)$.  Since
 we can identify $|\braket{\psi_0 | g_i}|^{2}$ in Eq.~\eqref{eq:cosavg} as the probability for configuration $g_i$ used in the average, a natural correspondence is given by $|\braket{\psi_0 | g_i}|^{2} \sim e^{-S(g_i)}/Z$ as well as $\cos(\varphi(g_i)) \sim \mathcal{O}(g_i)$, which can be used to define $\varphi(g_i)$.  However, one is allowed the freedom to reweigh parts of the observables into the weights, and \emph{vice versa}, to optimize. By rescaling, this method can compute averages in any finite range.

To actually compute Eq.~\eqref{eq:cosavg} requires preparing a state $|\psi_0\rangle$, and the use of QPE with $U_\varphi$.  We will present two separate implementations of QME using two different $|\psi_0\rangle$. One directly prepares the Boltzmann weights, and the other shifts the difficulties of preparing such states into the computation of observables through reweighting.

\subsection{Using state preparation}
\label{sec:state-prep}

We can use any state-preparation method, whether it be through efficient classical methods~\cite{grover:2002},  black-box methods~\cite{Wang_2022,Bausch_2022,PhysRevLett.122.020502}, using the quantum singular value transform~\cite{mcardle:2022}, or with state search~\cite{PhysRevLett.85.1334} to construct an appropriate $\ket{\psi_0}$.  Here, we prepare $\ket{\psi_0}$ using arithmetical oracles such that $|\braket{\psi_0|g_i}|^{2} = e^{-S(g_i)} / Z$.  We will denote this state preparation method as $\prep^{\text{SP}}$.

To prepare a quantum state with the desired probability distribution, let $0<w(g_i) \leq 1$ be the weights of the system, and $Z = \sum_{i}^M w(g_i)$ with $M$ the total number of configurations.  We can prepare a maximal superposition over all configurations, along with an ancilla in the state $|+\rangle$:
\begin{align}
    \ket{\psi} &= \frac{1}{\sqrt{M}}\ket{+}\sum_{i} \ket{g_i} = \frac{1}{\sqrt{2M}}(\ket{0}+\ket{1})\sum_{i} \ket{g_{i}}.
\end{align}
Next, we apply a controlled-$U_\phi$ with $\phi_i = \arccos(\sqrt{w_i})$ using the procedure from Sec.~\ref{sec:qme-overview},
\begin{align}
\label{eq:apply-phases}
    \ket{\psi} = \frac{1}{\sqrt{2M}}\left( \ket{0}\sum_{i} e^{i \phi_i} \ket{g_i} +  \ket{1} \sum_{i} e^{-i \phi_i} \ket{g_i} \right).
\end{align}
The phases used in state preparation are denoted with $\phi$, and those used in encoding observables with $\varphi$, since both use the same phase oracle.  
After applying a Hadamard to the ancilla we can rewrite the whole state as,
\begin{align}
    \ket{\psi} = \sqrt{\frac{Z}{M}} \ket{\psi_0} + \sqrt{\frac{M - Z}{M}}\ket{\psi_1}
\end{align}
where $\ket{\psi_0}$ is the desired state,
\begin{align}
\nonumber
    \ket{\psi_0} &= \frac{1}{\sqrt{Z}} \ket{0} \sum_i \cos(\phi_i) \ket{g_i} \\
    &=  \ket{0} \sum_i \sqrt{\frac{w_i}{Z}} \ket{g_i}
\end{align}
and $\ket{\psi_1}$ is an orthogonal state,
\begin{align}
\nonumber
    \ket{\psi_1} &= \frac{i}{\sqrt{M - Z}} \ket{1} \sum_i \sin(\phi_i) \ket{g_i} \\
    &= i  \ket{1} \sum_i \sqrt{\frac{1-w_{i}}{M - Z}} \ket{g_i}.
\end{align}
$\ket{\psi_0}$ contains the target probability distribution; however, $\ket{\psi_1}$ needs to be removed.  We use fixed-point oblivious amplitude amplification~\cite{yoder:2014,Berry:2014,gilyen:2019}, which requires a lower-bound on the amplitude, to isolate the zero-ancilla state.   
A crude lower bound is given by assuming every $g_i$ comes with the smallest possible weight, then,
\begin{align}
    Z = \sum_{i} w(g_i) > \sum_{i} w_{\text{min}} = M w_{\text{min}} > 0.
\end{align}
We find that $\sqrt{Z/M} > \sqrt{w_{\text{min}}}$, and we can use $\sqrt{w_{\text{min}}}$ as a lower bound for amplitude amplification.  Asymptotically the algorithm will take $O(\log(1/\epsilon) / \sqrt{w_{\text{min}}})$ amount of time to achieve a desired $\epsilon$ accuracy~\cite{yoder:2014}.  Deriving a tighter lower bound would reduce the algorithmic time; however, since the actual amplitude is $\sqrt{Z/M}$, at best we can expect $O(\sqrt{M}\log(1/\epsilon) / \sqrt{Z})$ time.  
The algorithm requires no conditional measurements, or ``repeat-until-success'' steps, and is captured by a unitary matrix.  After amplitude amplification we end with the desired state,
\begin{align}
\label{eq:want-state}
    \ket{\psi_0} = \ket{0} \sum_i \sqrt{\frac{w_i}{Z}} \ket{g_i}.
\end{align}
The QME algorithm requires an ancilla in the state $|+\rangle$, so after applying a Hadamard gate, the initial state is
\begin{align}
    \ket{\psi_0} = \ket{+} \sum_i \sqrt{\frac{w_i}{Z}} \ket{g_i}.
\end{align}
Having prepared $|\psi_0\rangle$, we can use QPE to perform the mean estimation, but this state preparation is potentially expensive; therefore, we discuss an alternative approach.

\subsection{Using reweighting}
\label{sec:reweigh}
Previously, we discussed how there exists freedom in associating the product $\mathcal{O}(g_i)e^{-S(g_i)} / Z$ to $|\braket{\psi_0|g_i}|^2$ and $\cos(\varphi(g_i))$.  The choices made determine the state preparation $\prep$ and the specific means computed. This \emph{reweighting}~\cite{Ferrenberg:1988yz} can be understood by introducing into Eq.~\eqref{eq:obsZ} a second probability distribution $q(g_i)$:
\begin{align}
\label{eq:ratavg}
    \avg{\obs} &= \frac{ \sum_{i} \obs(g_i) e^{-S(g_i)}}{\sum_{i} \;  e^{-S(g_i)}} = \frac{ \sum_{i} \frac{\obs(g_i)e^{-S(g_i)}}{q(g_i)}q(g_i)}{\sum_{i} \;  \frac{e^{-S(g_i)}}{q(g_i)}q(g_i)} \notag\\&= \frac{ \sum_{i} \frac{\obs(g_i)e^{-S(g_i)}}{q(g_i)}q(g_i)}{\sum_i q(g_i)}\frac{\sum_i q(g_i)}{\sum_{i} \;  \frac{e^{-S(g_i)}}{q(g_i)}q(g_i)}\notag\\&= \frac{\frac{1}{Z_q} \sum_{i} \tilde{\mathcal{O}}(g_i)q(g_i)}{\frac{1}{Z_q}\sum_{i} \;  \tilde{\mathcal{R}}\,q(g_i)}=\frac{\braket{\tilde{\mathcal{O}}}_q}{\braket{\tilde{\mathcal{R}}}_q}
\end{align}
where $Z_q=\sum_{i} q(g_i)$ and $\braket{\tilde{\mathcal{R}}}$ is a necessary reweighting factor. So instead of $\ket{\psi_0}$ with the distribution $e^{-S(g_i)}$, we can construct a different one by shuffling part of the distribution into the observable, and then computing the ratio of two means. One nice choice is the uniform distribution $q(g_i) = 1/M$ which can be prepared by $\prep^{\text{RW}} = H^{\otimes}$.  With two different state preparation methods described, we now discuss the role of QPE.

\subsection{Quantum phase estimation}

The QPE algorithm returns an estimate of a phase $\chi$ of an eigenvector $\ket{q}$ for a unitary $\grov$ such that $\grov \ket{q} = e^{i\chi}\ket{q}$.  Thus, QPE expects an input state, or a state preparation method $\prep$, a unitary matrix $\grov$ whose eigenvalues are of interest, and a result register of $N_r$ qubits prepared in a maximal superposition. It then returns the best, approximate phase with a probability $\geq 4/\pi^2$~\cite{Cleve_1998,nielsen_chuang_2010}.  

We can see the effect of the input state \emph{not} being an eigenstate of $\grov$ by inspecting its spectral decomposition,
\begin{align}
    \grov = \sum_{q} e^{i \chi(q)} \ket{q}\bra{q}.
\end{align}
For a generic state $\ket{\psi_{0}}$,
\begin{align}
    \grov\ket{\psi_0} = \sum_{q}e^{i \chi(q)} \braket{q | \psi_{0}} \ket{q},
\end{align}
and QPE operating on $\ket{\psi_0}$ returns angle $\chi(q)$ for each eigenstate into the result register in superposition with probability $|\braket{q | \psi_{0}}|^{2}$.
The accuracy of the returned phase and how likely that phase is to be measured, is specified by $N_r$. The possible phase values are restricted to the possible binary fractions expressible by $N_{r}$ qubits.

\begin{figure}
    \centering
    \includegraphics[width=0.6\linewidth]{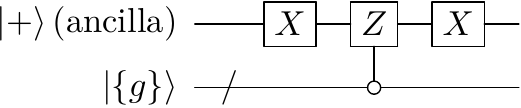}
    \caption{The circuit for $-S_{0}$ which performs the operation, $\one - 2\ket{0}\bra{0}$, which applies a negative sign to the ``all zero'' state.  The target qubit is defined as the least significant bit.}
    \label{fig:s0}
\end{figure}

\begin{figure}
    \centering
    \includegraphics[width=0.7\linewidth]{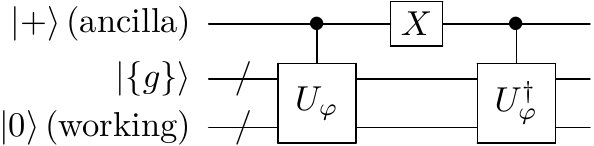}
    \caption{The circuit for $\po$ which describes the phase oracle used in QME.  The sub-circuit of $\uobs$ is provided in Fig.~\ref{fig:uobs}.}
    \label{fig:fatU}
\end{figure}

\begin{figure}
    \centering
    \includegraphics[width=0.7\linewidth]{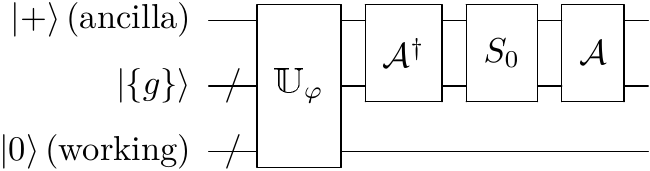}
    \caption{The quantum circuit for $\grov$.  Here the circuit is shown in application to the initial prepared state.}
    \label{fig:qcal}
\end{figure}

\begin{figure}
    \centering
    \includegraphics[width=\linewidth]{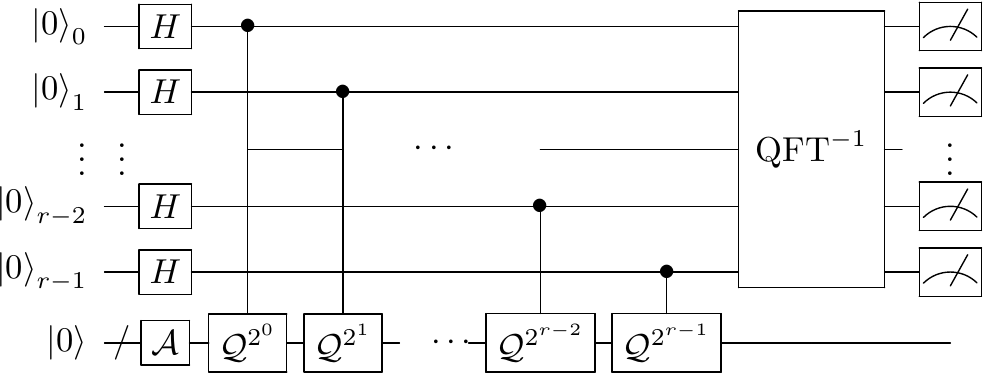}
    \caption{The quantum circuit for QPE with $N_r=r$.  $\text{QFT}^{-1}$ is the  inverse quantum Fourier transform}
    \label{fig:qpe}
\end{figure}

\subsection{Synthesis of parts}
Here we relate $\prep$ to $\grov$ to see how QME emerges from QPE.
In Secs.~\ref{sec:state-prep} and~\ref{sec:reweigh} we discussed two $\prep$ methods.  With those in mind, we will now construct $\grov$ used in QME.
The unitary $\grov$ is given by the product $\grov \equiv \iam \po$ with a phase oracle, $\po$, and the operator, $\iam$.
The form of $\iam$ is: $\iam \equiv \prep \snaught \prep^{\dagger}$, with $\snaught = 2\ket{0}\bra{0} - \one$.  The circuit for $-\snaught$ can be seen in Fig.~\ref{fig:s0}, and consists of applying a minus sign to the ``all zeros'' state.
The phase oracle of Ref.~\cite{Prasanth:2021} has two terms,
\begin{equation}
    \po =  \ket{0}\bra{0} \otimes \uobs + \ket{1}\bra{1} \otimes \uobs^{\dagger},
\end{equation}
and is responsible for applying the phases encoding $\obs$ onto the states.
The oracle $\uobs$ is described in Sec.~\ref{sec:qme-overview}.
The circuit for $\po$ can be seen in Fig.~\ref{fig:fatU}, and with it $\grov$ can be completely defined.  The circuit for $\grov$ can be seen in Fig.~\ref{fig:qcal}.  
The entire circuit for QPE---and through it QME---can be seen in Fig.~\ref{fig:qpe}.

With the relationship between $\prep$ and $\grov$ established the desired mean must come about from an eigenvalue of $\grov$.  To see this, let us revisit Eq.~\eqref{eq:cosavg}, and define the angle $\theta$ along with $\cos(\theta)$ through,
\begin{align}
\label{eq:theta-defined}
    \cos(\theta) &\equiv \sum_{i} |\braket{\psi_{0} | g_i}|^{2} \cos(\varphi(g_i)).
\end{align}
Written in this form we see that Eq.~\eqref{eq:theta-defined} computes the average value of cosine.  Now, consider the initial state $\ket{\psi_0} \equiv \prep \ket{0}$.  While not an eigenstate of $\grov$, it is a simple linear combination of eigenstates of $\grov$ (see Ref.~\cite{Prasanth:2021}),
\begin{align}
    \ket{\psi_0} = \frac{1}{\sqrt{2}} (\ket{\eta_{+}} - \ket{\eta_{-}})
\end{align}
with $\grov \ket{\eta_{\pm}} = e^{\pm i \theta} \ket{\eta_{\pm}}$.  Since $\ket{\psi_{0}}$ is constructed from eigenvectors of $\grov$, applying QPE to $\ket{\psi_{0}}$ returns either of the two phases associated with those states, $\theta$ or $2\pi - \theta$, with equal probability.  Then by taking the cosine of the output angle---which is insensitive to $\theta$ or $2\pi - \theta$---we recover the average in Eq.~\eqref{eq:theta-defined}.

To summarize,
\begin{itemize}
    \item[1)] Using a qubit encoding of $g_i$, construct $\prep$.  This can either be done by making a state with the correct probability distribution as in Sec.~\ref{sec:state-prep}, or by reweighting as in  Sec.~\ref{sec:reweigh}.
    \item[2)] Construct $\po$:
    \begin{itemize}
        \item Construct $\subo$.  This is model- and observable-dependent but consists of qubit arithmetic.  Define the phases, $\varphi(g_i)$, as needed.
        \item Construct $\uobs$ from $\subo$ and $e^{i \hat{\varphi}}$ using Fig.~\ref{fig:uobs}.
        \item Construct $\po$ from $\uobs$  using Fig.~\ref{fig:fatU}.
    \end{itemize}
    \item[3)] Construct $\grov$ from $\prep$, $\po$, and $\snaught$ using Fig.~\ref{fig:qcal}.
    \item[4)] Perform QPE with $\prep$ and $\grov$ as in Fig.~\ref{fig:qpe}.  The output is an estimated angle, $\omega$, where $\cos(\omega) \approx \sum_{i} |\braket{\psi_{0} | g_i}|^{2} \cos(\varphi(g_i))$.
\end{itemize}
With the method now explained, we proceed with some examples to help solidify each step above.

\section{Applications}
\label{sec:examples}

We consider several numerical examples: computing $\pi$ with and without negative weights, a toy U(1) gauge theory, and the two-dimensional Ising model. Then, a general QME is described for lattice gauge theories.  Finally, how $R_{z}$-gate synthesis errors affect QME is discussed.  All of the numerical results in this section were performed on the IBM \textsc{Qiskit} QASM simulator using 1024 shots.

\subsection{\texorpdfstring{$\pi$}{Pi} with and without a sign problem}
\label{sec:pi}
\begin{figure*}
    \centering
    \includegraphics[width=\linewidth]{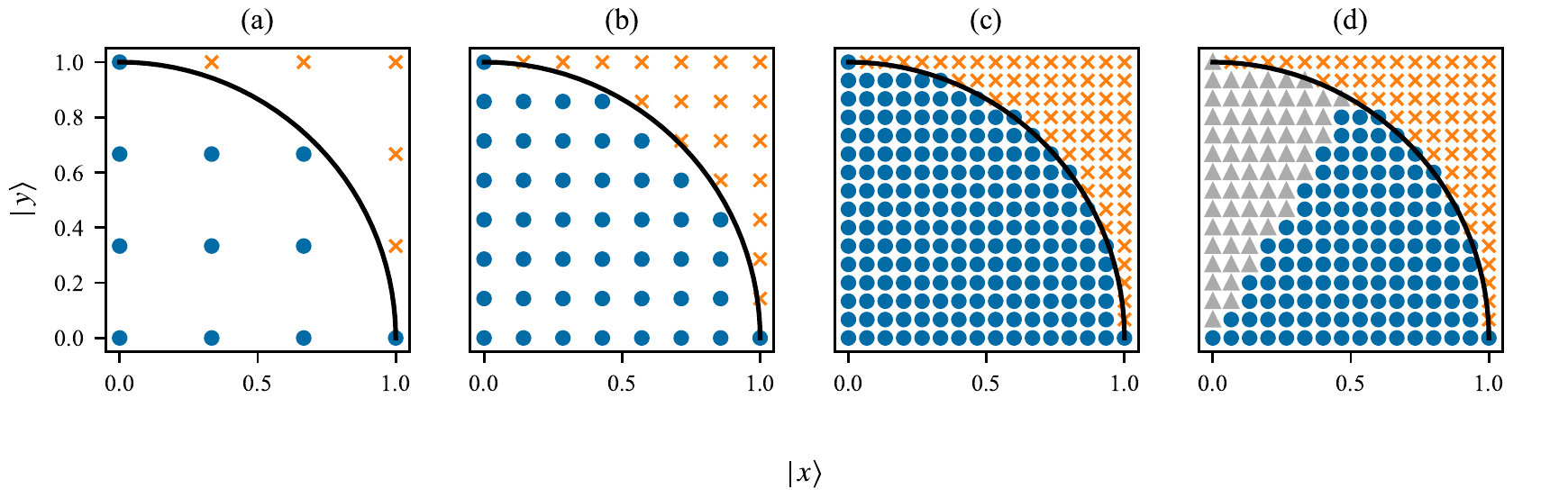}
    \caption{(a)-(c) Grids for $N_g=4, 6, 8$ from left to right, respectively, used in $\ket{x}$ and $\ket{y}$ registers. (d) Grid for $N_{g}=8$ in the case of a sign problem.  The circles have weight $+1$, while the triangles have weight $-1$.}
    \label{fig:pi-grid}
\end{figure*}

In the Monte Carlo estimate of $\pi$, one randomly chooses points, $x_g, y_g$ in the $x$-$y$ plane---which we assume to be a grid (see Fig.~\ref{fig:pi-grid})---such that $x_g, y_g \in [0,1)$.  A quarter circle in that plane has an area of $\pi / 4$, while the square has area one.  Then we expect that on average the fraction of points that fall within the boundary of the circle is $\approx \pi / 4$.  We can decide this for each $x_g$-$y_g$ pair if $x_{g}^{2} + y_{g}^{2} \leq 1$, and by counting the number of points that fall inside the circle we can compute $\pi$.

To compute $\pi$ using a quantum computer,
parameterize $x$ and $y$ using $N_{g} = 2 n$ qubits: $\ket{x} = H^{\otimes n} \ket{0}$, $\ket{y} = H^{\otimes n} \ket{0}$, which creates an equal superposition of all possible $x$-$y$ pairs.
The $x$ ($y$) coordinate is defined as the integer representation of the bit-string associated with $\ket{x}$ ($\ket{y}$) normalized by $2^n - 1$. For example, the state $ \ket{111}\ket{010} = \ket{7}\ket{2}$ when normalized corresponds to $\ket{7/7}\ket{2/7} = \ket{1}\ket{0.2857..}$.  With the addition of a single ancilla qubit prepared in the $\ket{+}$ state, this preparation of the $x$-$y$ coordinates constitutes $\prep=\prep^{\text{SP}} = \prep^{\text{RW}}$.

To construct $\grov$, we need to define the phase oracle, $\po$.  Let us consider a statistical mechanics average:
\begin{align}
\label{eq:piz}
    \frac{\pi}{4} = \frac{1}{Z} \iint_{0}^{1} dx \; dy \;  \Theta(1-r^{2}(x,y))
\end{align}
where $\Theta$ is the Heaviside function, $r^2(x,y) = x^2 + y^2$, and $Z \equiv \int_{0}^{1} dx \int_{0}^{1} dy$.  Going forward we will give the $x$ and $y$ dependence once, but omit it afterward.  We approximate this integral using a grid of points,
\begin{align}
\label{eq:approx-pi-over-4}
    \frac{\pi}{4} &\approx \frac{1}{Z} \sum_{x,y} \;  \Theta(1-r^{2})
\end{align}
where $Z = 2^{N_g}$, i.e. the number of points in the grid.

The phases for $\subo$ are then defined as $\varphi(x,y) = \arccos(\Theta(1-r^{2}))$, with the two cases: $\arccos(0) = \pi/2$ when $r^{2} > 1$, and $\arccos(1) = 0$ when $r^{2} \leq 1$.
Since the $x$-$y$ coordinates are in superposition we can demonstrate the action of $\subo$ on a single pair without loss of generality.  We will use quantum arithmetic gates~\cite{Ruiz-Perez2017,Seidel:2021} specifically an addition gate, $\add$, and a multiplication gate, $\mul$.  Further we require a comparator $\mathfrak{comp}$, that returns a boolean variable~\cite{https://doi.org/10.48550/arxiv.1611.07995}.  First we add $x$ and $y$ into two registers,
\begin{align}
    (\add\ket{x}\ket{0})(\add\ket{y}\ket{0}) = \ket{x}\ket{x}\ket{y}\ket{y}.
\end{align}
Then, we multiply the registers together,
\begin{align}
    (\mul\ket{x}\ket{x})(\mul\ket{y}\ket{y}) = \ket{x}\ket{x^2}\ket{y}\ket{y^2},
\end{align}
and add the products together
\begin{align}
    (\add\ket{x^2}\ket{y^2})(\ket{x}\ket{y}) = \ket{x^2}\ket{x^2 + y^2}\ket{x}\ket{y}.
\end{align}
Finally, we compare to a register$\ket{(2^{n}-1)^2}$,
\begin{align}
    \mathfrak{comp}\ket{x^2 + y^2}\ket{(2^n - 1)^2} = \ket{x^2 + y^2}\ket{r^2 \leq (2^n - 1)^2},
\end{align}
where the comparison register is either zero or one.  This defines $\subo$.  Using this register, a phase can be applied which gives $e^{i \pi / 2}$ if the comparison bit is zero, or $1$ if it is one.  The working registers can be uncomputed, defining the entire unitary $\uobs$.
With $\uobs$ defined, $\po$ is defined, and with $\prep$, $\grov$ is defined, and QPE can be performed.

\begin{figure}
    \centering
    \includegraphics[width=8.6cm]{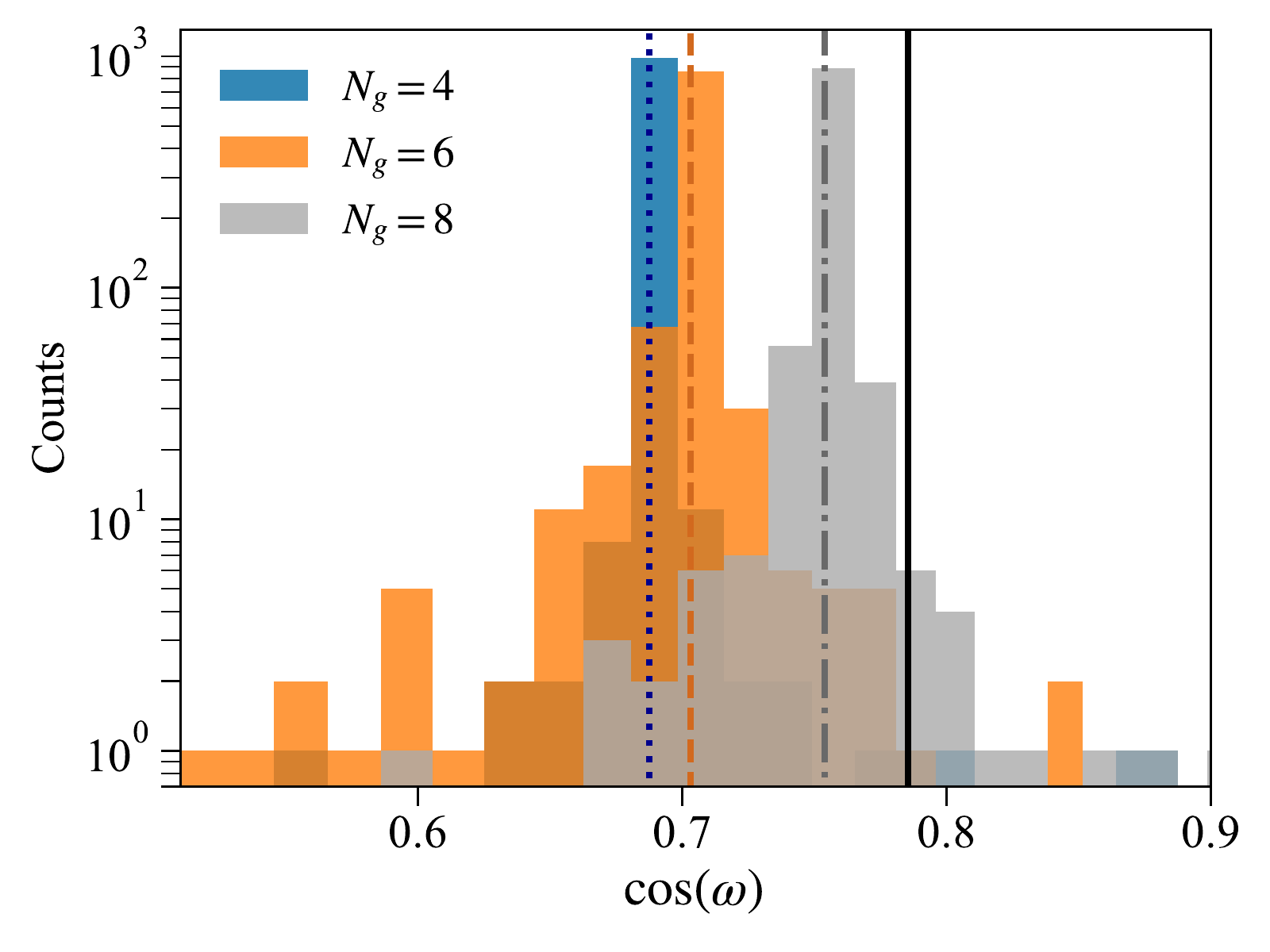}
    \caption{Count values from the QME of $\pi/4$ for $N_{g}=4,6,8$, at fixed $N_{r} = 8$.  From left to right, the lines correspond to the exact ratios for each grid, respectively $11/16$, $45/64$, and $193/256$ with the solid line indicating $\pi/4$.}
    \label{fig:pi-vary-ng}
\end{figure}

\begin{figure}
    \centering
    \includegraphics[width=8.6cm]{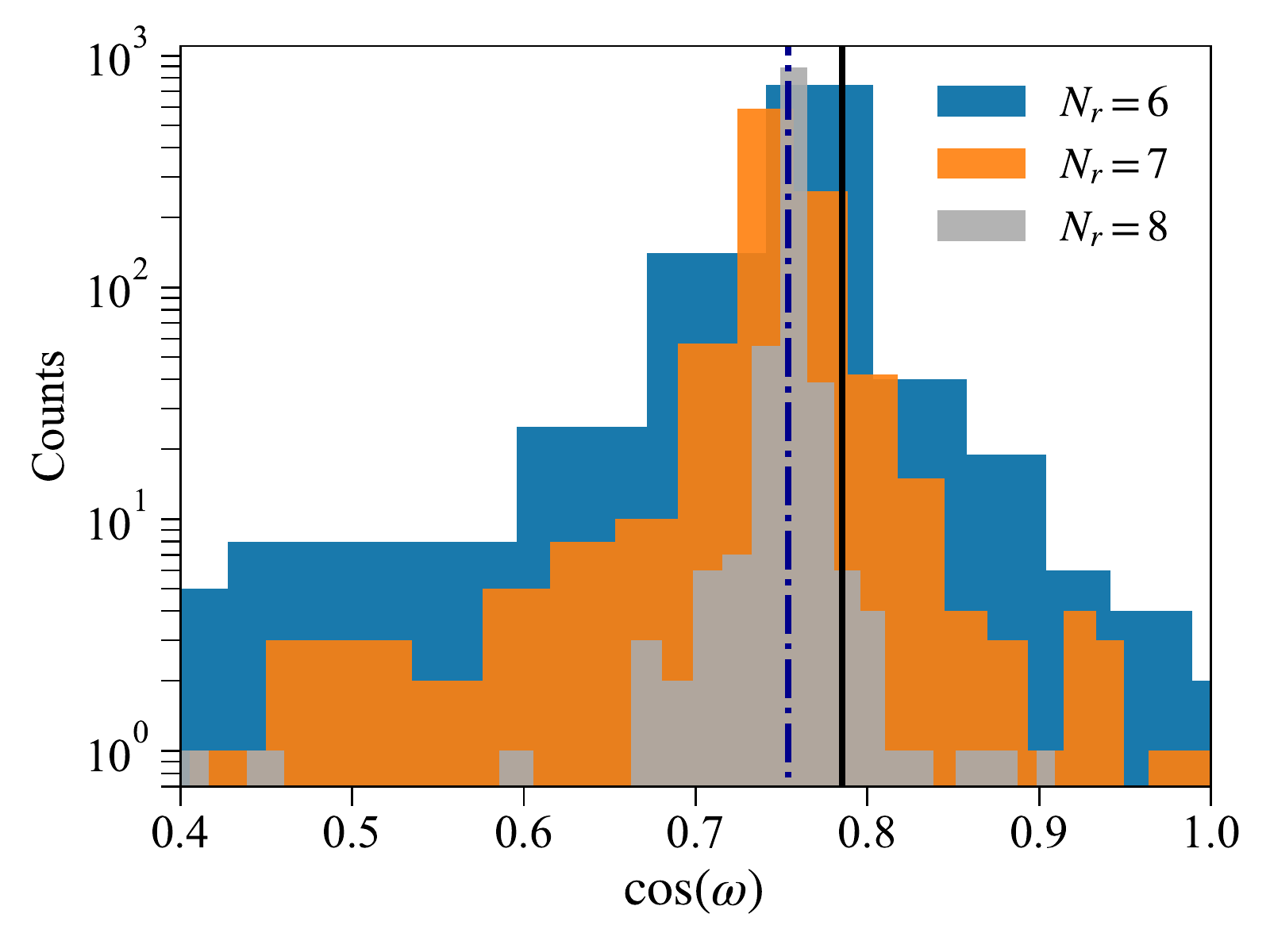}
    \caption{Count values from the QME of $\pi/4$ for $N_{r}=6,7,8$, at fixed $N_{g} = 8$.  The dash-dotted line indicates the exact ratio for $N_{g}=8$, $193/256$, while the solid line denotes $\pi/4$.}
    \label{fig:pi-vary-nr}
\end{figure}

Histograms showing counts of bit-strings using different $N_g$ and $N_r$ are in Figs.~\ref{fig:pi-vary-ng}, and~\ref{fig:pi-vary-nr}, respectively.  Along with the histograms are lines for the finite-grid ratios of inner-circle points to total points, as well the exact value of $\pi/4$.  In Fig.~\ref{fig:pi-vary-ng}, we see an increase in $N_{g}$ at fixed $N_{r} = 8$, leading towards $\pi/4$, with a constant bin width at no less than an exponential rate. In practice, one could extrapolate the measured value as a function of $N_g$ with potentially fewer resources than a single fine grid might require.  In Fig.~\ref{fig:pi-vary-nr}, we see varying $N_{r}$ while keeping $N_{g} = 8$ yields a most likely value at the finite-grid ratio, and the bin width diminishes as $2^{-N_r}$. Thus, larger $N_{g}$ improves accuracy, while larger $N_{r}$ improves precision.


We can include a sign problem by weighing locations in the quarter circle using positive and negative weights separated at $\theta=\pi / 3$ which corresponds to $(x/r)^2=1/4$ (See  Fig.~\ref{fig:pi-grid}~(d)). When $(x/r)^2\leq 1/4$, the weight is $-1$, and when $(x/r)^2\geq 1/4$, the weight is $+1$. This corresponds to the integral
\begin{align}
\label{eq:pisign}
    \frac{\pi}{12} &= \frac{1}{Z} \iint_{0}^{1} dx \; dy \;  \Theta(1-r^{2}) \left[ 2\Theta\left(\frac{x^2}{r^2} - \frac{1}{4} \right) - 1 \right].
\end{align}

To estimate $\pi$ from Eq.~\eqref{eq:pisign} we must modify $\po$ to also check if $x^2 / r^2 \geq 1/4$.  We compute $\ket{x^2}$ and $\ket{r^2}$ as before, and using a division oracle calculate $\ket{x^2 / r^2}$ which is then compared using $\mathfrak{comp}$ with $\ket{1/4}$.  We then apply a phase based on the compare register.  Therefore, with a sign problem there are two deciding qubits, $\ket{q_{\text{circle}}}$ and $\ket{q_{\text{angle}}}$,  which indicate if the point is inside the circle and $(x/r)^2 \geq 1/4$, respectively. From their values a phase $e^{i \varphi}$ is imparted, with $\varphi=\pi(1+q_{\text{circle}}-2q_{\text{angle}}) / 2$.
With the working register arithmetic outlined above, $\subo$, $\uobs$, and $\po$ are defined. 

\begin{figure}
    \centering
    \includegraphics[width=8.6cm]{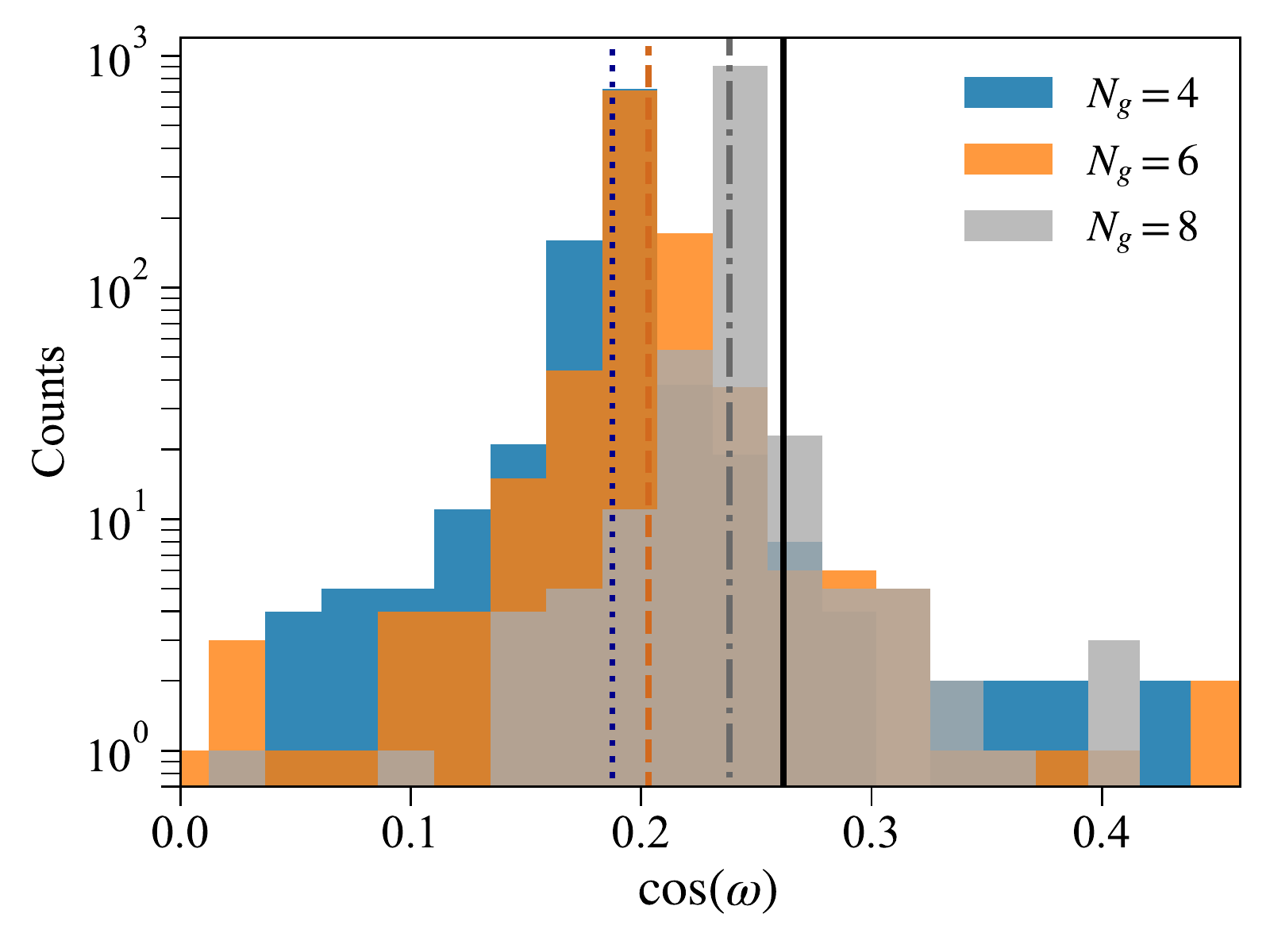}
    \caption{Count values from the QME of $\pi/12$ vs. $N_{g}$ at fixed $N_{r} = 8$.  The lines correspond to the exact ratios, respectively $3/16$, $13/64$, and $61/256$.  The solid line indicates $\pi/12$.}
    \label{fig:sign-vary-ng}
\end{figure}

\begin{figure}
    \centering
    \includegraphics[width=8.6cm]{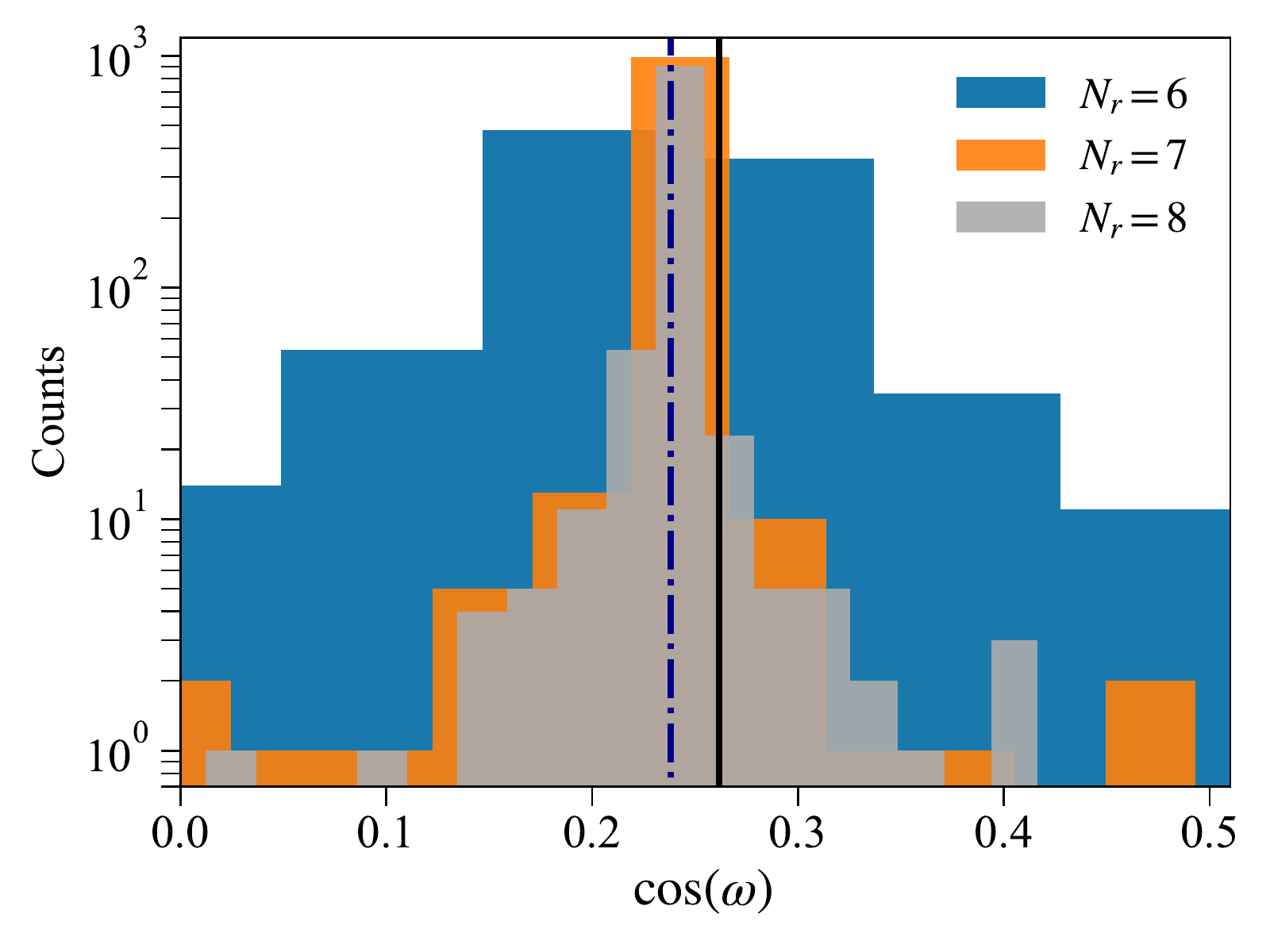}
    \caption{Count values from the QME of $\pi/12$ vs. $N_{r}$ at fixed $N_{g} = 8$.  The dash-dotted line indicates the exact $N_{g}=8$ value $61/256$, while the solid line labels $\pi/12$.  We see for increasing $N_r$, the distribution narrows.}
    \label{fig:sign-vary-nr}
\end{figure}

Analogous histograms to the case of $\pi$ without a sign problem can be seen in Figs.~\ref{fig:sign-vary-ng} and~\ref{fig:sign-vary-nr}.  We see in Fig.~\ref{fig:sign-vary-ng} as $N_g$ is varied with fixed $N_{r}$, the most likely value moves towards $\pi/12$ for larger values of $N_{g}$.  Likewise, in Fig.~\ref{fig:sign-vary-nr} as $N_r$ is increased we see the distribution narrows around the finite-grid ratio.  For QME this problem is as computationally difficult as without a sign problem.  One should note that the nonlinear transform from the measured $\omega$ to $\cos(\omega)$ results in non-constant bin width. This is acutely important for sign problems and reweighting when means are often near zero, where the bins are largest. Thus rescaling the result register to optimize bin locations could be valuable for reducing costs.


\subsection{U(1) gauge theory toy model}
\label{sec:u1-toy}

We introduce a toy model resembling the Villian~\cite{villain:1975,itzykson_drouffe_1989} approximation of a two-dimensional, Euclidean U(1) gauge theory, and the calculation of the Wilson loop. This has been used to investigate classical methods for ameliorating the sign problem~\cite{Kanwar:2021tkd}.  We aim to calculate
\begin{align}
    \mathcal{P}(n) &\equiv \frac{1}{Z} \iint_{0}^{1} dx\, dy \;  \Theta(1-r^2) \cos(n \theta)
    e^{- \theta^2}.
\end{align}
Here, $n$ controls the degree of oscillations in the integrand and can be any real number.  The value of $\mathcal{P}(n)$ is
\begin{align}
\label{eq:exact-pn}
    \mathcal{P}(n) = \frac{1}{4} \sqrt{\pi} e^{-n^2/4} \Re\left\{ \text{erf}\left[\frac{\pi + i n}{2}\right] \right\}.
\end{align}
We again approximate $x$ and $y$ by a grid and $Z = 2^{N_{g}}$.

State preparation is given by $\prep = \prep^{\text{SP}} = \prep^{\text{RW}}$.  We will encapsulate $\mathcal{O}(g_i)$ inside the phases and define $\varphi = \arccos\left(\Theta(1-r^2) \cos(n \theta) e^{-\theta^{2}} \right)$. 
Designing the circuit for computing the phases requires floating-point arithmetic, which we will not elaborate upon, but will assume exist. With the ability to create $\ket{\varphi}$, we can apply $\exp(i \hat{\varphi})$ to the phase register and extract the phase.  After uncomputing the phase register, this procedure defines $\uobs$ and $\po$.  

\begin{figure}
    \centering
    \includegraphics[width=8.6cm]{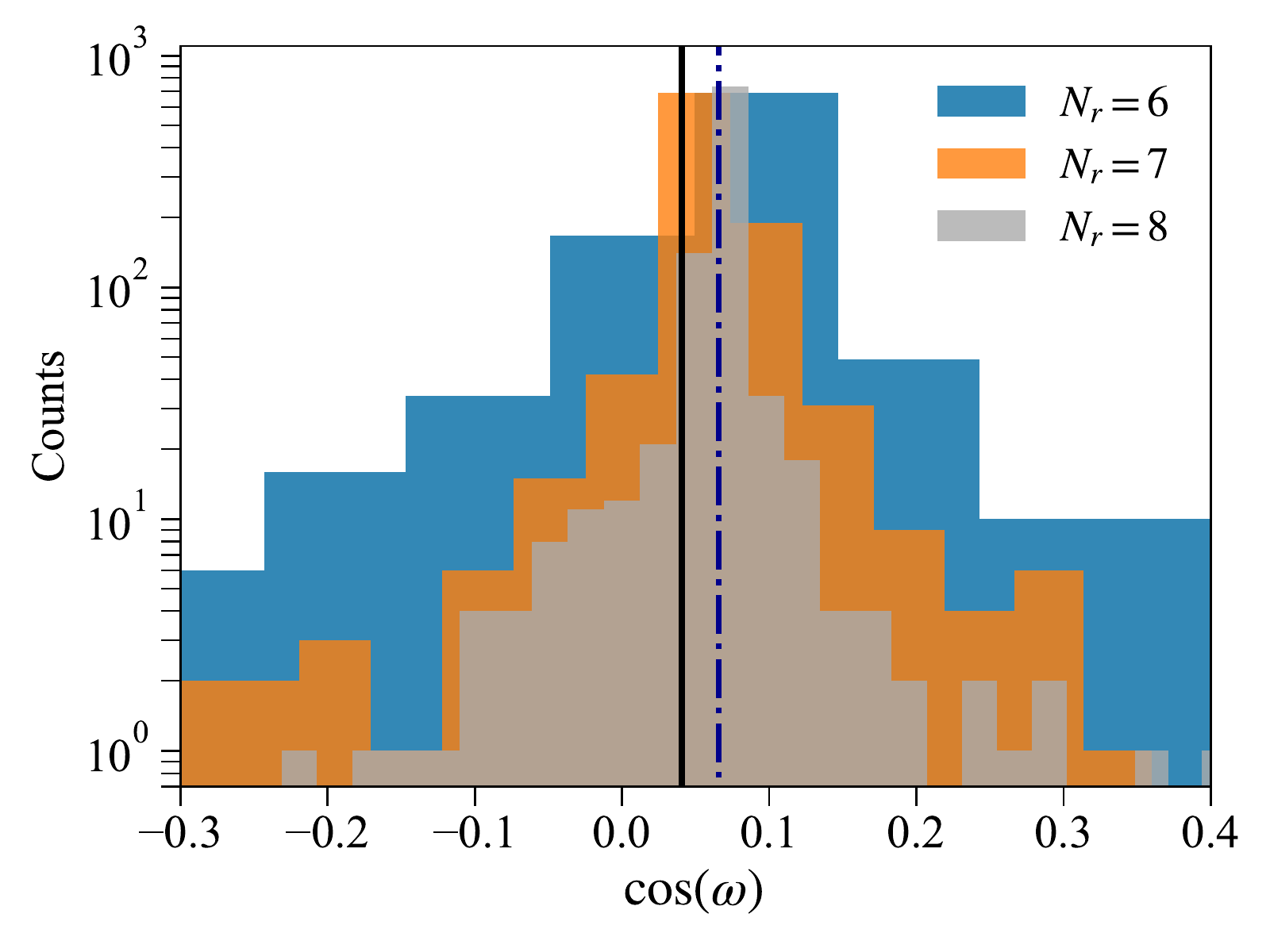}
    \caption{The total counts in the calculation of $\mathcal{P}(3)$ vs. $N_{r}$ at fixed $N_{g} = 8$.  The solid black line indicates the exact value from Eq.~\eqref{eq:exact-pn}, while the dotted-dashed line indicates the exact value calculated from the grid.}
    \label{fig:u1-vary-nr}
\end{figure}

\begin{figure}
    \centering
    \includegraphics[width=8.6cm]{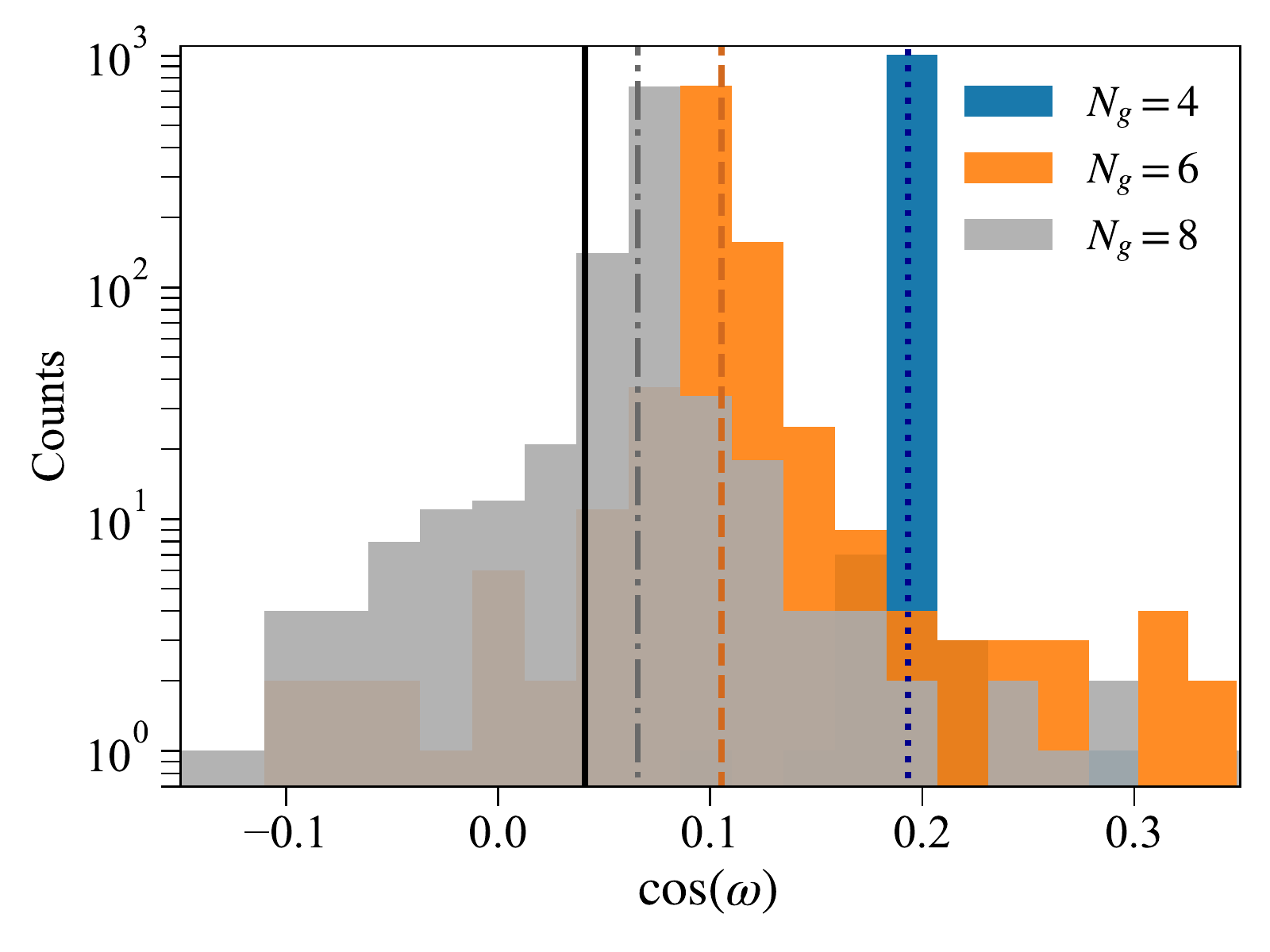}
    \caption{The total counts in the calculation of $\mathcal{P}(3)$ vs. $N_{g}$ at fixed $N_{r} = 8$.  The solid black line indicates the exact value from Eq.~\eqref{eq:exact-pn}, while the other lines are respectively the exact values at finite $N_g=4,6,8$.}
    \label{fig:u1-vary-ng}
\end{figure}

Results of these circuits for $\mathcal{P}(3)$ are shown in Figs.~\ref{fig:u1-vary-nr} and~\ref{fig:u1-vary-ng}.  In Fig.~\ref{fig:u1-vary-nr} we see that as $N_r$ is increased the distribution narrows around the exact value.  Figure~\ref{fig:u1-vary-ng}  shows the drift of the most likely value of $\mathcal{P}(3)$ towards the exact value of $\approx 0.041$ as $N_g$ increases at fixed $N_r$.

Having given circuits and explained QME for a variety of toy and demonstrative models, we now turn to an archetypal LFT, the Ising model.

\subsection{The Ising model}

\label{sec:ising}

We consider a two-dimensional Ising model on a $V=N_s\times N_t$ Euclidean lattice with periodic boundary conditions and spatial and temporal extents of  $N_{s}$ and $N_{t}$.  The action is 
\begin{align}
    S = \beta E = \frac{\beta}{2}\sum_{\langle i j \rangle} (1 - s_i s_j),
\end{align}
where $\sum_{\langle i j \rangle}$ sums over nearest-neighbor pairs, $s_i$ is a lattice spin taking values $\pm1$, and $\beta$ is a coupling constant commonly identified as an inverse temperature.  We can rewrite this as a sum over binary values, $n_i = 0,1$, using $s_i = 2n_i-1$.  Then
\begin{align}
\nonumber
    S &= \beta\sum_{\avg{ij}} [(n_i + n_j) - 2 n_i n_j] \\
    &= \beta \bigg( 4 \sum_{i} n_i - 2\sum_{\avg{ij}} n_i n_j \bigg).
\end{align}

As an observable, we will compute the square of the magnetization density
\begin{align}
    m^2 = \left(\frac{1}{V} \sum_{i} s_{i} \right)^2 = \left( \frac{2}{V}\sum_{i} n_i - 1 \right)^2,
\end{align}
which is bounded between zero and one.

On the quantum computer
each qubit in the system is associated with a spin on the lattice. 
To prepare the system, we put every qubit into the $\ket{+}$ state, $\ket{\{n\}} \equiv H^{\otimes V} \ket{0}$. This generates all spin configurations in superposition.
By appending one ancilla prepared in the $\ket{+}$ state we have the state preparation circuit $\prep^{\text{RW}}$, but must calculate a ratio of observables (see Sec.~\ref{sec:reweigh}).

Just as in the $\pi$ example, we require a phase oracle.  We will focus on a single spin configuration without loss of generality.  To compute the magnetization, we must sum the bit-string.  We do this by first summing every even-odd spin together in parallel.  We can then add the results of the first step again in even-odd pairs.  Repeating this we can compute the sum in $\sim \log_{2}(V)$ steps.  
The result of these sums is $\ket{\sum_{i} n_{i}}$.  By multiplying by two, dividing by the volume, and shifting by one we arrive at the magnetization density.  We can add this number to zero, and multiply the two $m$ registers to obtain $m^2$.

To compute the action, we already have the term containing $\sum_i n_i$.
To compute the nearest-neighbor interaction we can again compute the product of disjoint pairs of spins in each of the two directions in parallel, and sum the resulting collection of products.  This can be done in $\sim \log_{2}(2V)$ steps, resulting in a register $\ket{ \sum_{\avg{ij}} n_i n_j}$.  Multiplying by the appropriate factors and subtracting we can form the register $\ket{S}$.  Then using exponentiation, multiplication, and taking the arccosine we arrive at $\ket{\arccos(m^2 e^{-S})}$.  This defines $\subo$ for the numerator.  $\subo$ for the denominator can be calculated similarly without the factor of $m^2$.  Now by applying $\one \otimes e^{i \hat{\varphi}}$ we can extract the phase, then uncompute the phase register.  These steps define the unitary $\uobs$, and hence $\grov$, in QPE.

\begin{figure}
    \centering
    \includegraphics[width=8.6cm]{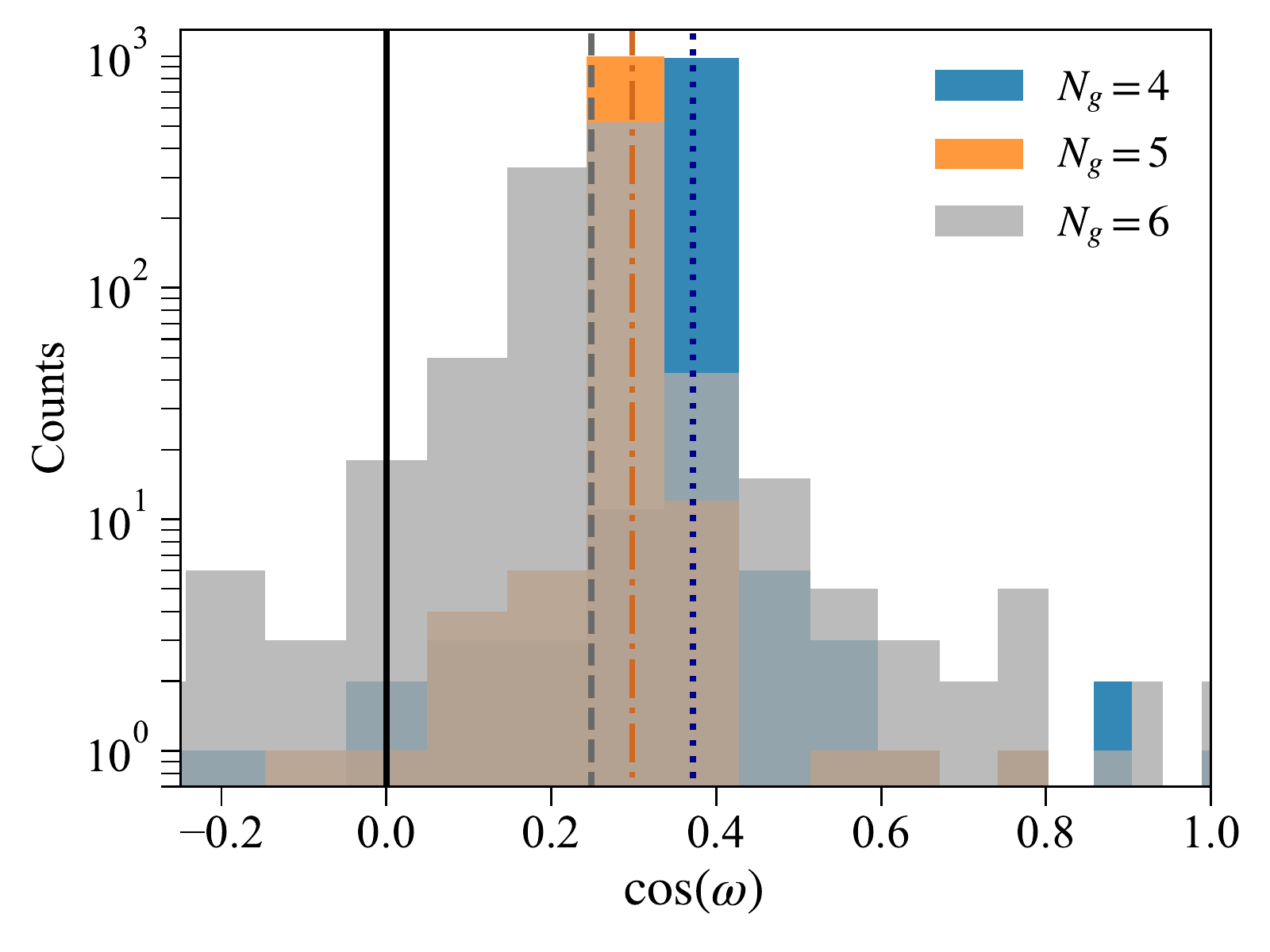}
    \caption{$\langle m^{2}\rangle$ vs. $N_{g}$, at a fixed $N_{r} = 6$ with $\beta = 0.4$. From right to left, the dotted, dash-dotted, and dashed lines indicate the exact values of $m^{2}$ for their respective volumes.  The solid black line is the infinite-volume value.}
    \label{fig:ising-vary-ng}
\end{figure}

\begin{figure}
    \centering
    \includegraphics[width=8.6cm]{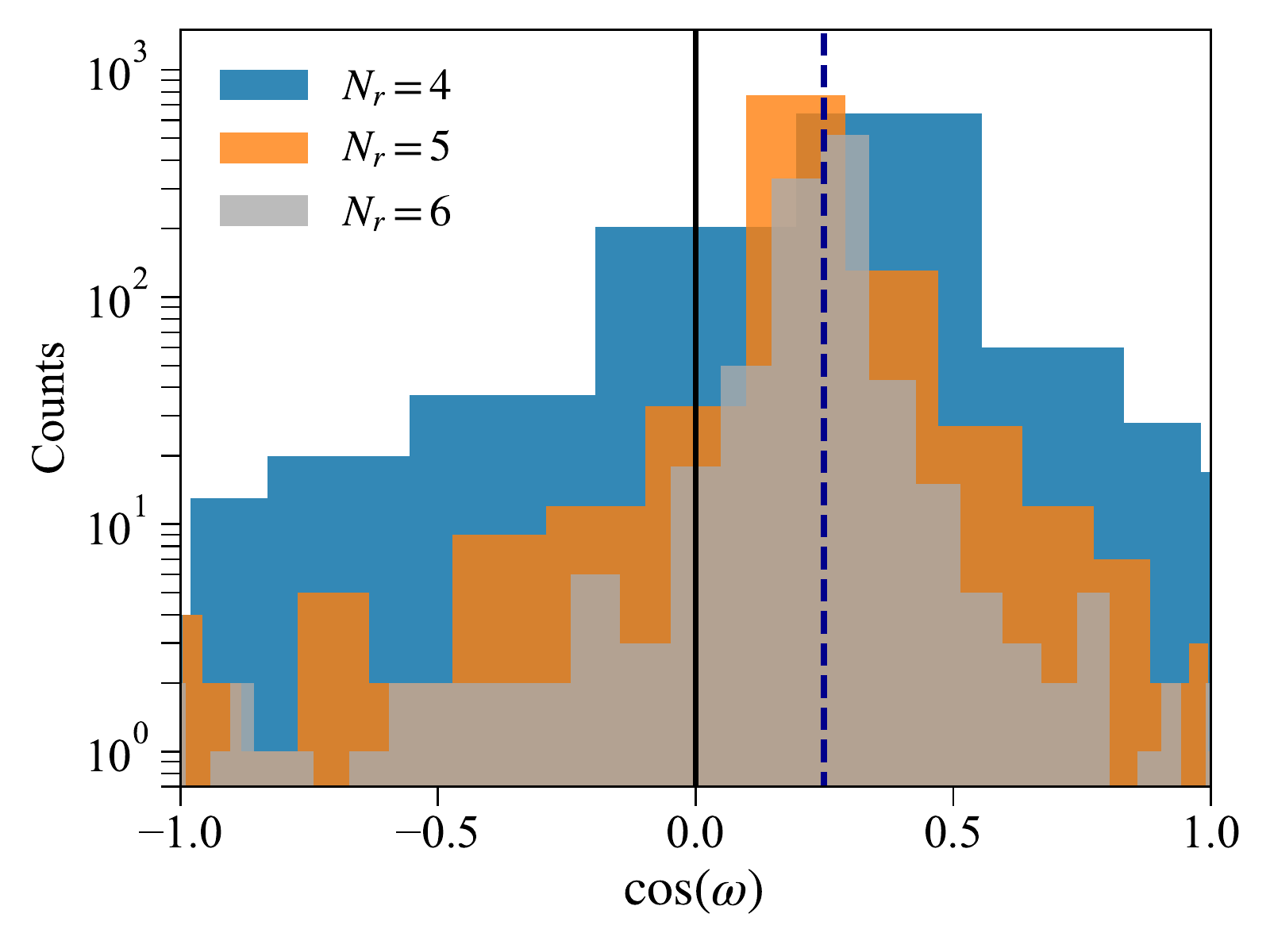}
    \caption{$\langle m^{2}\rangle$ vs. $N_{r}$ at a fixed $N_{g} = 6$ with $\beta = 0.4$.  We see with each increase in the register size the histogram bins narrow around the exact value---given by a dashed line, although the largest register straddles the exact answer tightly.}
    \label{fig:ising-vary-nr}
\end{figure}

In Figs.~\ref{fig:ising-vary-ng} and~\ref{fig:ising-vary-nr} we see histograms of $\langle m^{2}\rangle$ when varying $N_g$ and $N_r$ respectively, for the \emph{one}-dimensional, classical Ising chain at $\beta = 0.4$.  As $N_g$ is increased for fixed $N_r$, the mean tends towards the infinite-volume value.  Likewise, with $N_g$ fixed and $N_r$ increased, estimates converge to the exact finite-volume value.  

Fig.~\ref{fig:ising-vary-nr} demonstrates a desirable feature that results from 1) the bin-width providing a 100\% confidence interval of error, and  2) the maximal bin changing location.  That is that the ideal result must lie in the overlap between distributions. 
Therefore, the overlap between bins of different $N_r$ can estimate the mean more precisely than any single $N_r$ alone. This could be used to engineer sets of $N_r$ that maximize information gained.  In Fig.~\ref{fig:ising-vary-nr} we see a sliver of overlap between the $N_{r} = 5$ and $6$ maximal bins, which the exact answer must lie within.
Fig.~\ref{fig:u1-vary-nr} provides another example.  Having now seen QME used with simple examples, as well as the formulation and execution on a \emph{bona fide} LFT, we now give the formulation of QME for a generic lattice gauge theory.

\subsection{Generic lattice gauge theory}

Take a $D$ dimensional lattice with $V \equiv N_{s}^{D-1} N_{t}$ and periodic boundary conditions.  We consider the action
\begin{align}
    \label{eq:gaugeaction}
    S = \sum_{x} \sum_{1 \leq \mu \leq \nu \leq D}  \beta \left( 1 - \frac{1}{N} \Re[ \Tr[ g_{\mu\nu}(x)]] \right)
\end{align}
where $g_{\mu \nu}(x)= g_{\mu}(x) g_{\nu}(x+\hat{\mu}) g^{\dagger}_{\mu}(x+\hat{\nu}) g^{\dagger}_{\nu}(x)$ is a  plaquette, $g_{\mu}(x)$ is a gauge group element, and the sums are over lattice sites and the two directions $\mu$ and $\nu$ respectively .

Prepare a qubit register for each gauge link, $\ket{g_{\mu}(x)}$.  Let the operator $\mathcal{H}$ prepare a maximal superposition of all possible gauge link states, analogous to the Hadamard gate,
\begin{align}
    \mathcal{H}(x,\mu)\ket{0} = \frac{1}{\sqrt{|G|}} \sum_{g} \ket{g_{\mu}(x)}
\end{align}
where $|G|$ is the size of the local state space.
Then for state preparation we can apply $\mathcal{H}$ for every link register,
\begin{align}
\label{eq:hprep}
    \bigotimes_{x,\mu} \mathcal{H}(x,\mu) \ket{0} = \bigotimes_{x,\mu} \frac{1}{\sqrt{|G|}} \sum_{g} \ket{g_{\mu}(x)}.
\end{align}
This creates a superposition of all field configurations with equal probability.  We then append an ancilla qubit $\ket{+}$.  This constructs $\prep^{\text{RW}}$. For $\prep^{\text{SP}}$ we follow Sec.~\ref{sec:state-prep} with $\phi(g_i) =\arccos(e^{-S(g_i)/2})$.  When encoding observables one sets $\varphi(g_i) = \arccos(\mathcal{O}(g_i))$  in $\uobs$.
With these prescriptions of the phases, the operators for QME of lattice gauge theory are defined.  In the next section we will study the effects of noise on QME.

\subsection{The effects of noise}
\label{sec:noise}

On a fault-tolerant quantum computer the $R_{Z}$ gates are approximated with an infidelity $\varepsilon$ by interleaved $T$ and $H$ gates \cite{Eastin_2009,PhysRevLett.77.793,1996RSPSA.452.2551S,1996PhRvA..54.1098C,PhysRevA.54.4741,nielsen_chuang_2010,1997RuMaS..52.1191K,2020arXiv200505581M}. While one can derive a string of $T$ and $H$ gates that approximates any $R_{Z}$ gate, implementing this string and classically simulating it drastically extends the length of a quantum circuit, and becomes a computationally intensive problem.  Instead, to study the effects of nonzero $\varepsilon$, we approximate each $R_{Z}$ gate,
\begin{equation}
\label{eq:noisedrift}
    R_{Z}(\theta)\rightarrow \tilde{R}_{Z}(\theta; \varepsilon) = R_{Z}(\theta) R_{X}(\theta \varepsilon) R_Y(\theta \varepsilon).
\end{equation}
The parameter $\varepsilon$ approximates the infidelity in quantum computations, and drives coherent angle-dependent drift in the $R_{Z}$ gates. We tested variations on Eq.~\eqref{eq:noisedrift} by sending  $\theta \varepsilon \rightarrow -\theta \varepsilon$ in $R_{X}$ and found negligible effects. 

\begin{figure}
    \centering
    \includegraphics[width=8.6cm]{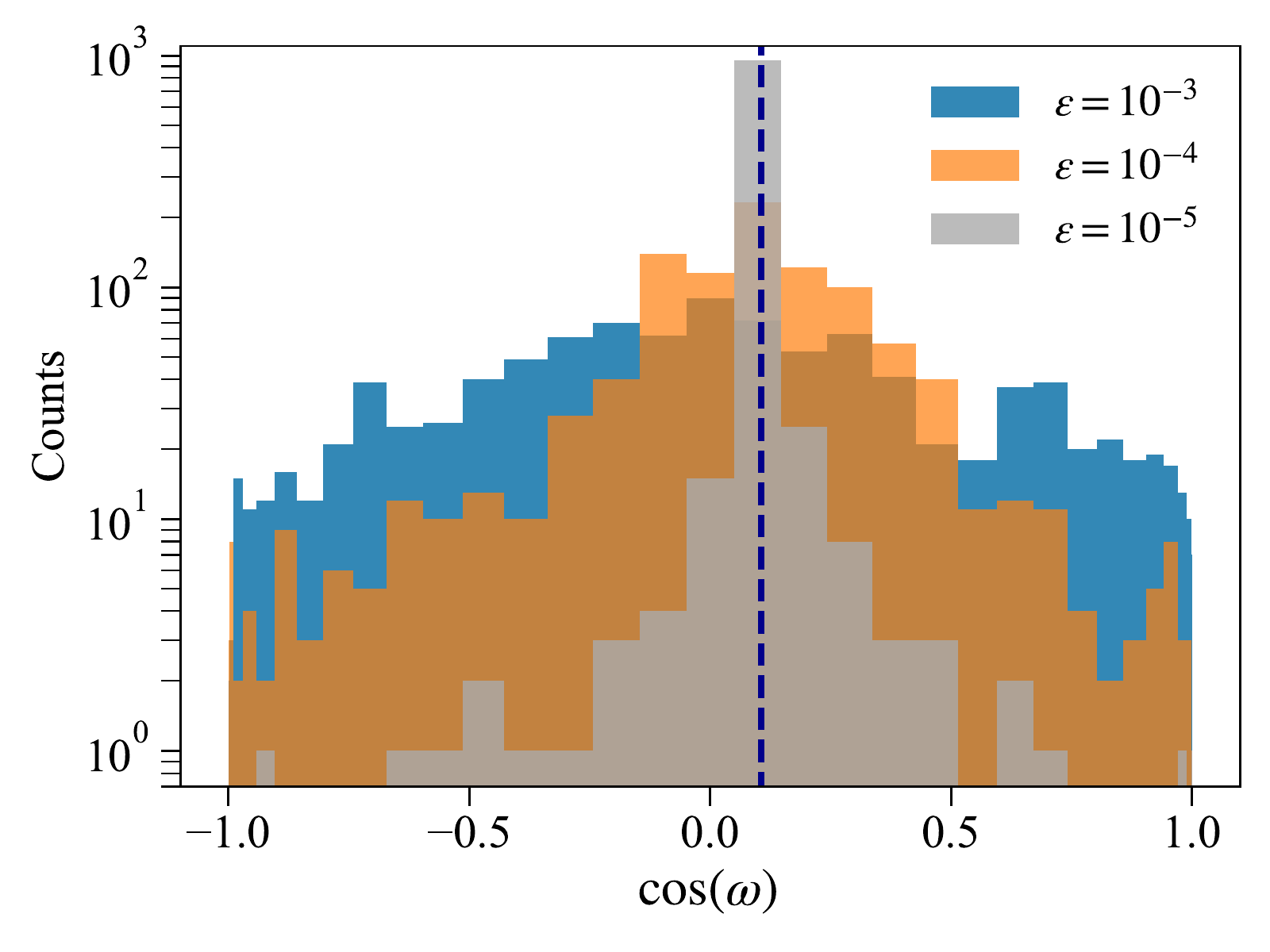}
    \caption{Results of noisy simulation of U(1) toy model vs noise rate $\varepsilon$.  The dashed line gives the ideal $N_{r} = N_{g} = 6$ value.}
    \label{fig:u1-noise}
\end{figure}

For a test case we consider the model from Sec.~\ref{sec:u1-toy}.  Using $N_{r} = N_{g} = 6$ we run the QME circuit for various $\varepsilon$.  This calculation requires synthesizing $N_{Z}\sim O(10^4)$ $R_Z$ gates. Fig.~\ref{fig:u1-noise} shows the histogram of the average returned from QME. For $\varepsilon \approx 10^{-4}$ a statistically clear signal for the most likely value consistent with the ideal value appears with $N_{\text{shots}}=1024$. 

In context, this indicates there is a threshold $\varepsilon$ for a given set of shots below which better synthesis of $R_Z$ provides no benefit at fixed $N_r$ and $N_g$.  Exact rotation gates are unnecessary, and imprecise rotation gates can give sufficient accuracy and precision. This is analogous to precision LFT calculations where half-precision floating point numbers are used to accelerate calculations~\cite{Clark:2009wm,doi:10.1137/S1064827599353865,Tu:2021dvv,Ishikawa:2021iqw}. Heuristically, the threshold is $\varepsilon\sim \sqrt{N_{\text{shots}}}/N_Z$.

\section{Quantum advantage}
\label{sec:cvq}

To appreciate the advantage provided by the QME algorithm, consider $\prep^{\text{SP}} \ket{0} = \ket{\psi_{0}}$.  At this stage, traditional sampling methods could be applied, and measuring $\ket{g_i}$ provides a single configuration.  By iterating this procedure one collects an ensemble of configurations from which averages can be taken.  
We expect $\sigma \sim 1/\sqrt{N}$, but $N$ is precisely the number of times $\prep^{\text{SP}}$ is run.  In this sense we can see $\prep^{\text{SP}}$ as the quantum mechanical analog to having  access to the partition function through, say, brute-force calculation.  

Now, without loss of generality, take $N$ to be a power of two, $N = 2^k$, implying $\sigma \sim 1/ 2^{k/2}$.
For the QME algorithm, we want the same precision of $1/2^{k/2}$.  Using QME we see this requires $N_{r} = k/2$ qubits.  Looking at Fig.~\ref{fig:qpe}, QPE executes $\grov$ a total of $\sum_{j=0}^{k/2 - 1} 2^j = 2^{k/2}-1$ times.  Since $\prep^{\text{SP}}$ appears in $\grov$ ${O}(1)$ times, $\prep^{\text{SP}}$ is similarly called ${O}(2^{k/2})$ times.  Thus we see the number of times $\prep^{\text{SP}}$ is needed in the QME algorithm ($2^{k/2}$) is quadratically less compared to the traditional sampling method ($2^k$).  Therefore rather than using the prepared state to sample and calculate expectation values, it is advantageous to pass the quantum state into the QME algorithm to avoid greater calls to the state preparation algorithm overall, while maintaining the same precision.

\begin{figure}
    \centering
    \includegraphics[width=8.6cm]{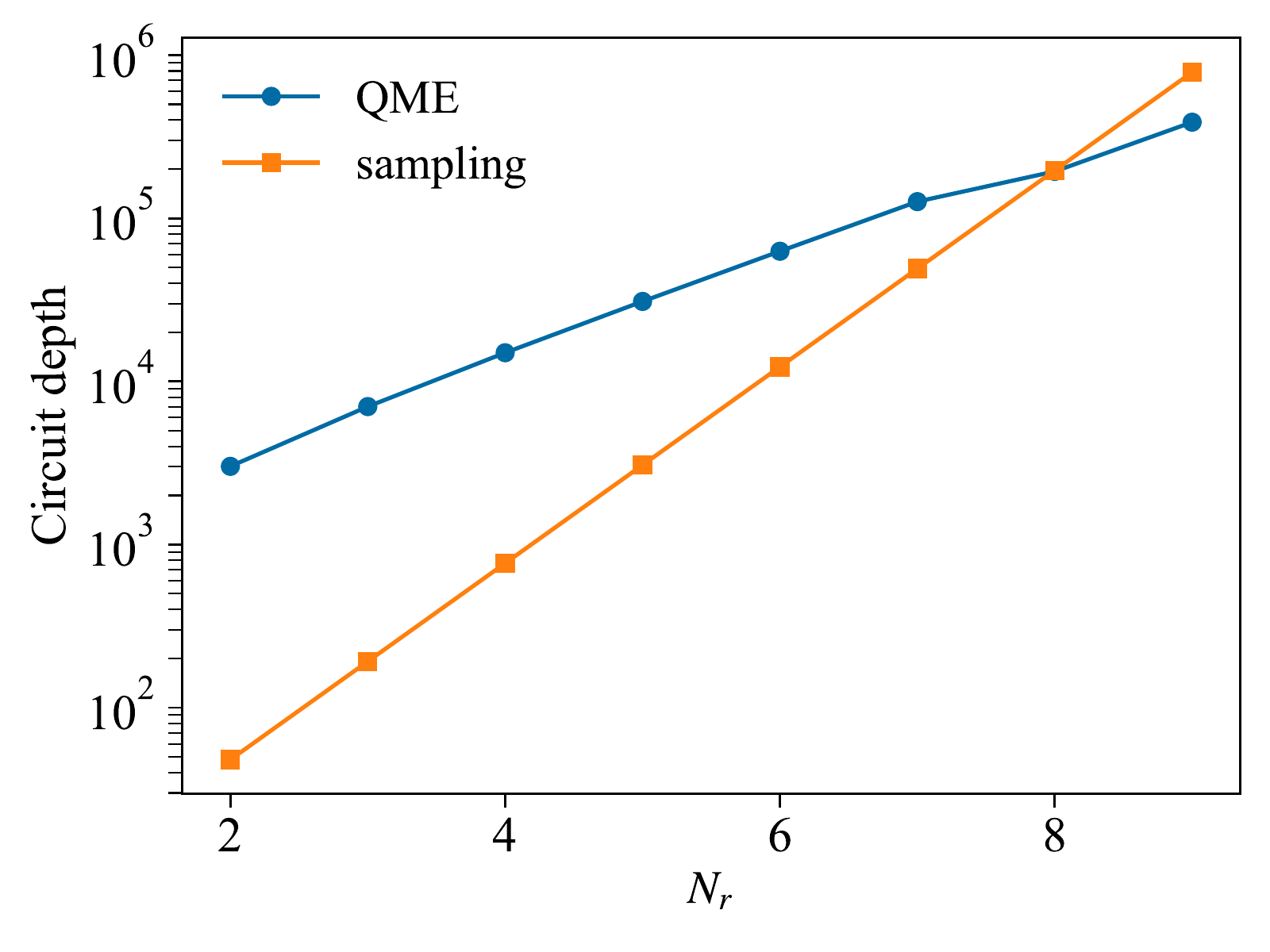}
    \caption{The circuit depth of QME and traditional sampling vs. $N_{r}$. For sampling, the precision is determined from $1/2^{N_{r}/2}$ while for QME, it is $1/2^{N_{r}}$.}
    \label{fig:depth-scaling}
\end{figure}

This increased efficiency in calls to $\prep^{\text{SP}}$ is seen in Fig.~\ref{fig:depth-scaling}, where QME is compared with the traditional sampling method to calculate $\pi$ for $N_g = 2$ and varying $N_r$.  Both the QME and sampling circuits are decomposed under the same basis gates.  We see the overhead for QME is large, however for a fixed accuracy, the scaling of QME is superior to sampling.  The ratio between the sampling and QME slopes is $\approx 2.09$, approaching the asymptotic quadratic value.  The sampling circuit depth is computed using the circuit depth for $\prep$, multiplied by the number of samples required to achieve a precision of $2^{-N_{r}}$ assuming the error converges like $2^{-N_{r}/2}$.  After $N_{r} = 8$, QME surpasses the traditional sampling method in efficiency.

What about in the case of a sign problem, or an exponentially small value to calculate?  This is relevant in the case of $\prep^{\text{RW}}$.  Consider the case where the degrees of freedom of the model are binary variables themselves.  Now the prefactor $Z/M$ in Sec.~\ref{sec:state-prep} can be rewritten as
\begin{align}
    \frac{Z}{M} = \frac{e^{f V}}{2^V} = \frac{e^{f V}}{e^{V \log(2)}} = e^{V(f-\log(2))},
\end{align}
where $f$ is the free energy density, $\log(Z)/V$. This same factor appears in the final ratio when computing $\braket{\obs}$ using Sec.~\ref{sec:reweigh} as well.  We see that this ratio depends exponentially on the number of qubits used in the system---note that $f$ is always less than $\log(2)$ here, since the Boltzmann weights will always be less than or equal to one, and $f=\log(2)$ when all Boltzmann weights are one.  Therefore, to compute a number as small as $2^{-V}$ using QME will require $O(V)$ qubits in the result register, which will entail $O(2^{V})$ calls to $\prep$.  Notice this is still quadratically faster than traditional sampling, since naively
\begin{align}
    \sigma \sim \frac{1}{\sqrt{N}} \sim 2^{-V} \implies N \sim 2^{2V}.
\end{align}
Therefore even in the case of a sign problem or an exponentially small signal, the QME algorithm is superior.

For many state preparation methods, QME depends exponentially on the system volume.  This is paid either in the amplitude amplification step, or in the precision of calculating the mean during QPE.  This dependence comes from producing the full probability distribution; however, one can instead approximate the distribution.  This is relevant to the case of classical sampling algorithms.  Using pseudo-random number generators, these algorithms sample from approximate probability distributions.  Since classical algorithms can be ported onto a quantum computer~\cite{Bennett:1973,nielsen_chuang_2010}, these distributions can be realized by an $\prep$ circuit.  Even still, QME  calls $\prep$ quadratically fewer times than would be required by sampling, and so calculations of $\langle \obs \rangle$ are accelerated---at least asymptotically.  Moreover, should classical algorithms improve in the future, QME can take those state preparation methods and use them, along with a further advantage.

\begin{figure}
    \centering
    \includegraphics[width=\linewidth]{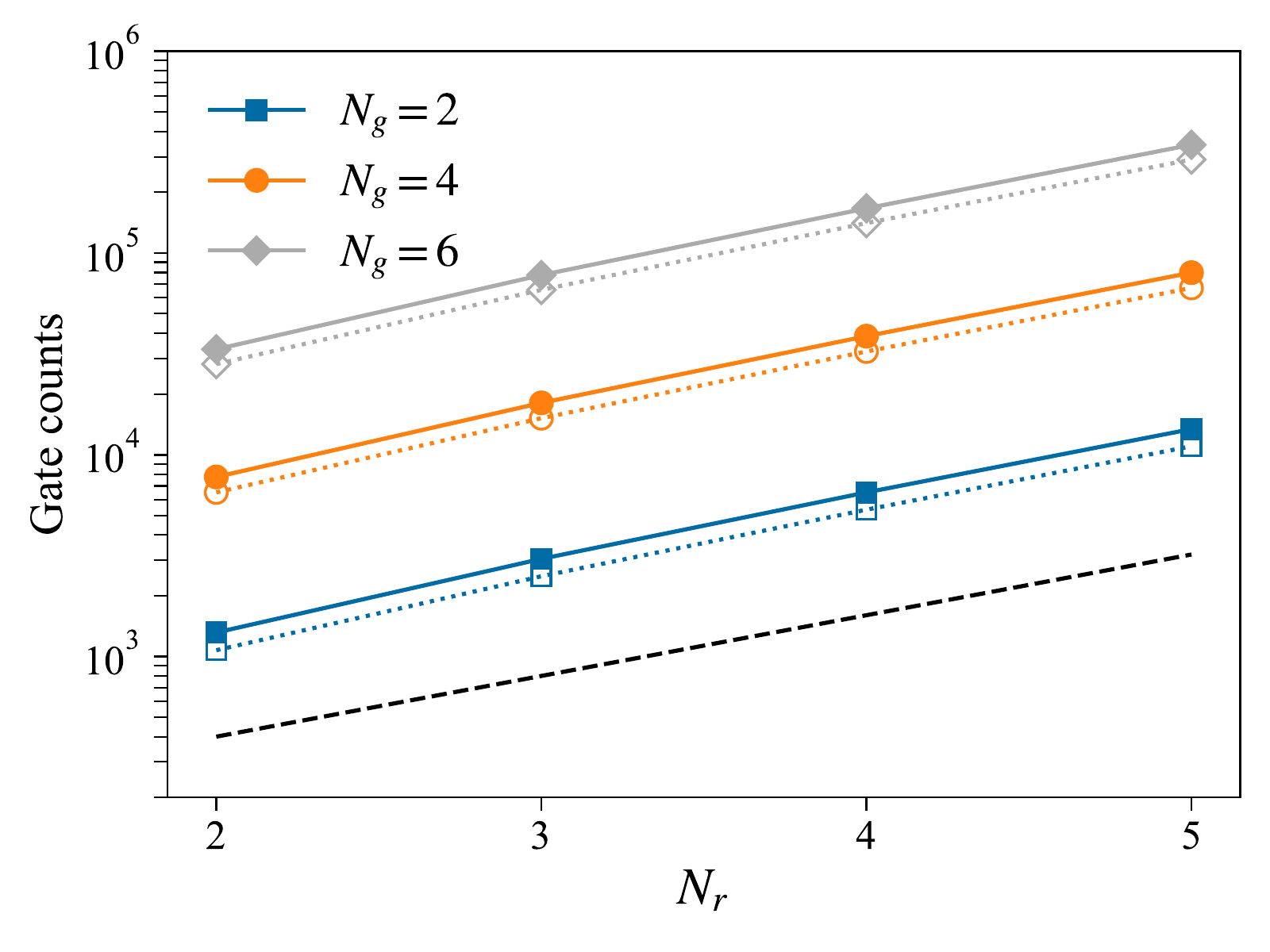}
    \caption{The gate counts of QME of $\pi$ vs. $N_{r}$.  Solid markers indicate $R_{Z}$ gates, and hollow markers indicate CNOT gates.  The dashed line corresponds to a scaling $\propto 2^{N_{r}}$.}
    \label{fig:gc-vs-nr}
\end{figure}

\begin{figure}
    \centering
    \includegraphics[width=\linewidth]{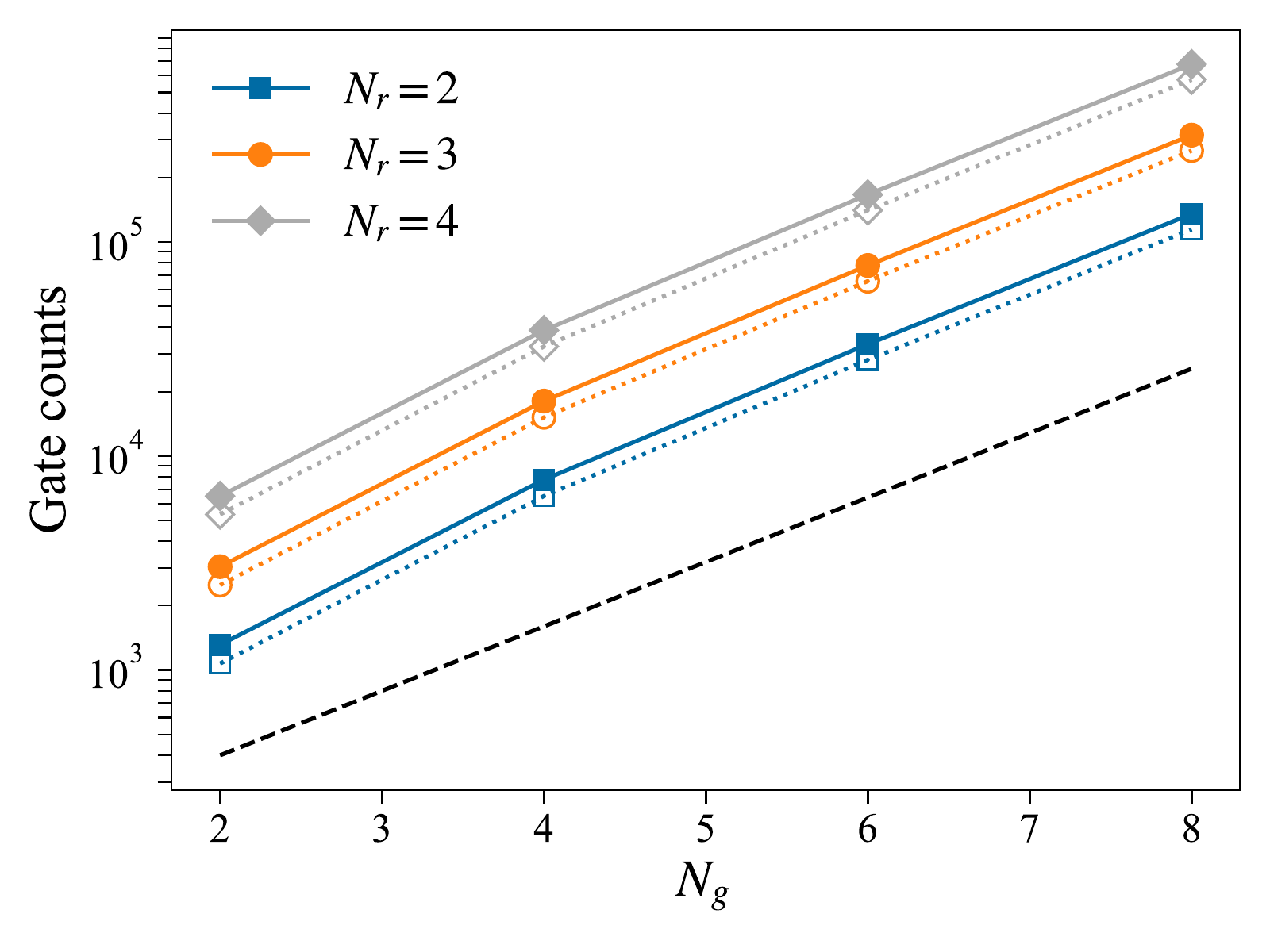}
    \caption{The gate counts in the QME of $\pi$ vs. $N_{g}$.  Solid markers indicate the $R_{Z}$ gates, and hollow markers indicate CNOT gates.  The dashed line corresponds to a scaling $\propto 2^{N_{g}}$.}
    \label{fig:gc-vs-ng}
\end{figure}

In light of this discussion, we can study the resource costs for QME.  We do this for the example of computing $\pi$ without a sign problem.  In Figs.~\ref{fig:gc-vs-nr} and~\ref{fig:gc-vs-ng} we see the total $R_{Z}$ and CNOT gate counts as a function of $N_{r}$, and $N_{g}$, respectively.  In both cases we see an exponential dependence. 
This can be understood as follows: an increase in $N_r$ exponentially increases the precision of the result in QPE.
Likewise, as we increase $N_{g}$ the grid fineness increases, and the accuracy improves exponentially.  The  growth in gate count is therefore a reflection of the  improvement in the accuracy and precision of the result with growth in both $N_{g}$ and $N_{r}$.

\section{Conclusion}
\label{sec:conclusion}

We have shown how QME can estimate Euclidean LFT observables.  The method relies on QPE and provides a quadratic speed-up over traditional methods. The possible improved scaling in the calculation of expectation values is tantalizing.  While the gate costs seem to require fault-tolerant quantum computers, further optimizations can reduce those costs.  Possible directions include improving QPE~\cite{Mohammadbagherpoor:2019,Smith:2022,Chakrabarti_2021} over the version used here, or advancements in encoding phases into the result register in other formats, \emph{e.g.} floating point.
Crucially, quantum state preparation of the probability distribution dominates the resource estimate.  However when using amplitude amplification, it is still quadratically faster than classically constructing the distribution. 
Further reductions may occur by fitting the measured phase distribution instead of taking only the most likely value.

Quantum mean estimation also provides an alternative perspective on sampling.  When calculating expectation values, the quantum system must be prepared, measured, and re-prepared.  Similarly, classical sampling algorithms follow this pattern, albeit obfuscated.  Quantum mean estimation instead embeds the preparation and measurement actions into the quantum algorithm itself via quantum superposition.  The average is returned with high probability, meaning, the only sampling necessary is used to distinguish the correct bit-string from others.
Moreover this speed-up persists even for sign problems and near criticality.  This advantage is appealing for LFT practitioners, where these issues are prohibitive. Conversely, the speed-up provided by QME is \emph{only} polynomial, and hence, the prefactors will dictate its utility.  Applying the algorithm on real quantum hardware with simple examples will elucidate its long-term scaling and practicality.

\begin{acknowledgments}
The authors thank Prasanth Shyamsundar for reading drafts of the manuscript, and Michael Wagman for stimulating discussions. This work is supported by the Department of Energy through the Fermilab QuantiSED program in the area of ``Intersections of QIS and Theoretical Particle Physics". Fermilab is operated by Fermi Research Alliance, LLC under contract number DE-AC02-07CH11359 with the United States Department of Energy.
\end{acknowledgments}

% \bibliography{refs}

\begin{thebibliography}{122}%
\makeatletter
\providecommand \@ifxundefined [1]{%
 \@ifx{#1\undefined}
}%
\providecommand \@ifnum [1]{%
 \ifnum #1\expandafter \@firstoftwo
 \else \expandafter \@secondoftwo
 \fi
}%
\providecommand \@ifx [1]{%
 \ifx #1\expandafter \@firstoftwo
 \else \expandafter \@secondoftwo
 \fi
}%
\providecommand \natexlab [1]{#1}%
\providecommand \enquote  [1]{``#1''}%
\providecommand \bibnamefont  [1]{#1}%
\providecommand \bibfnamefont [1]{#1}%
\providecommand \citenamefont [1]{#1}%
\providecommand \href@noop [0]{\@secondoftwo}%
\providecommand \href [0]{\begingroup \@sanitize@url \@href}%
\providecommand \@href[1]{\@@startlink{#1}\@@href}%
\providecommand \@@href[1]{\endgroup#1\@@endlink}%
\providecommand \@sanitize@url [0]{\catcode `\\12\catcode `\$12\catcode
  `\&12\catcode `\#12\catcode `\^12\catcode `\_12\catcode `\%12\relax}%
\providecommand \@@startlink[1]{}%
\providecommand \@@endlink[0]{}%
\providecommand \url  [0]{\begingroup\@sanitize@url \@url }%
\providecommand \@url [1]{\endgroup\@href {#1}{\urlprefix }}%
\providecommand \urlprefix  [0]{URL }%
\providecommand \Eprint [0]{\href }%
\providecommand \doibase [0]{https://doi.org/}%
\providecommand \selectlanguage [0]{\@gobble}%
\providecommand \bibinfo  [0]{\@secondoftwo}%
\providecommand \bibfield  [0]{\@secondoftwo}%
\providecommand \translation [1]{[#1]}%
\providecommand \BibitemOpen [0]{}%
\providecommand \bibitemStop [0]{}%
\providecommand \bibitemNoStop [0]{.\EOS\space}%
\providecommand \EOS [0]{\spacefactor3000\relax}%
\providecommand \BibitemShut  [1]{\csname bibitem#1\endcsname}%
\let\auto@bib@innerbib\@empty
%</preamble>
\bibitem [{\citenamefont {Montvay}\ and\ \citenamefont
  {M{\"{u}}nster}(1994)}]{montvay_munster_1994}%
  \BibitemOpen
  \bibfield  {author} {\bibinfo {author} {\bibfnamefont {I.}~\bibnamefont
  {Montvay}}\ and\ \bibinfo {author} {\bibfnamefont {G.}~\bibnamefont
  {M{\"{u}}nster}},\ }\href {https://doi.org/10.1017/CBO9780511470783} {\emph
  {\bibinfo {title} {Quantum Fields on a Lattice}}},\ Cambridge Monographs on
  Mathematical Physics\ (\bibinfo  {publisher} {Cambridge University Press},\
  \bibinfo {year} {1994})\BibitemShut {NoStop}%
\bibitem [{\citenamefont {Gattringer}\ and\ \citenamefont
  {Lang}(2010)}]{gattringer}%
  \BibitemOpen
  \bibfield  {author} {\bibinfo {author} {\bibfnamefont {C.}~\bibnamefont
  {Gattringer}}\ and\ \bibinfo {author} {\bibfnamefont {C.~B.}\ \bibnamefont
  {Lang}},\ }\href {https://doi.org/10.1007/978-3-642-01850-3} {\emph {\bibinfo
  {title} {Quantum Chromodynamics on the Lattice}}},\ Lecture Notes in Physics\
  (\bibinfo  {publisher} {Springer Berlin, Heidelberg},\ \bibinfo {year}
  {2010})\BibitemShut {NoStop}%
\bibitem [{\citenamefont {Kogut}(1979)}]{kogut:1979}%
  \BibitemOpen
  \bibfield  {author} {\bibinfo {author} {\bibfnamefont {J.~B.}\ \bibnamefont
  {Kogut}},\ }\bibfield  {title} {\bibinfo {title} {An introduction to lattice
  gauge theory and spin systems},\ }\href
  {https://doi.org/10.1103/RevModPhys.51.659} {\bibfield  {journal} {\bibinfo
  {journal} {Rev. Mod. Phys.}\ }\textbf {\bibinfo {volume} {51}},\ \bibinfo
  {pages} {659} (\bibinfo {year} {1979})}\BibitemShut {NoStop}%
\bibitem [{\citenamefont {Bazavov}\ \emph {et~al.}(2022)\citenamefont {Bazavov}
  \emph {et~al.}}]{FermilabLattice:2021cdg}%
  \BibitemOpen
  \bibfield  {author} {\bibinfo {author} {\bibfnamefont {A.}~\bibnamefont
  {Bazavov}} \emph {et~al.} (\bibinfo {collaboration} {Fermilab Lattice, MILC,
  Fermilab Lattice, MILC}),\ }\bibfield  {title} {\bibinfo {title}
  {{Semileptonic form factors for $B\rightarrow D^*\ell \nu $ at nonzero recoil
  from $2+1$-flavor lattice QCD: Fermilab Lattice~and~MILC~Collaborations}},\
  }\href {https://doi.org/10.1140/epjc/s10052-022-10984-9} {\bibfield
  {journal} {\bibinfo  {journal} {Eur. Phys. J. C}\ }\textbf {\bibinfo {volume}
  {82}},\ \bibinfo {pages} {1141} (\bibinfo {year} {2022})},\ \bibinfo {note}
  {[Erratum: Eur.Phys.J.C 83, 21 (2023)]},\ \Eprint
  {https://arxiv.org/abs/2105.14019} {arXiv:2105.14019 [hep-lat]} \BibitemShut
  {NoStop}%
\bibitem [{\citenamefont {Borsanyi}\ \emph {et~al.}(2021)\citenamefont
  {Borsanyi} \emph {et~al.}}]{Borsanyi:2020mff}%
  \BibitemOpen
  \bibfield  {author} {\bibinfo {author} {\bibfnamefont {S.}~\bibnamefont
  {Borsanyi}} \emph {et~al.},\ }\bibfield  {title} {\bibinfo {title} {{Leading
  hadronic contribution to the muon magnetic moment from lattice QCD}},\ }\href
  {https://doi.org/10.1038/s41586-021-03418-1} {\bibfield  {journal} {\bibinfo
  {journal} {Nature}\ }\textbf {\bibinfo {volume} {593}},\ \bibinfo {pages}
  {51} (\bibinfo {year} {2021})},\ \Eprint {https://arxiv.org/abs/2002.12347}
  {arXiv:2002.12347 [hep-lat]} \BibitemShut {NoStop}%
\bibitem [{\citenamefont {de~Forcrand}(2009)}]{deForcrand:2009zkb}%
  \BibitemOpen
  \bibfield  {author} {\bibinfo {author} {\bibfnamefont {P.}~\bibnamefont
  {de~Forcrand}},\ }\bibfield  {title} {\bibinfo {title} {{Simulating QCD at
  finite density}},\ }\href {https://doi.org/10.22323/1.091.0010} {\bibfield
  {journal} {\bibinfo  {journal} {PoS}\ }\textbf {\bibinfo {volume}
  {LAT2009}},\ \bibinfo {pages} {010} (\bibinfo {year} {2009})},\ \Eprint
  {https://arxiv.org/abs/1005.0539} {arXiv:1005.0539 [hep-lat]} \BibitemShut
  {NoStop}%
\bibitem [{\citenamefont {Tripolt}\ \emph {et~al.}(2019)\citenamefont
  {Tripolt}, \citenamefont {Gubler}, \citenamefont {Ulybyshev},\ and\
  \citenamefont {Von~Smekal}}]{Tripolt:2018xeo}%
  \BibitemOpen
  \bibfield  {author} {\bibinfo {author} {\bibfnamefont {R.-A.}\ \bibnamefont
  {Tripolt}}, \bibinfo {author} {\bibfnamefont {P.}~\bibnamefont {Gubler}},
  \bibinfo {author} {\bibfnamefont {M.}~\bibnamefont {Ulybyshev}},\ and\
  \bibinfo {author} {\bibfnamefont {L.}~\bibnamefont {Von~Smekal}},\ }\bibfield
   {title} {\bibinfo {title} {{Numerical analytic continuation of Euclidean
  data}},\ }\href {https://doi.org/10.1016/j.cpc.2018.11.012} {\bibfield
  {journal} {\bibinfo  {journal} {Comput. Phys. Commun.}\ }\textbf {\bibinfo
  {volume} {237}},\ \bibinfo {pages} {129} (\bibinfo {year} {2019})},\ \Eprint
  {https://arxiv.org/abs/1801.10348} {arXiv:1801.10348 [hep-ph]} \BibitemShut
  {NoStop}%
\bibitem [{\citenamefont {Schaefer}\ \emph {et~al.}(2011)\citenamefont
  {Schaefer}, \citenamefont {Sommer},\ and\ \citenamefont
  {Virotta}}]{Schaefer:2010hu}%
  \BibitemOpen
  \bibfield  {author} {\bibinfo {author} {\bibfnamefont {S.}~\bibnamefont
  {Schaefer}}, \bibinfo {author} {\bibfnamefont {R.}~\bibnamefont {Sommer}},\
  and\ \bibinfo {author} {\bibfnamefont {F.}~\bibnamefont {Virotta}} (\bibinfo
  {collaboration} {ALPHA}),\ }\bibfield  {title} {\bibinfo {title} {{Critical
  slowing down and error analysis in lattice QCD simulations}},\ }\href
  {https://doi.org/10.1016/j.nuclphysb.2010.11.020} {\bibfield  {journal}
  {\bibinfo  {journal} {Nucl. Phys. B}\ }\textbf {\bibinfo {volume} {845}},\
  \bibinfo {pages} {93} (\bibinfo {year} {2011})},\ \Eprint
  {https://arxiv.org/abs/1009.5228} {arXiv:1009.5228 [hep-lat]} \BibitemShut
  {NoStop}%
\bibitem [{\citenamefont {Philipsen}(2019)}]{Philipsen:2019rjq}%
  \BibitemOpen
  \bibfield  {author} {\bibinfo {author} {\bibfnamefont {O.}~\bibnamefont
  {Philipsen}},\ }\bibfield  {title} {\bibinfo {title} {{Constraining the phase
  diagram of QCD at finite temperature and density}},\ }\href
  {https://doi.org/10.22323/1.363.0273} {\bibfield  {journal} {\bibinfo
  {journal} {PoS}\ }\textbf {\bibinfo {volume} {LATTICE2019}},\ \bibinfo
  {pages} {273} (\bibinfo {year} {2019})},\ \Eprint
  {https://arxiv.org/abs/1912.04827} {arXiv:1912.04827 [hep-lat]} \BibitemShut
  {NoStop}%
\bibitem [{\citenamefont {Alexandru}\ \emph {et~al.}(2022)\citenamefont
  {Alexandru}, \citenamefont {Basar}, \citenamefont {Bedaque},\ and\
  \citenamefont {Warrington}}]{Alexandru:2020wrj}%
  \BibitemOpen
  \bibfield  {author} {\bibinfo {author} {\bibfnamefont {A.}~\bibnamefont
  {Alexandru}}, \bibinfo {author} {\bibfnamefont {G.}~\bibnamefont {Basar}},
  \bibinfo {author} {\bibfnamefont {P.~F.}\ \bibnamefont {Bedaque}},\ and\
  \bibinfo {author} {\bibfnamefont {N.~C.}\ \bibnamefont {Warrington}},\
  }\bibfield  {title} {\bibinfo {title} {{Complex paths around the sign
  problem}},\ }\href {https://doi.org/10.1103/RevModPhys.94.015006} {\bibfield
  {journal} {\bibinfo  {journal} {Rev. Mod. Phys.}\ }\textbf {\bibinfo {volume}
  {94}},\ \bibinfo {pages} {015006} (\bibinfo {year} {2022})},\ \Eprint
  {https://arxiv.org/abs/2007.05436} {arXiv:2007.05436 [hep-lat]} \BibitemShut
  {NoStop}%
\bibitem [{\citenamefont {Del~Debbio}\ \emph {et~al.}(2004)\citenamefont
  {Del~Debbio}, \citenamefont {Manca},\ and\ \citenamefont
  {Vicari}}]{DelDebbio:2004xh}%
  \BibitemOpen
  \bibfield  {author} {\bibinfo {author} {\bibfnamefont {L.}~\bibnamefont
  {Del~Debbio}}, \bibinfo {author} {\bibfnamefont {G.~M.}\ \bibnamefont
  {Manca}},\ and\ \bibinfo {author} {\bibfnamefont {E.}~\bibnamefont
  {Vicari}},\ }\bibfield  {title} {\bibinfo {title} {{Critical slowing down of
  topological modes}},\ }\href {https://doi.org/10.1016/j.physletb.2004.05.038}
  {\bibfield  {journal} {\bibinfo  {journal} {Phys. Lett. B}\ }\textbf
  {\bibinfo {volume} {594}},\ \bibinfo {pages} {315} (\bibinfo {year}
  {2004})},\ \Eprint {https://arxiv.org/abs/hep-lat/0403001}
  {arXiv:hep-lat/0403001} \BibitemShut {NoStop}%
\bibitem [{\citenamefont {Swendsen}\ and\ \citenamefont
  {Wang}(1987)}]{swendsen-wang:87}%
  \BibitemOpen
  \bibfield  {author} {\bibinfo {author} {\bibfnamefont {R.~H.}\ \bibnamefont
  {Swendsen}}\ and\ \bibinfo {author} {\bibfnamefont {J.-S.}\ \bibnamefont
  {Wang}},\ }\bibfield  {title} {\bibinfo {title} {Nonuniversal critical
  dynamics in monte carlo simulations},\ }\href
  {https://doi.org/10.1103/PhysRevLett.58.86} {\bibfield  {journal} {\bibinfo
  {journal} {Phys. Rev. Lett.}\ }\textbf {\bibinfo {volume} {58}},\ \bibinfo
  {pages} {86} (\bibinfo {year} {1987})}\BibitemShut {NoStop}%
\bibitem [{\citenamefont {Wolff}(1989)}]{wolff:89}%
  \BibitemOpen
  \bibfield  {author} {\bibinfo {author} {\bibfnamefont {U.}~\bibnamefont
  {Wolff}},\ }\bibfield  {title} {\bibinfo {title} {Collective monte carlo
  updating for spin systems},\ }\href
  {https://doi.org/10.1103/PhysRevLett.62.361} {\bibfield  {journal} {\bibinfo
  {journal} {Phys. Rev. Lett.}\ }\textbf {\bibinfo {volume} {62}},\ \bibinfo
  {pages} {361} (\bibinfo {year} {1989})}\BibitemShut {NoStop}%
\bibitem [{\citenamefont {Savit}(1980)}]{RevModPhys.52.453}%
  \BibitemOpen
  \bibfield  {author} {\bibinfo {author} {\bibfnamefont {R.}~\bibnamefont
  {Savit}},\ }\bibfield  {title} {\bibinfo {title} {Duality in field theory and
  statistical systems},\ }\href {https://doi.org/10.1103/RevModPhys.52.453}
  {\bibfield  {journal} {\bibinfo  {journal} {Rev. Mod. Phys.}\ }\textbf
  {\bibinfo {volume} {52}},\ \bibinfo {pages} {453} (\bibinfo {year}
  {1980})}\BibitemShut {NoStop}%
\bibitem [{\citenamefont {Marchis}\ and\ \citenamefont
  {Gattringer}(2018)}]{Marchis:2018}%
  \BibitemOpen
  \bibfield  {author} {\bibinfo {author} {\bibfnamefont {C.}~\bibnamefont
  {Marchis}}\ and\ \bibinfo {author} {\bibfnamefont {C.}~\bibnamefont
  {Gattringer}},\ }\bibfield  {title} {\bibinfo {title} {Dual representation of
  lattice qcd with worldlines and worldsheets of abelian color fluxes},\ }\href
  {https://doi.org/10.1103/PhysRevD.97.034508} {\bibfield  {journal} {\bibinfo
  {journal} {Phys. Rev. D}\ }\textbf {\bibinfo {volume} {97}},\ \bibinfo
  {pages} {034508} (\bibinfo {year} {2018})}\BibitemShut {NoStop}%
\bibitem [{\citenamefont {Or{\'u}s}(2019)}]{Orus2019}%
  \BibitemOpen
  \bibfield  {author} {\bibinfo {author} {\bibfnamefont {R.}~\bibnamefont
  {Or{\'u}s}},\ }\bibfield  {title} {\bibinfo {title} {Tensor networks for
  complex quantum systems},\ }\href {https://doi.org/10.1038/s42254-019-0086-7}
  {\bibfield  {journal} {\bibinfo  {journal} {Nature Reviews Physics}\ }\textbf
  {\bibinfo {volume} {1}},\ \bibinfo {pages} {538} (\bibinfo {year}
  {2019})}\BibitemShut {NoStop}%
\bibitem [{\citenamefont {Meurice}\ \emph {et~al.}(2022)\citenamefont
  {Meurice}, \citenamefont {Sakai},\ and\ \citenamefont
  {Unmuth-Yockey}}]{RevModPhys.94.025005}%
  \BibitemOpen
  \bibfield  {author} {\bibinfo {author} {\bibfnamefont {Y.}~\bibnamefont
  {Meurice}}, \bibinfo {author} {\bibfnamefont {R.}~\bibnamefont {Sakai}},\
  and\ \bibinfo {author} {\bibfnamefont {J.}~\bibnamefont {Unmuth-Yockey}},\
  }\bibfield  {title} {\bibinfo {title} {Tensor lattice field theory for
  renormalization and quantum computing},\ }\href
  {https://doi.org/10.1103/RevModPhys.94.025005} {\bibfield  {journal}
  {\bibinfo  {journal} {Rev. Mod. Phys.}\ }\textbf {\bibinfo {volume} {94}},\
  \bibinfo {pages} {025005} (\bibinfo {year} {2022})}\BibitemShut {NoStop}%
\bibitem [{\citenamefont {Langfeld}(2017)}]{klangfeld:2016}%
  \BibitemOpen
  \bibfield  {author} {\bibinfo {author} {\bibfnamefont {K.}~\bibnamefont
  {Langfeld}},\ }\href {https://doi.org/10.22323/1.256.0010} {\bibinfo {title}
  {{Density-of-states}}} (\bibinfo {year} {2017}),\ \Eprint
  {https://arxiv.org/abs/1610.09856} {arXiv:1610.09856 [hep-lat]} \BibitemShut
  {NoStop}%
\bibitem [{\citenamefont {Troyer}\ and\ \citenamefont
  {Wiese}(2005)}]{Troyer:2004ge}%
  \BibitemOpen
  \bibfield  {author} {\bibinfo {author} {\bibfnamefont {M.}~\bibnamefont
  {Troyer}}\ and\ \bibinfo {author} {\bibfnamefont {U.-J.}\ \bibnamefont
  {Wiese}},\ }\bibfield  {title} {\bibinfo {title} {{Computational complexity
  and fundamental limitations to fermionic quantum Monte Carlo simulations}},\
  }\href {https://doi.org/10.1103/PhysRevLett.94.170201} {\bibfield  {journal}
  {\bibinfo  {journal} {Phys. Rev. Lett.}\ }\textbf {\bibinfo {volume} {94}},\
  \bibinfo {pages} {170201} (\bibinfo {year} {2005})},\ \Eprint
  {https://arxiv.org/abs/cond-mat/0408370} {arXiv:cond-mat/0408370 [cond-mat]}
  \BibitemShut {NoStop}%
%%CITATION = COND-MAT/0408370;%%
\bibitem [{\citenamefont {Cleve}\ \emph {et~al.}(1998)\citenamefont {Cleve},
  \citenamefont {Ekert}, \citenamefont {Macchiavello},\ and\ \citenamefont
  {Mosca}}]{Cleve_1998}%
  \BibitemOpen
  \bibfield  {author} {\bibinfo {author} {\bibfnamefont {R.}~\bibnamefont
  {Cleve}}, \bibinfo {author} {\bibfnamefont {A.}~\bibnamefont {Ekert}},
  \bibinfo {author} {\bibfnamefont {C.}~\bibnamefont {Macchiavello}},\ and\
  \bibinfo {author} {\bibfnamefont {M.}~\bibnamefont {Mosca}},\ }\bibfield
  {title} {\bibinfo {title} {Quantum algorithms revisited},\ }\href
  {https://doi.org/10.1098/rspa.1998.0164} {\bibfield  {journal} {\bibinfo
  {journal} {Proceedings of the Royal Society of London. Series A:
  Mathematical, Physical and Engineering Sciences}\ }\textbf {\bibinfo {volume}
  {454}},\ \bibinfo {pages} {339} (\bibinfo {year} {1998})}\BibitemShut
  {NoStop}%
\bibitem [{\citenamefont {Griffiths}\ and\ \citenamefont
  {Niu}(1996)}]{PhysRevLett.76.3228}%
  \BibitemOpen
  \bibfield  {author} {\bibinfo {author} {\bibfnamefont {R.~B.}\ \bibnamefont
  {Griffiths}}\ and\ \bibinfo {author} {\bibfnamefont {C.-S.}\ \bibnamefont
  {Niu}},\ }\bibfield  {title} {\bibinfo {title} {Semiclassical fourier
  transform for quantum computation},\ }\href
  {https://doi.org/10.1103/PhysRevLett.76.3228} {\bibfield  {journal} {\bibinfo
   {journal} {Phys. Rev. Lett.}\ }\textbf {\bibinfo {volume} {76}},\ \bibinfo
  {pages} {3228} (\bibinfo {year} {1996})}\BibitemShut {NoStop}%
\bibitem [{\citenamefont {Coppersmith}(2002)}]{coppersmith2002approximate}%
  \BibitemOpen
  \bibfield  {author} {\bibinfo {author} {\bibfnamefont {D.}~\bibnamefont
  {Coppersmith}},\ }\href@noop {} {\bibinfo {title} {{An approximate Fourier
  transform useful in quantum factoring}}} (\bibinfo {year} {2002}),\ \Eprint
  {https://arxiv.org/abs/0201067} {arXiv:0201067 [quant-ph]} \BibitemShut
  {NoStop}%
\bibitem [{\citenamefont {Kitaev}(1995)}]{kitaev:1995}%
  \BibitemOpen
  \bibfield  {author} {\bibinfo {author} {\bibfnamefont {A.~Y.}\ \bibnamefont
  {Kitaev}},\ }\href@noop {} {\bibinfo {title} {{Quantum measurements and the
  Abelian stabilizer problem}}} (\bibinfo {year} {1995}),\ \Eprint
  {https://arxiv.org/abs/quant-ph/9511026} {arXiv:quant-ph/9511026}
  \BibitemShut {NoStop}%
\bibitem [{\citenamefont {Chapeau-Blondeau}\ and\ \citenamefont
  {Belin}(2020)}]{Chapeau-Blondeau2020}%
  \BibitemOpen
  \bibfield  {author} {\bibinfo {author} {\bibfnamefont {F.}~\bibnamefont
  {Chapeau-Blondeau}}\ and\ \bibinfo {author} {\bibfnamefont {E.}~\bibnamefont
  {Belin}},\ }\bibfield  {title} {\bibinfo {title} {Quantum signal processing
  for quantum phase estimation: Fourier transform versus maximum likelihood
  approaches},\ }\href {https://doi.org/10.1007/s12243-020-00803-1} {\bibfield
  {journal} {\bibinfo  {journal} {Annals of Telecommunications}\ }\textbf
  {\bibinfo {volume} {75}},\ \bibinfo {pages} {641} (\bibinfo {year}
  {2020})}\BibitemShut {NoStop}%
\bibitem [{\citenamefont {Smith}\ \emph {et~al.}(2022)\citenamefont {Smith},
  \citenamefont {Barnes},\ and\ \citenamefont {Arvidsson-Shukur}}]{Smith:2022}%
  \BibitemOpen
  \bibfield  {author} {\bibinfo {author} {\bibfnamefont {J.~G.}\ \bibnamefont
  {Smith}}, \bibinfo {author} {\bibfnamefont {C.~H.~W.}\ \bibnamefont
  {Barnes}},\ and\ \bibinfo {author} {\bibfnamefont {D.~R.~M.}\ \bibnamefont
  {Arvidsson-Shukur}},\ }\href@noop {} {\bibinfo {title} {{An iterative
  quantum-phase-estimation protocol for near-term quantum hardware}}} (\bibinfo
  {year} {2022}),\ \Eprint {https://arxiv.org/abs/2206.06392} {arXiv:2206.06392
  [quant-ph]} \BibitemShut {NoStop}%
\bibitem [{\citenamefont {Nielsen}\ and\ \citenamefont
  {Chuang}(2010)}]{nielsen_chuang_2010}%
  \BibitemOpen
  \bibfield  {author} {\bibinfo {author} {\bibfnamefont {M.~A.}\ \bibnamefont
  {Nielsen}}\ and\ \bibinfo {author} {\bibfnamefont {I.~L.}\ \bibnamefont
  {Chuang}},\ }\href {https://doi.org/10.1017/CBO9780511976667} {\emph
  {\bibinfo {title} {Quantum Computation and Quantum Information: 10th
  Anniversary Edition}}}\ (\bibinfo  {publisher} {Cambridge University Press},\
  \bibinfo {year} {2010})\BibitemShut {NoStop}%
\bibitem [{\citenamefont {Feynman}(1982)}]{Feynman:1981tf}%
  \BibitemOpen
  \bibfield  {author} {\bibinfo {author} {\bibfnamefont {R.~P.}\ \bibnamefont
  {Feynman}},\ }\bibfield  {title} {\bibinfo {title} {{Simulating physics with
  computers}},\ }\href {https://doi.org/10.1007/BF02650179} {\bibfield
  {journal} {\bibinfo  {journal} {Int. J. Theor. Phys.}\ }\textbf {\bibinfo
  {volume} {21}},\ \bibinfo {pages} {467} (\bibinfo {year} {1982})}\BibitemShut
  {NoStop}%
%%CITATION = IJTPB,21,467;%%
\bibitem [{\citenamefont {Lloyd}(1996)}]{Lloyd1073}%
  \BibitemOpen
  \bibfield  {author} {\bibinfo {author} {\bibfnamefont {S.}~\bibnamefont
  {Lloyd}},\ }\bibfield  {title} {\bibinfo {title} {Universal quantum
  simulators},\ }\href {https://doi.org/10.1126/science.273.5278.1073}
  {\bibfield  {journal} {\bibinfo  {journal} {Science}\ }\textbf {\bibinfo
  {volume} {273}},\ \bibinfo {pages} {1073} (\bibinfo {year}
  {1996})}\BibitemShut {NoStop}%
\bibitem [{\citenamefont {Jordan}\ \emph {et~al.}(2012)\citenamefont {Jordan},
  \citenamefont {Lee},\ and\ \citenamefont {Preskill}}]{Jordan:2011ne}%
  \BibitemOpen
  \bibfield  {author} {\bibinfo {author} {\bibfnamefont {S.~P.}\ \bibnamefont
  {Jordan}}, \bibinfo {author} {\bibfnamefont {K.~S.~M.}\ \bibnamefont {Lee}},\
  and\ \bibinfo {author} {\bibfnamefont {J.}~\bibnamefont {Preskill}},\
  }\bibfield  {title} {\bibinfo {title} {{Quantum Algorithms for Quantum Field
  Theories}},\ }\href {https://doi.org/10.1126/science.1217069} {\bibfield
  {journal} {\bibinfo  {journal} {Science}\ }\textbf {\bibinfo {volume}
  {336}},\ \bibinfo {pages} {1130} (\bibinfo {year} {2012})},\ \Eprint
  {https://arxiv.org/abs/1111.3633} {arXiv:1111.3633 [quant-ph]} \BibitemShut
  {NoStop}%
%%CITATION = ARXIV:1111.3633;%%
\bibitem [{\citenamefont {Jordan}\ \emph {et~al.}(2018)\citenamefont {Jordan},
  \citenamefont {Krovi}, \citenamefont {Lee},\ and\ \citenamefont
  {Preskill}}]{Jordan:2017lea}%
  \BibitemOpen
  \bibfield  {author} {\bibinfo {author} {\bibfnamefont {S.~P.}\ \bibnamefont
  {Jordan}}, \bibinfo {author} {\bibfnamefont {H.}~\bibnamefont {Krovi}},
  \bibinfo {author} {\bibfnamefont {K.~S.}\ \bibnamefont {Lee}},\ and\ \bibinfo
  {author} {\bibfnamefont {J.}~\bibnamefont {Preskill}},\ }\bibfield  {title}
  {\bibinfo {title} {{BQP-completeness of Scattering in Scalar Quantum Field
  Theory}},\ }\href {https://doi.org/10.22331/q-2018-01-08-44} {\bibfield
  {journal} {\bibinfo  {journal} {Quantum}\ }\textbf {\bibinfo {volume} {2}},\
  \bibinfo {pages} {44} (\bibinfo {year} {2018})},\ \Eprint
  {https://arxiv.org/abs/1703.00454} {arXiv:1703.00454 [quant-ph]} \BibitemShut
  {NoStop}%
\bibitem [{\citenamefont {Klco}\ \emph {et~al.}(2022)\citenamefont {Klco},
  \citenamefont {Roggero},\ and\ \citenamefont {Savage}}]{klco2021standard}%
  \BibitemOpen
  \bibfield  {author} {\bibinfo {author} {\bibfnamefont {N.}~\bibnamefont
  {Klco}}, \bibinfo {author} {\bibfnamefont {A.}~\bibnamefont {Roggero}},\ and\
  \bibinfo {author} {\bibfnamefont {M.~J.}\ \bibnamefont {Savage}},\ }\bibfield
   {title} {\bibinfo {title} {{Standard model physics and the digital quantum
  revolution: thoughts about the interface}},\ }\href
  {https://doi.org/10.1088/1361-6633/ac58a4} {\bibfield  {journal} {\bibinfo
  {journal} {Rept. Prog. Phys.}\ }\textbf {\bibinfo {volume} {85}},\ \bibinfo
  {pages} {064301} (\bibinfo {year} {2022})},\ \Eprint
  {https://arxiv.org/abs/2107.04769} {arXiv:2107.04769 [quant-ph]} \BibitemShut
  {NoStop}%
\bibitem [{\citenamefont {Bauer}\ \emph {et~al.}(2022)\citenamefont {Bauer}
  \emph {et~al.}}]{Bauer:2022hpo}%
  \BibitemOpen
  \bibfield  {author} {\bibinfo {author} {\bibfnamefont {C.~W.}\ \bibnamefont
  {Bauer}} \emph {et~al.},\ }\href@noop {} {\bibinfo {title} {{Quantum
  Simulation for High Energy Physics}}} (\bibinfo {year} {2022}),\ \Eprint
  {https://arxiv.org/abs/2204.03381} {arXiv:2204.03381 [quant-ph]} \BibitemShut
  {NoStop}%
\bibitem [{\citenamefont {Zohar}\ \emph {et~al.}(2012)\citenamefont {Zohar},
  \citenamefont {Cirac},\ and\ \citenamefont {Reznik}}]{Zohar:2012ay}%
  \BibitemOpen
  \bibfield  {author} {\bibinfo {author} {\bibfnamefont {E.}~\bibnamefont
  {Zohar}}, \bibinfo {author} {\bibfnamefont {J.~I.}\ \bibnamefont {Cirac}},\
  and\ \bibinfo {author} {\bibfnamefont {B.}~\bibnamefont {Reznik}},\
  }\bibfield  {title} {\bibinfo {title} {{Simulating Compact Quantum
  Electrodynamics with ultracold atoms: Probing confinement and nonperturbative
  effects}},\ }\href {https://doi.org/10.1103/PhysRevLett.109.125302}
  {\bibfield  {journal} {\bibinfo  {journal} {Phys. Rev. Lett.}\ }\textbf
  {\bibinfo {volume} {109}},\ \bibinfo {pages} {125302} (\bibinfo {year}
  {2012})},\ \Eprint {https://arxiv.org/abs/1204.6574} {arXiv:1204.6574
  [quant-ph]} \BibitemShut {NoStop}%
%%CITATION = ARXIV:1204.6574;%%
\bibitem [{\citenamefont {Zohar}\ \emph
  {et~al.}(2013{\natexlab{a}})\citenamefont {Zohar}, \citenamefont {Cirac},\
  and\ \citenamefont {Reznik}}]{Zohar:2012xf}%
  \BibitemOpen
  \bibfield  {author} {\bibinfo {author} {\bibfnamefont {E.}~\bibnamefont
  {Zohar}}, \bibinfo {author} {\bibfnamefont {J.~I.}\ \bibnamefont {Cirac}},\
  and\ \bibinfo {author} {\bibfnamefont {B.}~\bibnamefont {Reznik}},\
  }\bibfield  {title} {\bibinfo {title} {{Cold-Atom Quantum Simulator for SU(2)
  Yang-Mills Lattice Gauge Theory}},\ }\href
  {https://doi.org/10.1103/PhysRevLett.110.125304} {\bibfield  {journal}
  {\bibinfo  {journal} {Phys. Rev. Lett.}\ }\textbf {\bibinfo {volume} {110}},\
  \bibinfo {pages} {125304} (\bibinfo {year} {2013}{\natexlab{a}})},\ \Eprint
  {https://arxiv.org/abs/1211.2241} {arXiv:1211.2241 [quant-ph]} \BibitemShut
  {NoStop}%
%%CITATION = ARXIV:1211.2241;%%
\bibitem [{\citenamefont {Zohar}\ \emph
  {et~al.}(2013{\natexlab{b}})\citenamefont {Zohar}, \citenamefont {Cirac},\
  and\ \citenamefont {Reznik}}]{Zohar:2013zla}%
  \BibitemOpen
  \bibfield  {author} {\bibinfo {author} {\bibfnamefont {E.}~\bibnamefont
  {Zohar}}, \bibinfo {author} {\bibfnamefont {J.~I.}\ \bibnamefont {Cirac}},\
  and\ \bibinfo {author} {\bibfnamefont {B.}~\bibnamefont {Reznik}},\
  }\bibfield  {title} {\bibinfo {title} {{Quantum simulations of gauge theories
  with ultracold atoms: local gauge invariance from angular momentum
  conservation}},\ }\href {https://doi.org/10.1103/PhysRevA.88.023617}
  {\bibfield  {journal} {\bibinfo  {journal} {Phys. Rev.}\ }\textbf {\bibinfo
  {volume} {A88}},\ \bibinfo {pages} {023617} (\bibinfo {year}
  {2013}{\natexlab{b}})},\ \Eprint {https://arxiv.org/abs/1303.5040}
  {arXiv:1303.5040 [quant-ph]} \BibitemShut {NoStop}%
%%CITATION = ARXIV:1303.5040;%%
\bibitem [{\citenamefont {Zohar}\ and\ \citenamefont
  {Burrello}(2015)}]{Zohar:2014qma}%
  \BibitemOpen
  \bibfield  {author} {\bibinfo {author} {\bibfnamefont {E.}~\bibnamefont
  {Zohar}}\ and\ \bibinfo {author} {\bibfnamefont {M.}~\bibnamefont
  {Burrello}},\ }\bibfield  {title} {\bibinfo {title} {{Formulation of lattice
  gauge theories for quantum simulations}},\ }\href
  {https://doi.org/10.1103/PhysRevD.91.054506} {\bibfield  {journal} {\bibinfo
  {journal} {Phys. Rev.}\ }\textbf {\bibinfo {volume} {D91}},\ \bibinfo {pages}
  {054506} (\bibinfo {year} {2015})},\ \Eprint
  {https://arxiv.org/abs/1409.3085} {arXiv:1409.3085 [quant-ph]} \BibitemShut
  {NoStop}%
%%CITATION = ARXIV:1409.3085;%%
\bibitem [{\citenamefont {Zohar}\ \emph {et~al.}(2016)\citenamefont {Zohar},
  \citenamefont {Cirac},\ and\ \citenamefont {Reznik}}]{Zohar:2015hwa}%
  \BibitemOpen
  \bibfield  {author} {\bibinfo {author} {\bibfnamefont {E.}~\bibnamefont
  {Zohar}}, \bibinfo {author} {\bibfnamefont {J.~I.}\ \bibnamefont {Cirac}},\
  and\ \bibinfo {author} {\bibfnamefont {B.}~\bibnamefont {Reznik}},\
  }\bibfield  {title} {\bibinfo {title} {{Quantum Simulations of Lattice Gauge
  Theories using Ultracold Atoms in Optical Lattices}},\ }\href
  {https://doi.org/10.1088/0034-4885/79/1/014401} {\bibfield  {journal}
  {\bibinfo  {journal} {Rept. Prog. Phys.}\ }\textbf {\bibinfo {volume} {79}},\
  \bibinfo {pages} {014401} (\bibinfo {year} {2016})},\ \Eprint
  {https://arxiv.org/abs/1503.02312} {arXiv:1503.02312 [quant-ph]} \BibitemShut
  {NoStop}%
%%CITATION = ARXIV:1503.02312;%%
\bibitem [{\citenamefont {Zohar}\ \emph {et~al.}(2017)\citenamefont {Zohar},
  \citenamefont {Farace}, \citenamefont {Reznik},\ and\ \citenamefont
  {Cirac}}]{Zohar:2016iic}%
  \BibitemOpen
  \bibfield  {author} {\bibinfo {author} {\bibfnamefont {E.}~\bibnamefont
  {Zohar}}, \bibinfo {author} {\bibfnamefont {A.}~\bibnamefont {Farace}},
  \bibinfo {author} {\bibfnamefont {B.}~\bibnamefont {Reznik}},\ and\ \bibinfo
  {author} {\bibfnamefont {J.~I.}\ \bibnamefont {Cirac}},\ }\bibfield  {title}
  {\bibinfo {title} {{Digital lattice gauge theories}},\ }\href
  {https://doi.org/10.1103/PhysRevA.95.023604} {\bibfield  {journal} {\bibinfo
  {journal} {Phys. Rev.}\ }\textbf {\bibinfo {volume} {A95}},\ \bibinfo {pages}
  {023604} (\bibinfo {year} {2017})},\ \Eprint
  {https://arxiv.org/abs/1607.08121} {arXiv:1607.08121 [quant-ph]} \BibitemShut
  {NoStop}%
%%CITATION = ARXIV:1607.08121;%%
\bibitem [{\citenamefont {Klco}\ \emph {et~al.}(2020)\citenamefont {Klco},
  \citenamefont {Stryker},\ and\ \citenamefont {Savage}}]{Klco:2019evd}%
  \BibitemOpen
  \bibfield  {author} {\bibinfo {author} {\bibfnamefont {N.}~\bibnamefont
  {Klco}}, \bibinfo {author} {\bibfnamefont {J.~R.}\ \bibnamefont {Stryker}},\
  and\ \bibinfo {author} {\bibfnamefont {M.~J.}\ \bibnamefont {Savage}},\
  }\bibfield  {title} {\bibinfo {title} {{SU(2) non-Abelian gauge field theory
  in one dimension on digital quantum computers}},\ }\href
  {https://doi.org/10.1103/PhysRevD.101.074512} {\bibfield  {journal} {\bibinfo
   {journal} {Phys. Rev. D}\ }\textbf {\bibinfo {volume} {101}},\ \bibinfo
  {pages} {074512} (\bibinfo {year} {2020})},\ \Eprint
  {https://arxiv.org/abs/1908.06935} {arXiv:1908.06935 [quant-ph]} \BibitemShut
  {NoStop}%
\bibitem [{\citenamefont {Ciavarella}\ \emph {et~al.}(2021)\citenamefont
  {Ciavarella}, \citenamefont {Klco},\ and\ \citenamefont
  {Savage}}]{Ciavarella:2021nmj}%
  \BibitemOpen
  \bibfield  {author} {\bibinfo {author} {\bibfnamefont {A.}~\bibnamefont
  {Ciavarella}}, \bibinfo {author} {\bibfnamefont {N.}~\bibnamefont {Klco}},\
  and\ \bibinfo {author} {\bibfnamefont {M.~J.}\ \bibnamefont {Savage}},\
  }\href@noop {} {\bibinfo {title} {{A Trailhead for Quantum Simulation of
  SU(3) Yang-Mills Lattice Gauge Theory in the Local Multiplet Basis}}}
  (\bibinfo {year} {2021}),\ \Eprint {https://arxiv.org/abs/2101.10227}
  {arXiv:2101.10227 [quant-ph]} \BibitemShut {NoStop}%
\bibitem [{\citenamefont {Bender}\ \emph {et~al.}(2018)\citenamefont {Bender},
  \citenamefont {Zohar}, \citenamefont {Farace},\ and\ \citenamefont
  {Cirac}}]{Bender:2018rdp}%
  \BibitemOpen
  \bibfield  {author} {\bibinfo {author} {\bibfnamefont {J.}~\bibnamefont
  {Bender}}, \bibinfo {author} {\bibfnamefont {E.}~\bibnamefont {Zohar}},
  \bibinfo {author} {\bibfnamefont {A.}~\bibnamefont {Farace}},\ and\ \bibinfo
  {author} {\bibfnamefont {J.~I.}\ \bibnamefont {Cirac}},\ }\bibfield  {title}
  {\bibinfo {title} {{Digital quantum simulation of lattice gauge theories in
  three spatial dimensions}},\ }\href
  {https://doi.org/10.1088/1367-2630/aadb71} {\bibfield  {journal} {\bibinfo
  {journal} {New J. Phys.}\ }\textbf {\bibinfo {volume} {20}},\ \bibinfo
  {pages} {093001} (\bibinfo {year} {2018})},\ \Eprint
  {https://arxiv.org/abs/1804.02082} {arXiv:1804.02082 [quant-ph]} \BibitemShut
  {NoStop}%
%%CITATION = ARXIV:1804.02082;%%
\bibitem [{\citenamefont {Liu}\ and\ \citenamefont {Xin}(2020)}]{Liu:2020eoa}%
  \BibitemOpen
  \bibfield  {author} {\bibinfo {author} {\bibfnamefont {J.}~\bibnamefont
  {Liu}}\ and\ \bibinfo {author} {\bibfnamefont {Y.}~\bibnamefont {Xin}},\
  }\href@noop {} {\bibinfo {title} {{Quantum simulation of quantum field
  theories as quantum chemistry}}} (\bibinfo {year} {2020}),\ \Eprint
  {https://arxiv.org/abs/2004.13234} {arXiv:2004.13234 [hep-th]} \BibitemShut
  {NoStop}%
\bibitem [{\citenamefont {Hackett}\ \emph {et~al.}(2019)\citenamefont
  {Hackett}, \citenamefont {Howe}, \citenamefont {Hughes}, \citenamefont {Jay},
  \citenamefont {Neil},\ and\ \citenamefont {Simone}}]{Hackett:2018cel}%
  \BibitemOpen
  \bibfield  {author} {\bibinfo {author} {\bibfnamefont {D.~C.}\ \bibnamefont
  {Hackett}}, \bibinfo {author} {\bibfnamefont {K.}~\bibnamefont {Howe}},
  \bibinfo {author} {\bibfnamefont {C.}~\bibnamefont {Hughes}}, \bibinfo
  {author} {\bibfnamefont {W.}~\bibnamefont {Jay}}, \bibinfo {author}
  {\bibfnamefont {E.~T.}\ \bibnamefont {Neil}},\ and\ \bibinfo {author}
  {\bibfnamefont {J.~N.}\ \bibnamefont {Simone}},\ }\bibfield  {title}
  {\bibinfo {title} {{Digitizing Gauge Fields: Lattice Monte Carlo Results for
  Future Quantum Computers}},\ }\href
  {https://doi.org/10.1103/PhysRevA.99.062341} {\bibfield  {journal} {\bibinfo
  {journal} {Phys.\ Rev.\ A}\ }\textbf {\bibinfo {volume} {99}},\ \bibinfo
  {pages} {062341} (\bibinfo {year} {2019})},\ \Eprint
  {https://arxiv.org/abs/1811.03629} {arXiv:1811.03629 [quant-ph]} \BibitemShut
  {NoStop}%
\bibitem [{\citenamefont {Alexandru}\ \emph {et~al.}(2019)\citenamefont
  {Alexandru}, \citenamefont {Bedaque}, \citenamefont {Harmalkar},
  \citenamefont {Lamm}, \citenamefont {Lawrence},\ and\ \citenamefont
  {Warrington}}]{Alexandru:2019nsa}%
  \BibitemOpen
  \bibfield  {author} {\bibinfo {author} {\bibfnamefont {A.}~\bibnamefont
  {Alexandru}}, \bibinfo {author} {\bibfnamefont {P.~F.}\ \bibnamefont
  {Bedaque}}, \bibinfo {author} {\bibfnamefont {S.}~\bibnamefont {Harmalkar}},
  \bibinfo {author} {\bibfnamefont {H.}~\bibnamefont {Lamm}}, \bibinfo {author}
  {\bibfnamefont {S.}~\bibnamefont {Lawrence}},\ and\ \bibinfo {author}
  {\bibfnamefont {N.~C.}\ \bibnamefont {Warrington}} (\bibinfo {collaboration}
  {NuQS}),\ }\bibfield  {title} {\bibinfo {title} {Gluon field digitization for
  quantum computers},\ }\href {https://doi.org/10.1103/PhysRevD.100.114501}
  {\bibfield  {journal} {\bibinfo  {journal} {Phys.Rev.D}\ }\textbf {\bibinfo
  {volume} {100}},\ \bibinfo {pages} {114501} (\bibinfo {year} {2019})},\
  \Eprint {https://arxiv.org/abs/1906.11213} {arXiv:1906.11213 [hep-lat]}
  \BibitemShut {NoStop}%
\bibitem [{\citenamefont {Yamamoto}(2021)}]{Yamamoto:2020eqi}%
  \BibitemOpen
  \bibfield  {author} {\bibinfo {author} {\bibfnamefont {A.}~\bibnamefont
  {Yamamoto}},\ }\bibfield  {title} {\bibinfo {title} {{Real-time simulation of
  (2+1)-dimensional lattice gauge theory on qubits}},\ }\href
  {https://doi.org/10.1093/ptep/ptaa171} {\bibfield  {journal} {\bibinfo
  {journal} {PTEP}\ }\textbf {\bibinfo {volume} {2021}},\ \bibinfo {pages}
  {013B06} (\bibinfo {year} {2021})},\ \Eprint
  {https://arxiv.org/abs/2008.11395} {arXiv:2008.11395 [hep-lat]} \BibitemShut
  {NoStop}%
\bibitem [{\citenamefont {Haase}\ \emph {et~al.}(2021)\citenamefont {Haase},
  \citenamefont {Dellantonio}, \citenamefont {Celi}, \citenamefont {Paulson},
  \citenamefont {Kan}, \citenamefont {Jansen},\ and\ \citenamefont
  {Muschik}}]{Haase:2020kaj}%
  \BibitemOpen
  \bibfield  {author} {\bibinfo {author} {\bibfnamefont {J.~F.}\ \bibnamefont
  {Haase}}, \bibinfo {author} {\bibfnamefont {L.}~\bibnamefont {Dellantonio}},
  \bibinfo {author} {\bibfnamefont {A.}~\bibnamefont {Celi}}, \bibinfo {author}
  {\bibfnamefont {D.}~\bibnamefont {Paulson}}, \bibinfo {author} {\bibfnamefont
  {A.}~\bibnamefont {Kan}}, \bibinfo {author} {\bibfnamefont {K.}~\bibnamefont
  {Jansen}},\ and\ \bibinfo {author} {\bibfnamefont {C.~A.}\ \bibnamefont
  {Muschik}},\ }\bibfield  {title} {\bibinfo {title} {{A resource efficient
  approach for quantum and classical simulations of gauge theories in particle
  physics}},\ }\href {https://doi.org/10.22331/q-2021-02-04-393} {\bibfield
  {journal} {\bibinfo  {journal} {Quantum}\ }\textbf {\bibinfo {volume} {5}},\
  \bibinfo {pages} {393} (\bibinfo {year} {2021})},\ \Eprint
  {https://arxiv.org/abs/2006.14160} {arXiv:2006.14160 [quant-ph]} \BibitemShut
  {NoStop}%
\bibitem [{\citenamefont {Armon}\ \emph {et~al.}(2021)\citenamefont {Armon},
  \citenamefont {Ashkenazi}, \citenamefont {Garc\'\i{}a-Moreno}, \citenamefont
  {Gonz\'alez-Tudela},\ and\ \citenamefont {Zohar}}]{Armon:2021uqr}%
  \BibitemOpen
  \bibfield  {author} {\bibinfo {author} {\bibfnamefont {T.}~\bibnamefont
  {Armon}}, \bibinfo {author} {\bibfnamefont {S.}~\bibnamefont {Ashkenazi}},
  \bibinfo {author} {\bibfnamefont {G.}~\bibnamefont {Garc\'\i{}a-Moreno}},
  \bibinfo {author} {\bibfnamefont {A.}~\bibnamefont {Gonz\'alez-Tudela}},\
  and\ \bibinfo {author} {\bibfnamefont {E.}~\bibnamefont {Zohar}},\
  }\href@noop {} {\bibinfo {title} {{Photon-mediated Stroboscopic Quantum
  Simulation of a $\mathbb{Z}_{2}$ Lattice Gauge Theory}}} (\bibinfo {year}
  {2021}),\ \Eprint {https://arxiv.org/abs/2107.13024} {arXiv:2107.13024
  [quant-ph]} \BibitemShut {NoStop}%
\bibitem [{\citenamefont {Bazavov}\ \emph {et~al.}(2019)\citenamefont
  {Bazavov}, \citenamefont {Catterall}, \citenamefont {Jha},\ and\
  \citenamefont {Unmuth-Yockey}}]{PhysRevD.99.114507}%
  \BibitemOpen
  \bibfield  {author} {\bibinfo {author} {\bibfnamefont {A.}~\bibnamefont
  {Bazavov}}, \bibinfo {author} {\bibfnamefont {S.}~\bibnamefont {Catterall}},
  \bibinfo {author} {\bibfnamefont {R.~G.}\ \bibnamefont {Jha}},\ and\ \bibinfo
  {author} {\bibfnamefont {J.}~\bibnamefont {Unmuth-Yockey}},\ }\bibfield
  {title} {\bibinfo {title} {Tensor renormalization group study of the
  non-abelian higgs model in two dimensions},\ }\href
  {https://doi.org/10.1103/PhysRevD.99.114507} {\bibfield  {journal} {\bibinfo
  {journal} {Phys. Rev. D}\ }\textbf {\bibinfo {volume} {99}},\ \bibinfo
  {pages} {114507} (\bibinfo {year} {2019})}\BibitemShut {NoStop}%
\bibitem [{\citenamefont {Bazavov}\ \emph {et~al.}(2015)\citenamefont
  {Bazavov}, \citenamefont {Meurice}, \citenamefont {Tsai}, \citenamefont
  {Unmuth-Yockey},\ and\ \citenamefont {Zhang}}]{Bazavov:2015kka}%
  \BibitemOpen
  \bibfield  {author} {\bibinfo {author} {\bibfnamefont {A.}~\bibnamefont
  {Bazavov}}, \bibinfo {author} {\bibfnamefont {Y.}~\bibnamefont {Meurice}},
  \bibinfo {author} {\bibfnamefont {S.-W.}\ \bibnamefont {Tsai}}, \bibinfo
  {author} {\bibfnamefont {J.}~\bibnamefont {Unmuth-Yockey}},\ and\ \bibinfo
  {author} {\bibfnamefont {J.}~\bibnamefont {Zhang}},\ }\bibfield  {title}
  {\bibinfo {title} {{Gauge-invariant implementation of the Abelian Higgs model
  on optical lattices}},\ }\href {https://doi.org/10.1103/PhysRevD.92.076003}
  {\bibfield  {journal} {\bibinfo  {journal} {Phys. Rev.}\ }\textbf {\bibinfo
  {volume} {D92}},\ \bibinfo {pages} {076003} (\bibinfo {year} {2015})},\
  \Eprint {https://arxiv.org/abs/1503.08354} {arXiv:1503.08354 [hep-lat]}
  \BibitemShut {NoStop}%
%%CITATION = ARXIV:1503.08354;%%
\bibitem [{\citenamefont {Zhang}\ \emph {et~al.}(2018)\citenamefont {Zhang},
  \citenamefont {Unmuth-Yockey}, \citenamefont {Zeiher}, \citenamefont
  {Bazavov}, \citenamefont {Tsai},\ and\ \citenamefont
  {Meurice}}]{Zhang:2018ufj}%
  \BibitemOpen
  \bibfield  {author} {\bibinfo {author} {\bibfnamefont {J.}~\bibnamefont
  {Zhang}}, \bibinfo {author} {\bibfnamefont {J.}~\bibnamefont
  {Unmuth-Yockey}}, \bibinfo {author} {\bibfnamefont {J.}~\bibnamefont
  {Zeiher}}, \bibinfo {author} {\bibfnamefont {A.}~\bibnamefont {Bazavov}},
  \bibinfo {author} {\bibfnamefont {S.~W.}\ \bibnamefont {Tsai}},\ and\
  \bibinfo {author} {\bibfnamefont {Y.}~\bibnamefont {Meurice}},\ }\bibfield
  {title} {\bibinfo {title} {{Quantum simulation of the universal features of
  the Polyakov loop}},\ }\href {https://doi.org/10.1103/PhysRevLett.121.223201}
  {\bibfield  {journal} {\bibinfo  {journal} {Phys. Rev. Lett.}\ }\textbf
  {\bibinfo {volume} {121}},\ \bibinfo {pages} {223201} (\bibinfo {year}
  {2018})},\ \Eprint {https://arxiv.org/abs/1803.11166} {arXiv:1803.11166
  [hep-lat]} \BibitemShut {NoStop}%
%%CITATION = ARXIV:1803.11166;%%
\bibitem [{\citenamefont {Unmuth-Yockey}\ \emph {et~al.}(2018)\citenamefont
  {Unmuth-Yockey}, \citenamefont {Zhang}, \citenamefont {Bazavov},
  \citenamefont {Meurice},\ and\ \citenamefont {Tsai}}]{Unmuth-Yockey:2018ugm}%
  \BibitemOpen
  \bibfield  {author} {\bibinfo {author} {\bibfnamefont {J.}~\bibnamefont
  {Unmuth-Yockey}}, \bibinfo {author} {\bibfnamefont {J.}~\bibnamefont
  {Zhang}}, \bibinfo {author} {\bibfnamefont {A.}~\bibnamefont {Bazavov}},
  \bibinfo {author} {\bibfnamefont {Y.}~\bibnamefont {Meurice}},\ and\ \bibinfo
  {author} {\bibfnamefont {S.-W.}\ \bibnamefont {Tsai}},\ }\bibfield  {title}
  {\bibinfo {title} {{Universal features of the Abelian Polyakov loop in 1+1
  dimensions}},\ }\href {https://doi.org/10.1103/PhysRevD.98.094511} {\bibfield
   {journal} {\bibinfo  {journal} {Phys. Rev.}\ }\textbf {\bibinfo {volume}
  {D98}},\ \bibinfo {pages} {094511} (\bibinfo {year} {2018})},\ \Eprint
  {https://arxiv.org/abs/1807.09186} {arXiv:1807.09186 [hep-lat]} \BibitemShut
  {NoStop}%
%%CITATION = ARXIV:1807.09186;%%
\bibitem [{\citenamefont {Unmuth-Yockey}(2019)}]{Unmuth-Yockey:2018xak}%
  \BibitemOpen
  \bibfield  {author} {\bibinfo {author} {\bibfnamefont {J.~F.}\ \bibnamefont
  {Unmuth-Yockey}},\ }\bibfield  {title} {\bibinfo {title} {{Gauge-invariant
  rotor Hamiltonian from dual variables of 3D $U(1)$ gauge theory}},\ }\href
  {https://doi.org/10.1103/PhysRevD.99.074502} {\bibfield  {journal} {\bibinfo
  {journal} {Phys.\ Rev.\ D}\ }\textbf {\bibinfo {volume} {99}},\ \bibinfo
  {pages} {074502} (\bibinfo {year} {2019})},\ \Eprint
  {https://arxiv.org/abs/1811.05884} {arXiv:1811.05884 [hep-lat]} \BibitemShut
  {NoStop}%
\bibitem [{\citenamefont {Kreshchuk}\ \emph
  {et~al.}(2020{\natexlab{a}})\citenamefont {Kreshchuk}, \citenamefont {Kirby},
  \citenamefont {Goldstein}, \citenamefont {Beauchemin},\ and\ \citenamefont
  {Love}}]{Kreshchuk:2020dla}%
  \BibitemOpen
  \bibfield  {author} {\bibinfo {author} {\bibfnamefont {M.}~\bibnamefont
  {Kreshchuk}}, \bibinfo {author} {\bibfnamefont {W.~M.}\ \bibnamefont
  {Kirby}}, \bibinfo {author} {\bibfnamefont {G.}~\bibnamefont {Goldstein}},
  \bibinfo {author} {\bibfnamefont {H.}~\bibnamefont {Beauchemin}},\ and\
  \bibinfo {author} {\bibfnamefont {P.~J.}\ \bibnamefont {Love}},\ }\href@noop
  {} {\bibinfo {title} {{Quantum Simulation of Quantum Field Theory in the
  Light-Front Formulation}}} (\bibinfo {year} {2020}{\natexlab{a}}),\ \Eprint
  {https://arxiv.org/abs/2002.04016} {arXiv:2002.04016 [quant-ph]} \BibitemShut
  {NoStop}%
\bibitem [{\citenamefont {Kreshchuk}\ \emph
  {et~al.}(2020{\natexlab{b}})\citenamefont {Kreshchuk}, \citenamefont {Jia},
  \citenamefont {Kirby}, \citenamefont {Goldstein}, \citenamefont {Vary},\ and\
  \citenamefont {Love}}]{Kreshchuk:2020aiq}%
  \BibitemOpen
  \bibfield  {author} {\bibinfo {author} {\bibfnamefont {M.}~\bibnamefont
  {Kreshchuk}}, \bibinfo {author} {\bibfnamefont {S.}~\bibnamefont {Jia}},
  \bibinfo {author} {\bibfnamefont {W.~M.}\ \bibnamefont {Kirby}}, \bibinfo
  {author} {\bibfnamefont {G.}~\bibnamefont {Goldstein}}, \bibinfo {author}
  {\bibfnamefont {J.~P.}\ \bibnamefont {Vary}},\ and\ \bibinfo {author}
  {\bibfnamefont {P.~J.}\ \bibnamefont {Love}},\ }\href@noop {} {\bibinfo
  {title} {{Simulating Hadronic Physics on NISQ devices using Basis Light-Front
  Quantization}}} (\bibinfo {year} {2020}{\natexlab{b}}),\ \Eprint
  {https://arxiv.org/abs/2011.13443} {arXiv:2011.13443 [quant-ph]} \BibitemShut
  {NoStop}%
\bibitem [{\citenamefont {Raychowdhury}\ and\ \citenamefont
  {Stryker}(2018)}]{Raychowdhury:2018osk}%
  \BibitemOpen
  \bibfield  {author} {\bibinfo {author} {\bibfnamefont {I.}~\bibnamefont
  {Raychowdhury}}\ and\ \bibinfo {author} {\bibfnamefont {J.~R.}\ \bibnamefont
  {Stryker}},\ }\href@noop {} {\bibinfo {title} {{Solving Gauss's Law on
  Digital Quantum Computers with Loop-String-Hadron Digitization}}} (\bibinfo
  {year} {2018}),\ \Eprint {https://arxiv.org/abs/1812.07554} {arXiv:1812.07554
  [hep-lat]} \BibitemShut {NoStop}%
\bibitem [{\citenamefont {Raychowdhury}\ and\ \citenamefont
  {Stryker}(2020)}]{Raychowdhury:2019iki}%
  \BibitemOpen
  \bibfield  {author} {\bibinfo {author} {\bibfnamefont {I.}~\bibnamefont
  {Raychowdhury}}\ and\ \bibinfo {author} {\bibfnamefont {J.~R.}\ \bibnamefont
  {Stryker}},\ }\bibfield  {title} {\bibinfo {title} {{Loop, String, and Hadron
  Dynamics in SU(2) Hamiltonian Lattice Gauge Theories}},\ }\href
  {https://doi.org/10.1103/PhysRevD.101.114502} {\bibfield  {journal} {\bibinfo
   {journal} {Phys. Rev. D}\ }\textbf {\bibinfo {volume} {101}},\ \bibinfo
  {pages} {114502} (\bibinfo {year} {2020})},\ \Eprint
  {https://arxiv.org/abs/1912.06133} {arXiv:1912.06133 [hep-lat]} \BibitemShut
  {NoStop}%
\bibitem [{\citenamefont {Davoudi}\ \emph {et~al.}(2020)\citenamefont
  {Davoudi}, \citenamefont {Raychowdhury},\ and\ \citenamefont
  {Shaw}}]{Davoudi:2020yln}%
  \BibitemOpen
  \bibfield  {author} {\bibinfo {author} {\bibfnamefont {Z.}~\bibnamefont
  {Davoudi}}, \bibinfo {author} {\bibfnamefont {I.}~\bibnamefont
  {Raychowdhury}},\ and\ \bibinfo {author} {\bibfnamefont {A.}~\bibnamefont
  {Shaw}},\ }\href@noop {} {\bibinfo {title} {{Search for Efficient
  Formulations for Hamiltonian Simulation of non-Abelian Lattice Gauge
  Theories}}} (\bibinfo {year} {2020}),\ \Eprint
  {https://arxiv.org/abs/2009.11802} {arXiv:2009.11802 [hep-lat]} \BibitemShut
  {NoStop}%
\bibitem [{\citenamefont {Wiese}(2014)}]{Wiese:2014rla}%
  \BibitemOpen
  \bibfield  {author} {\bibinfo {author} {\bibfnamefont {U.-J.}\ \bibnamefont
  {Wiese}},\ }\bibfield  {title} {\bibinfo {title} {{Towards Quantum Simulating
  QCD}},\ }\bibfield  {booktitle} {\emph {\bibinfo {booktitle} {{Proceedings,
  24th International Conference on Ultra-Relativistic Nucleus-Nucleus
  Collisions (Quark Matter 2014): Darmstadt, Germany, May 19-24, 2014}}},\
  }\href {https://doi.org/10.1016/j.nuclphysa.2014.09.102} {\bibfield
  {journal} {\bibinfo  {journal} {Nucl. Phys.}\ }\textbf {\bibinfo {volume}
  {A931}},\ \bibinfo {pages} {246} (\bibinfo {year} {2014})},\ \Eprint
  {https://arxiv.org/abs/1409.7414} {arXiv:1409.7414 [hep-th]} \BibitemShut
  {NoStop}%
%%CITATION = ARXIV:1409.7414;%%
\bibitem [{\citenamefont {Luo}\ \emph {et~al.}(2019)\citenamefont {Luo},
  \citenamefont {Shen}, \citenamefont {Highman}, \citenamefont {Clark},
  \citenamefont {DeMarco}, \citenamefont {El-Khadra},\ and\ \citenamefont
  {Gadway}}]{Luo:2019vmi}%
  \BibitemOpen
  \bibfield  {author} {\bibinfo {author} {\bibfnamefont {D.}~\bibnamefont
  {Luo}}, \bibinfo {author} {\bibfnamefont {J.}~\bibnamefont {Shen}}, \bibinfo
  {author} {\bibfnamefont {M.}~\bibnamefont {Highman}}, \bibinfo {author}
  {\bibfnamefont {B.~K.}\ \bibnamefont {Clark}}, \bibinfo {author}
  {\bibfnamefont {B.}~\bibnamefont {DeMarco}}, \bibinfo {author} {\bibfnamefont
  {A.~X.}\ \bibnamefont {El-Khadra}},\ and\ \bibinfo {author} {\bibfnamefont
  {B.}~\bibnamefont {Gadway}},\ }\href@noop {} {\bibinfo {title} {{A Framework
  for Simulating Gauge Theories with Dipolar Spin Systems}}} (\bibinfo {year}
  {2019}),\ \Eprint {https://arxiv.org/abs/1912.11488} {arXiv:1912.11488
  [quant-ph]} \BibitemShut {NoStop}%
\bibitem [{\citenamefont {Brower}\ \emph {et~al.}(2019)\citenamefont {Brower},
  \citenamefont {Berenstein},\ and\ \citenamefont {Kawai}}]{Brower:2020huh}%
  \BibitemOpen
  \bibfield  {author} {\bibinfo {author} {\bibfnamefont {R.~C.}\ \bibnamefont
  {Brower}}, \bibinfo {author} {\bibfnamefont {D.}~\bibnamefont {Berenstein}},\
  and\ \bibinfo {author} {\bibfnamefont {H.}~\bibnamefont {Kawai}},\ }\bibfield
   {title} {\bibinfo {title} {{Lattice Gauge Theory for a Quantum Computer}},\
  }\href@noop {} {\bibfield  {journal} {\bibinfo  {journal} {PoS}\ }\textbf
  {\bibinfo {volume} {LATTICE2019}},\ \bibinfo {pages} {112} (\bibinfo {year}
  {2019})},\ \Eprint {https://arxiv.org/abs/2002.10028} {arXiv:2002.10028
  [hep-lat]} \BibitemShut {NoStop}%
\bibitem [{\citenamefont {Mathis}\ \emph {et~al.}(2020)\citenamefont {Mathis},
  \citenamefont {Mazzola},\ and\ \citenamefont {Tavernelli}}]{Mathis:2020fuo}%
  \BibitemOpen
  \bibfield  {author} {\bibinfo {author} {\bibfnamefont {S.~V.}\ \bibnamefont
  {Mathis}}, \bibinfo {author} {\bibfnamefont {G.}~\bibnamefont {Mazzola}},\
  and\ \bibinfo {author} {\bibfnamefont {I.}~\bibnamefont {Tavernelli}},\
  }\bibfield  {title} {\bibinfo {title} {{Toward scalable simulations of
  Lattice Gauge Theories on quantum computers}},\ }\href
  {https://doi.org/10.1103/PhysRevD.102.094501} {\bibfield  {journal} {\bibinfo
   {journal} {Phys. Rev. D}\ }\textbf {\bibinfo {volume} {102}},\ \bibinfo
  {pages} {094501} (\bibinfo {year} {2020})},\ \Eprint
  {https://arxiv.org/abs/2005.10271} {arXiv:2005.10271 [quant-ph]} \BibitemShut
  {NoStop}%
\bibitem [{\citenamefont {Singh}(2019)}]{Singh:2019jog}%
  \BibitemOpen
  \bibfield  {author} {\bibinfo {author} {\bibfnamefont {H.}~\bibnamefont
  {Singh}},\ }\href@noop {} {\bibinfo {title} {{Qubit $O(N)$ nonlinear sigma
  models}}} (\bibinfo {year} {2019}),\ \Eprint
  {https://arxiv.org/abs/1911.12353} {arXiv:1911.12353 [hep-lat]} \BibitemShut
  {NoStop}%
\bibitem [{\citenamefont {Singh}\ and\ \citenamefont
  {Chandrasekharan}(2019)}]{Singh:2019uwd}%
  \BibitemOpen
  \bibfield  {author} {\bibinfo {author} {\bibfnamefont {H.}~\bibnamefont
  {Singh}}\ and\ \bibinfo {author} {\bibfnamefont {S.}~\bibnamefont
  {Chandrasekharan}},\ }\bibfield  {title} {\bibinfo {title} {{Qubit
  regularization of the $O(3)$ sigma model}},\ }\href
  {https://doi.org/10.1103/PhysRevD.100.054505} {\bibfield  {journal} {\bibinfo
   {journal} {Phys. Rev. D}\ }\textbf {\bibinfo {volume} {100}},\ \bibinfo
  {pages} {054505} (\bibinfo {year} {2019})},\ \Eprint
  {https://arxiv.org/abs/1905.13204} {arXiv:1905.13204 [hep-lat]} \BibitemShut
  {NoStop}%
\bibitem [{\citenamefont {Buser}\ \emph {et~al.}(2020)\citenamefont {Buser},
  \citenamefont {Bhattacharya}, \citenamefont {Cincio},\ and\ \citenamefont
  {Gupta}}]{Buser:2020uzs}%
  \BibitemOpen
  \bibfield  {author} {\bibinfo {author} {\bibfnamefont {A.~J.}\ \bibnamefont
  {Buser}}, \bibinfo {author} {\bibfnamefont {T.}~\bibnamefont {Bhattacharya}},
  \bibinfo {author} {\bibfnamefont {L.}~\bibnamefont {Cincio}},\ and\ \bibinfo
  {author} {\bibfnamefont {R.}~\bibnamefont {Gupta}},\ }\href@noop {} {\bibinfo
  {title} {{Quantum simulation of the qubit-regularized O(3)-sigma model}}}
  (\bibinfo {year} {2020}),\ \Eprint {https://arxiv.org/abs/2006.15746}
  {arXiv:2006.15746 [quant-ph]} \BibitemShut {NoStop}%
\bibitem [{\citenamefont {Bhattacharya}\ \emph {et~al.}(2020)\citenamefont
  {Bhattacharya}, \citenamefont {Buser}, \citenamefont {Chandrasekharan},
  \citenamefont {Gupta},\ and\ \citenamefont {Singh}}]{Bhattacharya:2020gpm}%
  \BibitemOpen
  \bibfield  {author} {\bibinfo {author} {\bibfnamefont {T.}~\bibnamefont
  {Bhattacharya}}, \bibinfo {author} {\bibfnamefont {A.~J.}\ \bibnamefont
  {Buser}}, \bibinfo {author} {\bibfnamefont {S.}~\bibnamefont
  {Chandrasekharan}}, \bibinfo {author} {\bibfnamefont {R.}~\bibnamefont
  {Gupta}},\ and\ \bibinfo {author} {\bibfnamefont {H.}~\bibnamefont {Singh}},\
  }\href@noop {} {\bibinfo {title} {{Qubit regularization of asymptotic
  freedom}}} (\bibinfo {year} {2020}),\ \Eprint
  {https://arxiv.org/abs/2012.02153} {arXiv:2012.02153 [hep-lat]} \BibitemShut
  {NoStop}%
\bibitem [{\citenamefont {Barata}\ \emph {et~al.}(2020)\citenamefont {Barata},
  \citenamefont {Mueller}, \citenamefont {Tarasov},\ and\ \citenamefont
  {Venugopalan}}]{Barata:2020jtq}%
  \BibitemOpen
  \bibfield  {author} {\bibinfo {author} {\bibfnamefont {J.~a.}\ \bibnamefont
  {Barata}}, \bibinfo {author} {\bibfnamefont {N.}~\bibnamefont {Mueller}},
  \bibinfo {author} {\bibfnamefont {A.}~\bibnamefont {Tarasov}},\ and\ \bibinfo
  {author} {\bibfnamefont {R.}~\bibnamefont {Venugopalan}},\ }\href@noop {}
  {\bibinfo {title} {{Single-particle digitization strategy for quantum
  computation of a $\phi^4$ scalar field theory}}} (\bibinfo {year} {2020}),\
  \Eprint {https://arxiv.org/abs/2012.00020} {arXiv:2012.00020 [hep-th]}
  \BibitemShut {NoStop}%
\bibitem [{\citenamefont {Kreshchuk}\ \emph
  {et~al.}(2020{\natexlab{c}})\citenamefont {Kreshchuk}, \citenamefont {Jia},
  \citenamefont {Kirby}, \citenamefont {Goldstein}, \citenamefont {Vary},\ and\
  \citenamefont {Love}}]{Kreshchuk:2020kcz}%
  \BibitemOpen
  \bibfield  {author} {\bibinfo {author} {\bibfnamefont {M.}~\bibnamefont
  {Kreshchuk}}, \bibinfo {author} {\bibfnamefont {S.}~\bibnamefont {Jia}},
  \bibinfo {author} {\bibfnamefont {W.~M.}\ \bibnamefont {Kirby}}, \bibinfo
  {author} {\bibfnamefont {G.}~\bibnamefont {Goldstein}}, \bibinfo {author}
  {\bibfnamefont {J.~P.}\ \bibnamefont {Vary}},\ and\ \bibinfo {author}
  {\bibfnamefont {P.~J.}\ \bibnamefont {Love}},\ }\href@noop {} {\bibinfo
  {title} {{Light-Front Field Theory on Current Quantum Computers}}} (\bibinfo
  {year} {2020}{\natexlab{c}}),\ \Eprint {https://arxiv.org/abs/2009.07885}
  {arXiv:2009.07885 [quant-ph]} \BibitemShut {NoStop}%
\bibitem [{\citenamefont {Ji}\ \emph {et~al.}(2020)\citenamefont {Ji},
  \citenamefont {Lamm},\ and\ \citenamefont {Zhu}}]{Ji:2020kjk}%
  \BibitemOpen
  \bibfield  {author} {\bibinfo {author} {\bibfnamefont {Y.}~\bibnamefont
  {Ji}}, \bibinfo {author} {\bibfnamefont {H.}~\bibnamefont {Lamm}},\ and\
  \bibinfo {author} {\bibfnamefont {S.}~\bibnamefont {Zhu}} (\bibinfo
  {collaboration} {NuQS}),\ }\bibfield  {title} {\bibinfo {title} {{Gluon Field
  Digitization via Group Space Decimation for Quantum Computers}},\ }\href
  {https://doi.org/10.1103/PhysRevD.102.114513} {\bibfield  {journal} {\bibinfo
   {journal} {Phys. Rev. D}\ }\textbf {\bibinfo {volume} {102}},\ \bibinfo
  {pages} {114513} (\bibinfo {year} {2020})},\ \Eprint
  {https://arxiv.org/abs/2005.14221} {arXiv:2005.14221 [hep-lat]} \BibitemShut
  {NoStop}%
\bibitem [{\citenamefont {Bauer}\ and\ \citenamefont
  {Grabowska}(2021)}]{Bauer:2021gek}%
  \BibitemOpen
  \bibfield  {author} {\bibinfo {author} {\bibfnamefont {C.~W.}\ \bibnamefont
  {Bauer}}\ and\ \bibinfo {author} {\bibfnamefont {D.~M.}\ \bibnamefont
  {Grabowska}},\ }\href@noop {} {\bibinfo {title} {{Efficient Representation
  for Simulating U(1) Gauge Theories on Digital Quantum Computers at All Values
  of the Coupling}}} (\bibinfo {year} {2021}),\ \Eprint
  {https://arxiv.org/abs/2111.08015} {arXiv:2111.08015 [hep-ph]} \BibitemShut
  {NoStop}%
\bibitem [{\citenamefont {Gustafson}(2021)}]{Gustafson:2021qbt}%
  \BibitemOpen
  \bibfield  {author} {\bibinfo {author} {\bibfnamefont {E.}~\bibnamefont
  {Gustafson}},\ }\bibfield  {title} {\bibinfo {title} {{Prospects for
  Simulating a Qudit Based Model of (1+1)d Scalar QED}},\ }\href
  {https://doi.org/10.1103/PhysRevD.103.114505} {\bibfield  {journal} {\bibinfo
   {journal} {Phys. Rev. D}\ }\textbf {\bibinfo {volume} {103}},\ \bibinfo
  {pages} {114505} (\bibinfo {year} {2021})},\ \Eprint
  {https://arxiv.org/abs/2104.10136} {arXiv:2104.10136 [quant-ph]} \BibitemShut
  {NoStop}%
\bibitem [{\citenamefont {Hartung}\ \emph {et~al.}(2022)\citenamefont
  {Hartung}, \citenamefont {Jakobs}, \citenamefont {Jansen}, \citenamefont
  {Ostmeyer},\ and\ \citenamefont {Urbach}}]{Hartung:2022hoz}%
  \BibitemOpen
  \bibfield  {author} {\bibinfo {author} {\bibfnamefont {T.}~\bibnamefont
  {Hartung}}, \bibinfo {author} {\bibfnamefont {T.}~\bibnamefont {Jakobs}},
  \bibinfo {author} {\bibfnamefont {K.}~\bibnamefont {Jansen}}, \bibinfo
  {author} {\bibfnamefont {J.}~\bibnamefont {Ostmeyer}},\ and\ \bibinfo
  {author} {\bibfnamefont {C.}~\bibnamefont {Urbach}},\ }\bibfield  {title}
  {\bibinfo {title} {{Digitising SU(2) gauge fields and the freezing
  transition}},\ }\href {https://doi.org/10.1140/epjc/s10052-022-10192-5}
  {\bibfield  {journal} {\bibinfo  {journal} {Eur. Phys. J. C}\ }\textbf
  {\bibinfo {volume} {82}},\ \bibinfo {pages} {237} (\bibinfo {year} {2022})},\
  \Eprint {https://arxiv.org/abs/2201.09625} {arXiv:2201.09625 [hep-lat]}
  \BibitemShut {NoStop}%
\bibitem [{\citenamefont {Grabowska}\ \emph {et~al.}(2022)\citenamefont
  {Grabowska}, \citenamefont {Kane}, \citenamefont {Nachman},\ and\
  \citenamefont {Bauer}}]{Grabowska:2022uos}%
  \BibitemOpen
  \bibfield  {author} {\bibinfo {author} {\bibfnamefont {D.~M.}\ \bibnamefont
  {Grabowska}}, \bibinfo {author} {\bibfnamefont {C.}~\bibnamefont {Kane}},
  \bibinfo {author} {\bibfnamefont {B.}~\bibnamefont {Nachman}},\ and\ \bibinfo
  {author} {\bibfnamefont {C.~W.}\ \bibnamefont {Bauer}},\ }\href@noop {}
  {\bibinfo {title} {{Overcoming exponential scaling with system size in
  Trotter-Suzuki implementations of constrained Hamiltonians: 2+1 U(1) lattice
  gauge theories}}} (\bibinfo {year} {2022}),\ \Eprint
  {https://arxiv.org/abs/2208.03333} {arXiv:2208.03333 [quant-ph]} \BibitemShut
  {NoStop}%
\bibitem [{\citenamefont {Murairi}\ \emph {et~al.}(2022)\citenamefont
  {Murairi}, \citenamefont {Cervia}, \citenamefont {Kumar}, \citenamefont
  {Bedaque},\ and\ \citenamefont {Alexandru}}]{Murairi:2022zdg}%
  \BibitemOpen
  \bibfield  {author} {\bibinfo {author} {\bibfnamefont {E.~M.}\ \bibnamefont
  {Murairi}}, \bibinfo {author} {\bibfnamefont {M.~J.}\ \bibnamefont {Cervia}},
  \bibinfo {author} {\bibfnamefont {H.}~\bibnamefont {Kumar}}, \bibinfo
  {author} {\bibfnamefont {P.~F.}\ \bibnamefont {Bedaque}},\ and\ \bibinfo
  {author} {\bibfnamefont {A.}~\bibnamefont {Alexandru}},\ }\href@noop {}
  {\bibinfo {title} {{How many quantum gates do gauge theories require?}}}
  (\bibinfo {year} {2022}),\ \Eprint {https://arxiv.org/abs/2208.11789}
  {arXiv:2208.11789 [hep-lat]} \BibitemShut {NoStop}%
\bibitem [{\citenamefont {Ji}\ \emph {et~al.}(2022)\citenamefont {Ji},
  \citenamefont {Lamm},\ and\ \citenamefont {Zhu}}]{Ji:2022qvr}%
  \BibitemOpen
  \bibfield  {author} {\bibinfo {author} {\bibfnamefont {Y.}~\bibnamefont
  {Ji}}, \bibinfo {author} {\bibfnamefont {H.}~\bibnamefont {Lamm}},\ and\
  \bibinfo {author} {\bibfnamefont {S.}~\bibnamefont {Zhu}},\ }\href@noop {}
  {\bibinfo {title} {{Gluon Digitization via Character Expansion for Quantum
  Computers}}} (\bibinfo {year} {2022}),\ \Eprint
  {https://arxiv.org/abs/2203.02330} {arXiv:2203.02330 [hep-lat]} \BibitemShut
  {NoStop}%
\bibitem [{\citenamefont {Alexandru}\ \emph {et~al.}(2021)\citenamefont
  {Alexandru}, \citenamefont {Bedaque}, \citenamefont {Brett},\ and\
  \citenamefont {Lamm}}]{Alexandru:2021jpm}%
  \BibitemOpen
  \bibfield  {author} {\bibinfo {author} {\bibfnamefont {A.}~\bibnamefont
  {Alexandru}}, \bibinfo {author} {\bibfnamefont {P.~F.}\ \bibnamefont
  {Bedaque}}, \bibinfo {author} {\bibfnamefont {R.}~\bibnamefont {Brett}},\
  and\ \bibinfo {author} {\bibfnamefont {H.}~\bibnamefont {Lamm}},\ }\href@noop
  {} {\bibinfo {title} {{The spectrum of qubitized QCD: glueballs in a
  $S(1080)$ gauge theory}}} (\bibinfo {year} {2021}),\ \Eprint
  {https://arxiv.org/abs/2112.08482} {arXiv:2112.08482 [hep-lat]} \BibitemShut
  {NoStop}%
\bibitem [{\citenamefont {Gustafson}(2022)}]{Gustafson:2022xlj}%
  \BibitemOpen
  \bibfield  {author} {\bibinfo {author} {\bibfnamefont {E.}~\bibnamefont
  {Gustafson}},\ }\href@noop {} {\bibinfo {title} {{Noise Improvements in
  Quantum Simulations of sQED using Qutrits}}} (\bibinfo {year} {2022}),\
  \Eprint {https://arxiv.org/abs/2201.04546} {arXiv:2201.04546 [quant-ph]}
  \BibitemShut {NoStop}%
\bibitem [{\citenamefont {Gustafson}\ \emph {et~al.}(2019)\citenamefont
  {Gustafson}, \citenamefont {Meurice},\ and\ \citenamefont
  {Unmuth-Yockey}}]{Gustafson:2019mpk}%
  \BibitemOpen
  \bibfield  {author} {\bibinfo {author} {\bibfnamefont {E.}~\bibnamefont
  {Gustafson}}, \bibinfo {author} {\bibfnamefont {Y.}~\bibnamefont {Meurice}},\
  and\ \bibinfo {author} {\bibfnamefont {J.}~\bibnamefont {Unmuth-Yockey}},\
  }\href@noop {} {\bibinfo {title} {{Quantum simulation of scattering in the
  quantum Ising model}}} (\bibinfo {year} {2019}),\ \Eprint
  {https://arxiv.org/abs/1901.05944} {arXiv:1901.05944 [hep-lat]} \BibitemShut
  {NoStop}%
%%CITATION = ARXIV:1901.05944;%%
\bibitem [{\citenamefont {Lamm}\ \emph {et~al.}(2019)\citenamefont {Lamm},
  \citenamefont {Lawrence},\ and\ \citenamefont {Yamauchi}}]{Lamm:2019bik}%
  \BibitemOpen
  \bibfield  {author} {\bibinfo {author} {\bibfnamefont {H.}~\bibnamefont
  {Lamm}}, \bibinfo {author} {\bibfnamefont {S.}~\bibnamefont {Lawrence}},\
  and\ \bibinfo {author} {\bibfnamefont {Y.}~\bibnamefont {Yamauchi}} (\bibinfo
  {collaboration} {NuQS}),\ }\bibfield  {title} {\bibinfo {title} {{General
  Methods for Digital Quantum Simulation of Gauge Theories}},\ }\href
  {https://doi.org/10.1103/PhysRevD.100.034518} {\bibfield  {journal} {\bibinfo
   {journal} {Phys. Rev.}\ }\textbf {\bibinfo {volume} {D100}},\ \bibinfo
  {pages} {034518} (\bibinfo {year} {2019})},\ \Eprint
  {https://arxiv.org/abs/1903.08807} {arXiv:1903.08807 [hep-lat]} \BibitemShut
  {NoStop}%
%%CITATION = ARXIV:1903.08807;%%
\bibitem [{\citenamefont {Alam}\ \emph {et~al.}(2022)\citenamefont {Alam},
  \citenamefont {Hadfield}, \citenamefont {Lamm},\ and\ \citenamefont
  {Li}}]{Alam:2021uuq}%
  \BibitemOpen
  \bibfield  {author} {\bibinfo {author} {\bibfnamefont {M.~S.}\ \bibnamefont
  {Alam}}, \bibinfo {author} {\bibfnamefont {S.}~\bibnamefont {Hadfield}},
  \bibinfo {author} {\bibfnamefont {H.}~\bibnamefont {Lamm}},\ and\ \bibinfo
  {author} {\bibfnamefont {A.~C.~Y.}\ \bibnamefont {Li}} (\bibinfo
  {collaboration} {SQMS}),\ }\bibfield  {title} {\bibinfo {title} {{Primitive
  quantum gates for dihedral gauge theories}},\ }\href
  {https://doi.org/10.1103/PhysRevD.105.114501} {\bibfield  {journal} {\bibinfo
   {journal} {Phys. Rev. D}\ }\textbf {\bibinfo {volume} {105}},\ \bibinfo
  {pages} {114501} (\bibinfo {year} {2022})},\ \Eprint
  {https://arxiv.org/abs/2108.13305} {arXiv:2108.13305 [quant-ph]} \BibitemShut
  {NoStop}%
\bibitem [{\citenamefont {Fromm}\ \emph {et~al.}(2022)\citenamefont {Fromm},
  \citenamefont {Philipsen},\ and\ \citenamefont {Winterowd}}]{Fromm:2022vaj}%
  \BibitemOpen
  \bibfield  {author} {\bibinfo {author} {\bibfnamefont {M.}~\bibnamefont
  {Fromm}}, \bibinfo {author} {\bibfnamefont {O.}~\bibnamefont {Philipsen}},\
  and\ \bibinfo {author} {\bibfnamefont {C.}~\bibnamefont {Winterowd}},\
  }\href@noop {} {\bibinfo {title} {{Dihedral Lattice Gauge Theories on a
  Quantum Annealer}}} (\bibinfo {year} {2022}),\ \Eprint
  {https://arxiv.org/abs/2206.14679} {arXiv:2206.14679 [hep-lat]} \BibitemShut
  {NoStop}%
\bibitem [{\citenamefont {Yeter-Aydeniz}\ \emph {et~al.}(2018)\citenamefont
  {Yeter-Aydeniz}, \citenamefont {Dumitrescu}, \citenamefont {McCaskey},
  \citenamefont {Bennink}, \citenamefont {Pooser},\ and\ \citenamefont
  {Siopsis}}]{Yeter-Aydeniz:2018mix}%
  \BibitemOpen
  \bibfield  {author} {\bibinfo {author} {\bibfnamefont {K.}~\bibnamefont
  {Yeter-Aydeniz}}, \bibinfo {author} {\bibfnamefont {E.~F.}\ \bibnamefont
  {Dumitrescu}}, \bibinfo {author} {\bibfnamefont {A.~J.}\ \bibnamefont
  {McCaskey}}, \bibinfo {author} {\bibfnamefont {R.~S.}\ \bibnamefont
  {Bennink}}, \bibinfo {author} {\bibfnamefont {R.~C.}\ \bibnamefont
  {Pooser}},\ and\ \bibinfo {author} {\bibfnamefont {G.}~\bibnamefont
  {Siopsis}},\ }\href@noop {} {\bibinfo {title} {{Scalar Quantum Field Theories
  as a Benchmark for Near-Term Quantum Computers}}} (\bibinfo {year} {2018}),\
  \Eprint {https://arxiv.org/abs/1811.12332} {arXiv:1811.12332 [quant-ph]}
  \BibitemShut {NoStop}%
%%CITATION = ARXIV:1811.12332;%%
\bibitem [{\citenamefont {Gustafson}\ \emph {et~al.}(2022)\citenamefont
  {Gustafson}, \citenamefont {Lamm}, \citenamefont {Lovelace},\ and\
  \citenamefont {Musk}}]{Gustafson:2022xdt}%
  \BibitemOpen
  \bibfield  {author} {\bibinfo {author} {\bibfnamefont {E.~J.}\ \bibnamefont
  {Gustafson}}, \bibinfo {author} {\bibfnamefont {H.}~\bibnamefont {Lamm}},
  \bibinfo {author} {\bibfnamefont {F.}~\bibnamefont {Lovelace}},\ and\
  \bibinfo {author} {\bibfnamefont {D.}~\bibnamefont {Musk}},\ }\bibfield
  {title} {\bibinfo {title} {{Primitive quantum gates for an SU(2) discrete
  subgroup: Binary tetrahedral}},\ }\href
  {https://doi.org/10.1103/PhysRevD.106.114501} {\bibfield  {journal} {\bibinfo
   {journal} {Phys. Rev. D}\ }\textbf {\bibinfo {volume} {106}},\ \bibinfo
  {pages} {114501} (\bibinfo {year} {2022})},\ \Eprint
  {https://arxiv.org/abs/2208.12309} {arXiv:2208.12309 [quant-ph]} \BibitemShut
  {NoStop}%
\bibitem [{\citenamefont {Carena}\ \emph {et~al.}(2022)\citenamefont {Carena},
  \citenamefont {Lamm}, \citenamefont {Li},\ and\ \citenamefont
  {Liu}}]{Carena:2022kpg}%
  \BibitemOpen
  \bibfield  {author} {\bibinfo {author} {\bibfnamefont {M.}~\bibnamefont
  {Carena}}, \bibinfo {author} {\bibfnamefont {H.}~\bibnamefont {Lamm}},
  \bibinfo {author} {\bibfnamefont {Y.-Y.}\ \bibnamefont {Li}},\ and\ \bibinfo
  {author} {\bibfnamefont {W.}~\bibnamefont {Liu}},\ }\href@noop {} {\bibinfo
  {title} {{Improved Hamiltonians for Quantum Simulations}}} (\bibinfo {year}
  {2022}),\ \Eprint {https://arxiv.org/abs/2203.02823} {arXiv:2203.02823
  [hep-lat]} \BibitemShut {NoStop}%
\bibitem [{\citenamefont {Tang}(2019)}]{10.1145/3313276.3316310}%
  \BibitemOpen
  \bibfield  {author} {\bibinfo {author} {\bibfnamefont {E.}~\bibnamefont
  {Tang}},\ }\bibfield  {title} {\bibinfo {title} {A quantum-inspired classical
  algorithm for recommendation systems},\ }in\ \href
  {https://doi.org/10.1145/3313276.3316310} {\emph {\bibinfo {booktitle}
  {Proceedings of the 51st Annual ACM SIGACT Symposium on Theory of
  Computing}}},\ \bibinfo {series and number} {STOC 2019}\ (\bibinfo
  {publisher} {Association for Computing Machinery},\ \bibinfo {address} {New
  York, NY, USA},\ \bibinfo {year} {2019})\ p.\ \bibinfo {pages}
  {217–228}\BibitemShut {NoStop}%
\bibitem [{\citenamefont {Arrazola}\ \emph {et~al.}(2020)\citenamefont
  {Arrazola}, \citenamefont {Delgado}, \citenamefont {Bardhan},\ and\
  \citenamefont {Lloyd}}]{Arrazola2020quantuminspired}%
  \BibitemOpen
  \bibfield  {author} {\bibinfo {author} {\bibfnamefont {J.~M.}\ \bibnamefont
  {Arrazola}}, \bibinfo {author} {\bibfnamefont {A.}~\bibnamefont {Delgado}},
  \bibinfo {author} {\bibfnamefont {B.~R.}\ \bibnamefont {Bardhan}},\ and\
  \bibinfo {author} {\bibfnamefont {S.}~\bibnamefont {Lloyd}},\ }\bibfield
  {title} {\bibinfo {title} {Quantum-inspired algorithms in practice},\ }\href
  {https://doi.org/10.22331/q-2020-08-13-307} {\bibfield  {journal} {\bibinfo
  {journal} {{Quantum}}\ }\textbf {\bibinfo {volume} {4}},\ \bibinfo {pages}
  {307} (\bibinfo {year} {2020})}\BibitemShut {NoStop}%
\bibitem [{\citenamefont {Kothari}\ and\ \citenamefont
  {O'Donnell}(2022)}]{kothari:2022}%
  \BibitemOpen
  \bibfield  {author} {\bibinfo {author} {\bibfnamefont {R.}~\bibnamefont
  {Kothari}}\ and\ \bibinfo {author} {\bibfnamefont {R.}~\bibnamefont
  {O'Donnell}},\ }\href@noop {} {\bibinfo {title} {{Mean estimation when you
  have the source code; or, quantum Monte Carlo methods}}} (\bibinfo {year}
  {2022}),\ \Eprint {https://arxiv.org/abs/2208.07544} {arXiv:2208.07544
  [quant-ph]} \BibitemShut {NoStop}%
\bibitem [{\citenamefont {Shyamsundar}(2021)}]{Prasanth:2021}%
  \BibitemOpen
  \bibfield  {author} {\bibinfo {author} {\bibfnamefont {P.}~\bibnamefont
  {Shyamsundar}},\ }\href@noop {} {\bibinfo {title} {Non-boolean quantum
  amplitude amplification and quantum mean estimation}} (\bibinfo {year}
  {2021}),\ \Eprint {https://arxiv.org/abs/2102.04975} {arXiv:2102.04975
  [quant-ph]} \BibitemShut {NoStop}%
\bibitem [{\citenamefont {Hamoudi}(2021)}]{Ham21}%
  \BibitemOpen
  \bibfield  {author} {\bibinfo {author} {\bibfnamefont {Y.}~\bibnamefont
  {Hamoudi}},\ }\emph {\bibinfo {title} {Quantum Algorithms for the Monte Carlo
  Method}},\ \href@noop {} {Ph.D. thesis},\ \bibinfo  {school} {Universit\'{e}
  de Paris} (\bibinfo {year} {2021})\BibitemShut {NoStop}%
\bibitem [{\citenamefont {Dob\ifmmode \check{s}\else
  \v{s}\fi{}\'{\i}\ifmmode~\check{c}\else \v{c}\fi{}ek}\ \emph
  {et~al.}(2007)\citenamefont {Dob\ifmmode \check{s}\else
  \v{s}\fi{}\'{\i}\ifmmode~\check{c}\else \v{c}\fi{}ek}, \citenamefont
  {Johansson}, \citenamefont {Shumeiko},\ and\ \citenamefont
  {Wendin}}]{PhysRevA.76.030306}%
  \BibitemOpen
  \bibfield  {author} {\bibinfo {author} {\bibfnamefont {M.}~\bibnamefont
  {Dob\ifmmode \check{s}\else \v{s}\fi{}\'{\i}\ifmmode~\check{c}\else
  \v{c}\fi{}ek}}, \bibinfo {author} {\bibfnamefont {G.}~\bibnamefont
  {Johansson}}, \bibinfo {author} {\bibfnamefont {V.}~\bibnamefont
  {Shumeiko}},\ and\ \bibinfo {author} {\bibfnamefont {G.}~\bibnamefont
  {Wendin}},\ }\bibfield  {title} {\bibinfo {title} {Arbitrary accuracy
  iterative quantum phase estimation algorithm using a single ancillary qubit:
  A two-qubit benchmark},\ }\href {https://doi.org/10.1103/PhysRevA.76.030306}
  {\bibfield  {journal} {\bibinfo  {journal} {Phys. Rev. A}\ }\textbf {\bibinfo
  {volume} {76}},\ \bibinfo {pages} {030306} (\bibinfo {year}
  {2007})}\BibitemShut {NoStop}%
\bibitem [{\citenamefont {Grinko}\ \emph {et~al.}(2021)\citenamefont {Grinko},
  \citenamefont {Gacon}, \citenamefont {Zoufal},\ and\ \citenamefont
  {Woerner}}]{Grinko_2021}%
  \BibitemOpen
  \bibfield  {author} {\bibinfo {author} {\bibfnamefont {D.}~\bibnamefont
  {Grinko}}, \bibinfo {author} {\bibfnamefont {J.}~\bibnamefont {Gacon}},
  \bibinfo {author} {\bibfnamefont {C.}~\bibnamefont {Zoufal}},\ and\ \bibinfo
  {author} {\bibfnamefont {S.}~\bibnamefont {Woerner}},\ }\bibfield  {title}
  {\bibinfo {title} {Iterative quantum amplitude estimation},\ }\bibfield
  {journal} {\bibinfo  {journal} {npj Quantum Information}\ }\textbf {\bibinfo
  {volume} {7}},\ \href {https://doi.org/10.1038/s41534-021-00379-1}
  {10.1038/s41534-021-00379-1} (\bibinfo {year} {2021})\BibitemShut {NoStop}%
\bibitem [{\citenamefont {Montanaro}(2015)}]{montanaro:2015}%
  \BibitemOpen
  \bibfield  {author} {\bibinfo {author} {\bibfnamefont {A.}~\bibnamefont
  {Montanaro}},\ }\bibfield  {title} {\bibinfo {title} {Quantum speedup of
  monte carlo methods},\ }\href {https://doi.org/10.1098/rspa.2015.0301}
  {\bibfield  {journal} {\bibinfo  {journal} {Proceedings of the Royal Society
  A: Mathematical, Physical and Engineering Sciences}\ }\textbf {\bibinfo
  {volume} {471}},\ \bibinfo {pages} {20150301} (\bibinfo {year}
  {2015})}\BibitemShut {NoStop}%
\bibitem [{\citenamefont {Brassard}\ \emph {et~al.}(2002)\citenamefont
  {Brassard}, \citenamefont {Hoyer}, \citenamefont {Mosca},\ and\ \citenamefont
  {Tapp}}]{Brassard_2002}%
  \BibitemOpen
  \bibfield  {author} {\bibinfo {author} {\bibfnamefont {G.}~\bibnamefont
  {Brassard}}, \bibinfo {author} {\bibfnamefont {P.}~\bibnamefont {Hoyer}},
  \bibinfo {author} {\bibfnamefont {M.}~\bibnamefont {Mosca}},\ and\ \bibinfo
  {author} {\bibfnamefont {A.}~\bibnamefont {Tapp}},\ }\bibfield  {title}
  {\bibinfo {title} {Quantum amplitude amplification and estimation},\ }\href
  {https://doi.org/10.1090/conm/305/05215} {\bibfield  {journal} {\bibinfo
  {journal} {Contemporary Mathematics}\ }\textbf {\bibinfo {volume} {305}},\
  \bibinfo {pages} {53} (\bibinfo {year} {2002})}\BibitemShut {NoStop}%
\bibitem [{\citenamefont {Grover}\ and\ \citenamefont
  {Rudolph}(2002)}]{grover:2002}%
  \BibitemOpen
  \bibfield  {author} {\bibinfo {author} {\bibfnamefont {L.}~\bibnamefont
  {Grover}}\ and\ \bibinfo {author} {\bibfnamefont {T.}~\bibnamefont
  {Rudolph}},\ }\href {https://doi.org/10.48550/ARXIV.QUANT-PH/0208112}
  {\bibinfo {title} {Creating superpositions that correspond to efficiently
  integrable probability distributions}} (\bibinfo {year} {2002})\BibitemShut
  {NoStop}%
\bibitem [{\citenamefont {Wang}\ \emph {et~al.}(2022)\citenamefont {Wang},
  \citenamefont {Wang}, \citenamefont {He}, \citenamefont {Shi}, \citenamefont
  {Cui}, \citenamefont {Shang}, \citenamefont {Li}, \citenamefont {Li},
  \citenamefont {Li}, \citenamefont {Wei},\ and\ \citenamefont
  {Gu}}]{Wang_2022}%
  \BibitemOpen
  \bibfield  {author} {\bibinfo {author} {\bibfnamefont {S.}~\bibnamefont
  {Wang}}, \bibinfo {author} {\bibfnamefont {Z.}~\bibnamefont {Wang}}, \bibinfo
  {author} {\bibfnamefont {R.}~\bibnamefont {He}}, \bibinfo {author}
  {\bibfnamefont {S.}~\bibnamefont {Shi}}, \bibinfo {author} {\bibfnamefont
  {G.}~\bibnamefont {Cui}}, \bibinfo {author} {\bibfnamefont {R.}~\bibnamefont
  {Shang}}, \bibinfo {author} {\bibfnamefont {J.}~\bibnamefont {Li}}, \bibinfo
  {author} {\bibfnamefont {Y.}~\bibnamefont {Li}}, \bibinfo {author}
  {\bibfnamefont {W.}~\bibnamefont {Li}}, \bibinfo {author} {\bibfnamefont
  {Z.}~\bibnamefont {Wei}},\ and\ \bibinfo {author} {\bibfnamefont
  {Y.}~\bibnamefont {Gu}},\ }\bibfield  {title} {\bibinfo {title}
  {Inverse-coefficient black-box quantum state preparation},\ }\href
  {https://doi.org/10.1088/1367-2630/ac93a8} {\bibfield  {journal} {\bibinfo
  {journal} {New Journal of Physics}\ }\textbf {\bibinfo {volume} {24}},\
  \bibinfo {pages} {103004} (\bibinfo {year} {2022})}\BibitemShut {NoStop}%
\bibitem [{\citenamefont {Bausch}(2022)}]{Bausch_2022}%
  \BibitemOpen
  \bibfield  {author} {\bibinfo {author} {\bibfnamefont {J.}~\bibnamefont
  {Bausch}},\ }\bibfield  {title} {\bibinfo {title} {Fast black-box quantum
  state preparation},\ }\href {https://doi.org/10.22331/q-2022-08-04-773}
  {\bibfield  {journal} {\bibinfo  {journal} {Quantum}\ }\textbf {\bibinfo
  {volume} {6}},\ \bibinfo {pages} {773} (\bibinfo {year} {2022})}\BibitemShut
  {NoStop}%
\bibitem [{\citenamefont {Sanders}\ \emph {et~al.}(2019)\citenamefont
  {Sanders}, \citenamefont {Low}, \citenamefont {Scherer},\ and\ \citenamefont
  {Berry}}]{PhysRevLett.122.020502}%
  \BibitemOpen
  \bibfield  {author} {\bibinfo {author} {\bibfnamefont {Y.~R.}\ \bibnamefont
  {Sanders}}, \bibinfo {author} {\bibfnamefont {G.~H.}\ \bibnamefont {Low}},
  \bibinfo {author} {\bibfnamefont {A.}~\bibnamefont {Scherer}},\ and\ \bibinfo
  {author} {\bibfnamefont {D.~W.}\ \bibnamefont {Berry}},\ }\bibfield  {title}
  {\bibinfo {title} {Black-box quantum state preparation without arithmetic},\
  }\href {https://doi.org/10.1103/PhysRevLett.122.020502} {\bibfield  {journal}
  {\bibinfo  {journal} {Phys. Rev. Lett.}\ }\textbf {\bibinfo {volume} {122}},\
  \bibinfo {pages} {020502} (\bibinfo {year} {2019})}\BibitemShut {NoStop}%
\bibitem [{\citenamefont {McArdle}\ \emph {et~al.}(2022)\citenamefont
  {McArdle}, \citenamefont {Gily\'en},\ and\ \citenamefont
  {Berta}}]{mcardle:2022}%
  \BibitemOpen
  \bibfield  {author} {\bibinfo {author} {\bibfnamefont {S.}~\bibnamefont
  {McArdle}}, \bibinfo {author} {\bibfnamefont {A.}~\bibnamefont {Gily\'en}},\
  and\ \bibinfo {author} {\bibfnamefont {M.}~\bibnamefont {Berta}},\
  }\href@noop {} {\bibinfo {title} {{Quantum state preparation without coherent
  arithmetic}}} (\bibinfo {year} {2022}),\ \Eprint
  {https://arxiv.org/abs/2210.14892} {arXiv:2210.14892 [quant-ph]} \BibitemShut
  {NoStop}%
\bibitem [{\citenamefont {Grover}(2000)}]{PhysRevLett.85.1334}%
  \BibitemOpen
  \bibfield  {author} {\bibinfo {author} {\bibfnamefont {L.~K.}\ \bibnamefont
  {Grover}},\ }\bibfield  {title} {\bibinfo {title} {Synthesis of quantum
  superpositions by quantum computation},\ }\href
  {https://doi.org/10.1103/PhysRevLett.85.1334} {\bibfield  {journal} {\bibinfo
   {journal} {Phys. Rev. Lett.}\ }\textbf {\bibinfo {volume} {85}},\ \bibinfo
  {pages} {1334} (\bibinfo {year} {2000})}\BibitemShut {NoStop}%
\bibitem [{\citenamefont {Yoder}\ \emph {et~al.}(2014)\citenamefont {Yoder},
  \citenamefont {Low},\ and\ \citenamefont {Chuang}}]{yoder:2014}%
  \BibitemOpen
  \bibfield  {author} {\bibinfo {author} {\bibfnamefont {T.~J.}\ \bibnamefont
  {Yoder}}, \bibinfo {author} {\bibfnamefont {G.~H.}\ \bibnamefont {Low}},\
  and\ \bibinfo {author} {\bibfnamefont {I.~L.}\ \bibnamefont {Chuang}},\
  }\bibfield  {title} {\bibinfo {title} {Fixed-point quantum search with an
  optimal number of queries},\ }\href
  {https://doi.org/10.1103/PhysRevLett.113.210501} {\bibfield  {journal}
  {\bibinfo  {journal} {Phys. Rev. Lett.}\ }\textbf {\bibinfo {volume} {113}},\
  \bibinfo {pages} {210501} (\bibinfo {year} {2014})}\BibitemShut {NoStop}%
\bibitem [{\citenamefont {Berry}\ \emph {et~al.}(2014)\citenamefont {Berry},
  \citenamefont {Childs}, \citenamefont {Cleve}, \citenamefont {Kothari},\ and\
  \citenamefont {Somma}}]{Berry:2014}%
  \BibitemOpen
  \bibfield  {author} {\bibinfo {author} {\bibfnamefont {D.~W.}\ \bibnamefont
  {Berry}}, \bibinfo {author} {\bibfnamefont {A.~M.}\ \bibnamefont {Childs}},
  \bibinfo {author} {\bibfnamefont {R.}~\bibnamefont {Cleve}}, \bibinfo
  {author} {\bibfnamefont {R.}~\bibnamefont {Kothari}},\ and\ \bibinfo {author}
  {\bibfnamefont {R.~D.}\ \bibnamefont {Somma}},\ }\bibfield  {title} {\bibinfo
  {title} {Exponential improvement in precision for simulating sparse
  hamiltonians},\ }in\ \href {https://doi.org/10.1145/2591796.2591854} {\emph
  {\bibinfo {booktitle} {Proceedings of the Forty-Sixth Annual ACM Symposium on
  Theory of Computing}}},\ \bibinfo {series and number} {STOC '14}\ (\bibinfo
  {publisher} {Association for Computing Machinery},\ \bibinfo {address} {New
  York, NY, USA},\ \bibinfo {year} {2014})\ p.\ \bibinfo {pages}
  {283–292}\BibitemShut {NoStop}%
\bibitem [{\citenamefont {Gily\'{e}n}\ \emph {et~al.}(2019)\citenamefont
  {Gily\'{e}n}, \citenamefont {Su}, \citenamefont {Low},\ and\ \citenamefont
  {Wiebe}}]{gilyen:2019}%
  \BibitemOpen
  \bibfield  {author} {\bibinfo {author} {\bibfnamefont {A.}~\bibnamefont
  {Gily\'{e}n}}, \bibinfo {author} {\bibfnamefont {Y.}~\bibnamefont {Su}},
  \bibinfo {author} {\bibfnamefont {G.~H.}\ \bibnamefont {Low}},\ and\ \bibinfo
  {author} {\bibfnamefont {N.}~\bibnamefont {Wiebe}},\ }\bibfield  {title}
  {\bibinfo {title} {Quantum singular value transformation and beyond:
  Exponential improvements for quantum matrix arithmetics},\ }in\ \href
  {https://doi.org/10.1145/3313276.3316366} {\emph {\bibinfo {booktitle}
  {Proceedings of the 51st Annual ACM SIGACT Symposium on Theory of
  Computing}}},\ \bibinfo {series and number} {STOC 2019}\ (\bibinfo
  {publisher} {Association for Computing Machinery},\ \bibinfo {address} {New
  York, NY, USA},\ \bibinfo {year} {2019})\ p.\ \bibinfo {pages}
  {193–204}\BibitemShut {NoStop}%
\bibitem [{\citenamefont {Ferrenberg}\ and\ \citenamefont
  {Swendsen}(1988)}]{Ferrenberg:1988yz}%
  \BibitemOpen
  \bibfield  {author} {\bibinfo {author} {\bibfnamefont {A.~M.}\ \bibnamefont
  {Ferrenberg}}\ and\ \bibinfo {author} {\bibfnamefont {R.~H.}\ \bibnamefont
  {Swendsen}},\ }\bibfield  {title} {\bibinfo {title} {{New Monte Carlo
  Technique for Studying Phase Transitions}},\ }\href
  {https://doi.org/10.1103/PhysRevLett.61.2635} {\bibfield  {journal} {\bibinfo
   {journal} {Phys. Rev. Lett.}\ }\textbf {\bibinfo {volume} {61}},\ \bibinfo
  {pages} {2635} (\bibinfo {year} {1988})}\BibitemShut {NoStop}%
\bibitem [{\citenamefont {Ruiz-Perez}\ and\ \citenamefont
  {Garcia-Escartin}(2017)}]{Ruiz-Perez2017}%
  \BibitemOpen
  \bibfield  {author} {\bibinfo {author} {\bibfnamefont {L.}~\bibnamefont
  {Ruiz-Perez}}\ and\ \bibinfo {author} {\bibfnamefont {J.~C.}\ \bibnamefont
  {Garcia-Escartin}},\ }\bibfield  {title} {\bibinfo {title} {Quantum
  arithmetic with the quantum fourier transform},\ }\href
  {https://doi.org/10.1007/s11128-017-1603-1} {\bibfield  {journal} {\bibinfo
  {journal} {Quantum Information Processing}\ }\textbf {\bibinfo {volume}
  {16}},\ \bibinfo {pages} {152} (\bibinfo {year} {2017})}\BibitemShut
  {NoStop}%
\bibitem [{\citenamefont {Seidel}\ \emph {et~al.}(2021)\citenamefont {Seidel},
  \citenamefont {Tcholtchev}, \citenamefont {Bock}, \citenamefont {Becker},\
  and\ \citenamefont {Hauswirth}}]{Seidel:2021}%
  \BibitemOpen
  \bibfield  {author} {\bibinfo {author} {\bibfnamefont {R.}~\bibnamefont
  {Seidel}}, \bibinfo {author} {\bibfnamefont {N.}~\bibnamefont {Tcholtchev}},
  \bibinfo {author} {\bibfnamefont {S.}~\bibnamefont {Bock}}, \bibinfo {author}
  {\bibfnamefont {C.~K.-U.}\ \bibnamefont {Becker}},\ and\ \bibinfo {author}
  {\bibfnamefont {M.}~\bibnamefont {Hauswirth}},\ }\href@noop {} {\bibinfo
  {title} {{Efficient Floating Point Arithmetic for Quantum Computers}}}
  (\bibinfo {year} {2021}),\ \Eprint {https://arxiv.org/abs/2112.10537}
  {arXiv:2112.10537 [quant-ph]} \BibitemShut {NoStop}%
\bibitem [{\citenamefont {Häner}\ \emph {et~al.}(2016)\citenamefont {Häner},
  \citenamefont {Roetteler},\ and\ \citenamefont
  {Svore}}]{https://doi.org/10.48550/arxiv.1611.07995}%
  \BibitemOpen
  \bibfield  {author} {\bibinfo {author} {\bibfnamefont {T.}~\bibnamefont
  {Häner}}, \bibinfo {author} {\bibfnamefont {M.}~\bibnamefont {Roetteler}},\
  and\ \bibinfo {author} {\bibfnamefont {K.~M.}\ \bibnamefont {Svore}},\
  }\href@noop {} {\bibinfo {title} {Factoring using 2n+2 qubits with toffoli
  based modular multiplication}} (\bibinfo {year} {2016}),\ \Eprint
  {https://arxiv.org/abs/1611.07995} {arXiv:1611.07995 [quant-ph]} \BibitemShut
  {NoStop}%
\bibitem [{\citenamefont {Villain}(1975)}]{villain:1975}%
  \BibitemOpen
  \bibfield  {author} {\bibinfo {author} {\bibfnamefont {J.}~\bibnamefont
  {Villain}},\ }\bibfield  {title} {\bibinfo {title} {Theory of one- and
  two-dimensional magnets with an easy magnetization plane. ii. the planar,
  classical, two-dimensional magnet},\ }\href
  {https://doi.org/10.1051/jphys:01975003606058100} {\bibfield  {journal}
  {\bibinfo  {journal} {J. Phys. France}\ }\textbf {\bibinfo {volume} {36}},\
  \bibinfo {pages} {581} (\bibinfo {year} {1975})}\BibitemShut {NoStop}%
\bibitem [{\citenamefont {Itzykson}\ and\ \citenamefont
  {Drouffe}(1989)}]{itzykson_drouffe_1989}%
  \BibitemOpen
  \bibfield  {author} {\bibinfo {author} {\bibfnamefont {C.}~\bibnamefont
  {Itzykson}}\ and\ \bibinfo {author} {\bibfnamefont {J.-M.}\ \bibnamefont
  {Drouffe}},\ }\href {https://doi.org/10.1017/CBO9780511622779} {\emph
  {\bibinfo {title} {Statistical Field Theory}}},\ \bibinfo {series} {Cambridge
  Monographs on Mathematical Physics}, Vol.~\bibinfo {volume} {1}\ (\bibinfo
  {publisher} {Cambridge University Press},\ \bibinfo {year}
  {1989})\BibitemShut {NoStop}%
\bibitem [{\citenamefont {Kanwar}\ and\ \citenamefont
  {Wagman}(2021)}]{Kanwar:2021tkd}%
  \BibitemOpen
  \bibfield  {author} {\bibinfo {author} {\bibfnamefont {G.}~\bibnamefont
  {Kanwar}}\ and\ \bibinfo {author} {\bibfnamefont {M.~L.}\ \bibnamefont
  {Wagman}},\ }\href@noop {} {\bibinfo {title} {{Real-time lattice gauge theory
  actions: unitarity, convergence, and path integral contour deformations}}}
  (\bibinfo {year} {2021}),\ \Eprint {https://arxiv.org/abs/2103.02602}
  {arXiv:2103.02602 [hep-lat]} \BibitemShut {NoStop}%
\bibitem [{\citenamefont {Eastin}\ and\ \citenamefont
  {Knill}(2009)}]{Eastin_2009}%
  \BibitemOpen
  \bibfield  {author} {\bibinfo {author} {\bibfnamefont {B.}~\bibnamefont
  {Eastin}}\ and\ \bibinfo {author} {\bibfnamefont {E.}~\bibnamefont {Knill}},\
  }\bibfield  {title} {\bibinfo {title} {Restrictions on transversal encoded
  quantum gate sets},\ }\href {https://doi.org/10.1103/PhysRevLett.102.110502}
  {\bibfield  {journal} {\bibinfo  {journal} {Phys. Rev. Lett.}\ }\textbf
  {\bibinfo {volume} {102}},\ \bibinfo {pages} {110502} (\bibinfo {year}
  {2009})}\BibitemShut {NoStop}%
\bibitem [{\citenamefont {Steane}(1996)}]{PhysRevLett.77.793}%
  \BibitemOpen
  \bibfield  {author} {\bibinfo {author} {\bibfnamefont {A.~M.}\ \bibnamefont
  {Steane}},\ }\bibfield  {title} {\bibinfo {title} {Error correcting codes in
  quantum theory},\ }\href {https://doi.org/10.1103/PhysRevLett.77.793}
  {\bibfield  {journal} {\bibinfo  {journal} {Phys. Rev. Lett.}\ }\textbf
  {\bibinfo {volume} {77}},\ \bibinfo {pages} {793} (\bibinfo {year}
  {1996})}\BibitemShut {NoStop}%
\bibitem [{\citenamefont {{Steane}}(1996)}]{1996RSPSA.452.2551S}%
  \BibitemOpen
  \bibfield  {author} {\bibinfo {author} {\bibfnamefont {A.}~\bibnamefont
  {{Steane}}},\ }\bibfield  {title} {\bibinfo {title} {{Multiple-Particle
  Interference and Quantum Error Correction}},\ }\href
  {https://doi.org/10.1098/rspa.1996.0136} {\bibfield  {journal} {\bibinfo
  {journal} {Proceedings of the Royal Society of London Series A}\ }\textbf
  {\bibinfo {volume} {452}},\ \bibinfo {pages} {2551} (\bibinfo {year}
  {1996})},\ \Eprint {https://arxiv.org/abs/quant-ph/9601029}
  {arXiv:quant-ph/9601029 [quant-ph]} \BibitemShut {NoStop}%
\bibitem [{\citenamefont {{Calderbank}}\ and\ \citenamefont
  {{Shor}}(1996)}]{1996PhRvA..54.1098C}%
  \BibitemOpen
  \bibfield  {author} {\bibinfo {author} {\bibfnamefont {A.~R.}\ \bibnamefont
  {{Calderbank}}}\ and\ \bibinfo {author} {\bibfnamefont {P.~W.}\ \bibnamefont
  {{Shor}}},\ }\bibfield  {title} {\bibinfo {title} {{Good quantum
  error-correcting codes exist}},\ }\href
  {https://doi.org/10.1103/PhysRevA.54.1098} {\bibfield  {journal} {\bibinfo
  {journal} {\pra}\ }\textbf {\bibinfo {volume} {54}},\ \bibinfo {pages} {1098}
  (\bibinfo {year} {1996})},\ \Eprint {https://arxiv.org/abs/quant-ph/9512032}
  {arXiv:quant-ph/9512032 [quant-ph]} \BibitemShut {NoStop}%
\bibitem [{\citenamefont {Steane}(1996)}]{PhysRevA.54.4741}%
  \BibitemOpen
  \bibfield  {author} {\bibinfo {author} {\bibfnamefont {A.~M.}\ \bibnamefont
  {Steane}},\ }\bibfield  {title} {\bibinfo {title} {Simple quantum
  error-correcting codes},\ }\href {https://doi.org/10.1103/PhysRevA.54.4741}
  {\bibfield  {journal} {\bibinfo  {journal} {Phys. Rev. A}\ }\textbf {\bibinfo
  {volume} {54}},\ \bibinfo {pages} {4741} (\bibinfo {year}
  {1996})}\BibitemShut {NoStop}%
\bibitem [{\citenamefont {{Kitaev}}(1997)}]{1997RuMaS..52.1191K}%
  \BibitemOpen
  \bibfield  {author} {\bibinfo {author} {\bibfnamefont {A.~Y.}\ \bibnamefont
  {{Kitaev}}},\ }\bibfield  {title} {\bibinfo {title} {{Quantum computations:
  algorithms and error correction}},\ }\href
  {https://doi.org/10.1070/RM1997v052n06ABEH002155} {\bibfield  {journal}
  {\bibinfo  {journal} {Russian Mathematical Surveys}\ }\textbf {\bibinfo
  {volume} {52}},\ \bibinfo {pages} {1191} (\bibinfo {year}
  {1997})}\BibitemShut {NoStop}%
\bibitem [{\citenamefont {Mooney}\ \emph {et~al.}(2021)\citenamefont {Mooney},
  \citenamefont {Hill},\ and\ \citenamefont
  {Hollenberg}}]{2020arXiv200505581M}%
  \BibitemOpen
  \bibfield  {author} {\bibinfo {author} {\bibfnamefont {G.~J.}\ \bibnamefont
  {Mooney}}, \bibinfo {author} {\bibfnamefont {C.~D.}\ \bibnamefont {Hill}},\
  and\ \bibinfo {author} {\bibfnamefont {L.~C.~L.}\ \bibnamefont
  {Hollenberg}},\ }\bibfield  {title} {\bibinfo {title} {Cost-optimal
  single-qubit gate synthesis in the {C}lifford hierarchy},\ }\href
  {https://doi.org/10.22331/q-2021-02-15-396} {\bibfield  {journal} {\bibinfo
  {journal} {{Quantum}}\ }\textbf {\bibinfo {volume} {5}},\ \bibinfo {pages}
  {396} (\bibinfo {year} {2021})}\BibitemShut {NoStop}%
\bibitem [{\citenamefont {Clark}\ \emph {et~al.}(2010)\citenamefont {Clark},
  \citenamefont {Babich}, \citenamefont {Barros}, \citenamefont {Brower},\ and\
  \citenamefont {Rebbi}}]{Clark:2009wm}%
  \BibitemOpen
  \bibfield  {author} {\bibinfo {author} {\bibfnamefont {M.~A.}\ \bibnamefont
  {Clark}}, \bibinfo {author} {\bibfnamefont {R.}~\bibnamefont {Babich}},
  \bibinfo {author} {\bibfnamefont {K.}~\bibnamefont {Barros}}, \bibinfo
  {author} {\bibfnamefont {R.~C.}\ \bibnamefont {Brower}},\ and\ \bibinfo
  {author} {\bibfnamefont {C.}~\bibnamefont {Rebbi}},\ }\bibfield  {title}
  {\bibinfo {title} {{Solving Lattice QCD systems of equations using mixed
  precision solvers on GPUs}},\ }\href
  {https://doi.org/10.1016/j.cpc.2010.05.002} {\bibfield  {journal} {\bibinfo
  {journal} {Comput. Phys. Commun.}\ }\textbf {\bibinfo {volume} {181}},\
  \bibinfo {pages} {1517} (\bibinfo {year} {2010})},\ \Eprint
  {https://arxiv.org/abs/0911.3191} {arXiv:0911.3191 [hep-lat]} \BibitemShut
  {NoStop}%
\bibitem [{\citenamefont {van~der Vorst}\ and\ \citenamefont
  {Ye}(2000)}]{doi:10.1137/S1064827599353865}%
  \BibitemOpen
  \bibfield  {author} {\bibinfo {author} {\bibfnamefont {H.~A.}\ \bibnamefont
  {van~der Vorst}}\ and\ \bibinfo {author} {\bibfnamefont {Q.}~\bibnamefont
  {Ye}},\ }\bibfield  {title} {\bibinfo {title} {Residual replacement
  strategies for krylov subspace iterative methods for the convergence of true
  residuals},\ }\href {https://doi.org/10.1137/S1064827599353865} {\bibfield
  {journal} {\bibinfo  {journal} {SIAM Journal on Scientific Computing}\
  }\textbf {\bibinfo {volume} {22}},\ \bibinfo {pages} {835} (\bibinfo {year}
  {2000})}\BibitemShut {NoStop}%
\bibitem [{\citenamefont {Tu}\ \emph {et~al.}(2021)\citenamefont {Tu},
  \citenamefont {Clark}, \citenamefont {Jung},\ and\ \citenamefont
  {Mawhinney}}]{Tu:2021dvv}%
  \BibitemOpen
  \bibfield  {author} {\bibinfo {author} {\bibfnamefont {J.}~\bibnamefont
  {Tu}}, \bibinfo {author} {\bibfnamefont {M.~A.}\ \bibnamefont {Clark}},
  \bibinfo {author} {\bibfnamefont {C.}~\bibnamefont {Jung}},\ and\ \bibinfo
  {author} {\bibfnamefont {R.}~\bibnamefont {Mawhinney}},\ }\href
  {https://doi.org/10.1145/3468267.3470613} {\bibinfo {title} {{Solving DWF
  Dirac Equation Using Multi-splitting Preconditioned Conjugate Gradient with
  Tensor Cores on NVIDIA GPUs}}} (\bibinfo {year} {2021}),\ \Eprint
  {https://arxiv.org/abs/2104.05615} {arXiv:2104.05615 [hep-lat]} \BibitemShut
  {NoStop}%
\bibitem [{\citenamefont {Ishikawa}\ \emph {et~al.}(2023)\citenamefont
  {Ishikawa}, \citenamefont {Kanamori}, \citenamefont {Matsufuru},
  \citenamefont {Miyoshi}, \citenamefont {Mukai}, \citenamefont {Nakamura},
  \citenamefont {Nitadori},\ and\ \citenamefont {Tsuji}}]{Ishikawa:2021iqw}%
  \BibitemOpen
  \bibfield  {author} {\bibinfo {author} {\bibfnamefont {K.-I.}\ \bibnamefont
  {Ishikawa}}, \bibinfo {author} {\bibfnamefont {I.}~\bibnamefont {Kanamori}},
  \bibinfo {author} {\bibfnamefont {H.}~\bibnamefont {Matsufuru}}, \bibinfo
  {author} {\bibfnamefont {I.}~\bibnamefont {Miyoshi}}, \bibinfo {author}
  {\bibfnamefont {Y.}~\bibnamefont {Mukai}}, \bibinfo {author} {\bibfnamefont
  {Y.}~\bibnamefont {Nakamura}}, \bibinfo {author} {\bibfnamefont
  {K.}~\bibnamefont {Nitadori}},\ and\ \bibinfo {author} {\bibfnamefont
  {M.}~\bibnamefont {Tsuji}},\ }\bibfield  {title} {\bibinfo {title} {{102
  PFLOPS lattice QCD quark solver on Fugaku}},\ }\href
  {https://doi.org/10.1016/j.cpc.2022.108510} {\bibfield  {journal} {\bibinfo
  {journal} {Comput. Phys. Commun.}\ }\textbf {\bibinfo {volume} {282}},\
  \bibinfo {pages} {108510} (\bibinfo {year} {2023})},\ \Eprint
  {https://arxiv.org/abs/2109.10687} {arXiv:2109.10687 [hep-lat]} \BibitemShut
  {NoStop}%
\bibitem [{\citenamefont {Bennett}(1973)}]{Bennett:1973}%
  \BibitemOpen
  \bibfield  {author} {\bibinfo {author} {\bibfnamefont {C.~H.}\ \bibnamefont
  {Bennett}},\ }\bibfield  {title} {\bibinfo {title} {Logical reversibility of
  computation},\ }\href {https://doi.org/10.1147/rd.176.0525} {\bibfield
  {journal} {\bibinfo  {journal} {IBM Journal of Research and Development}\
  }\textbf {\bibinfo {volume} {17}},\ \bibinfo {pages} {525} (\bibinfo {year}
  {1973})}\BibitemShut {NoStop}%
\bibitem [{\citenamefont {Mohammadbagherpoor}\ \emph
  {et~al.}(2019)\citenamefont {Mohammadbagherpoor}, \citenamefont {Oh},
  \citenamefont {Dreher}, \citenamefont {Singh}, \citenamefont {Yu},\ and\
  \citenamefont {Rindos}}]{Mohammadbagherpoor:2019}%
  \BibitemOpen
  \bibfield  {author} {\bibinfo {author} {\bibfnamefont {H.}~\bibnamefont
  {Mohammadbagherpoor}}, \bibinfo {author} {\bibfnamefont {Y.-H.}\ \bibnamefont
  {Oh}}, \bibinfo {author} {\bibfnamefont {P.}~\bibnamefont {Dreher}}, \bibinfo
  {author} {\bibfnamefont {A.}~\bibnamefont {Singh}}, \bibinfo {author}
  {\bibfnamefont {X.}~\bibnamefont {Yu}},\ and\ \bibinfo {author}
  {\bibfnamefont {A.~J.}\ \bibnamefont {Rindos}},\ }\href@noop {} {\bibinfo
  {title} {An improved implementation approach for quantum phase estimation on
  quantum computers}} (\bibinfo {year} {2019}),\ \Eprint
  {https://arxiv.org/abs/1910.11696} {arXiv:1910.11696 [quant-ph]} \BibitemShut
  {NoStop}%
\bibitem [{\citenamefont {Chakrabarti}\ \emph {et~al.}(2021)\citenamefont
  {Chakrabarti}, \citenamefont {Krishnakumar}, \citenamefont {Mazzola},
  \citenamefont {Stamatopoulos}, \citenamefont {Woerner},\ and\ \citenamefont
  {Zeng}}]{Chakrabarti_2021}%
  \BibitemOpen
  \bibfield  {author} {\bibinfo {author} {\bibfnamefont {S.}~\bibnamefont
  {Chakrabarti}}, \bibinfo {author} {\bibfnamefont {R.}~\bibnamefont
  {Krishnakumar}}, \bibinfo {author} {\bibfnamefont {G.}~\bibnamefont
  {Mazzola}}, \bibinfo {author} {\bibfnamefont {N.}~\bibnamefont
  {Stamatopoulos}}, \bibinfo {author} {\bibfnamefont {S.}~\bibnamefont
  {Woerner}},\ and\ \bibinfo {author} {\bibfnamefont {W.~J.}\ \bibnamefont
  {Zeng}},\ }\bibfield  {title} {\bibinfo {title} {A threshold for quantum
  advantage in derivative pricing},\ }\href
  {https://doi.org/10.22331/q-2021-06-01-463} {\bibfield  {journal} {\bibinfo
  {journal} {Quantum}\ }\textbf {\bibinfo {volume} {5}},\ \bibinfo {pages}
  {463} (\bibinfo {year} {2021})}\BibitemShut {NoStop}%
\end{thebibliography}
%

\end{document}